\documentstyle[preprint,tighten,aps,floats,epsf,psfig,amssymb]{revtex}
\newcommand\0{\phantom{0}}

\begin{document}
\draft
\title{
Critical behavior of the three-dimensional XY universality class
}

\author{Massimo Campostrini,${}^1$ Martin Hasenbusch,${}^2$
        Andrea Pelissetto,${}^3$\\
        Paolo Rossi,${}^1$ and Ettore Vicari${}^1$ }
\address{${}^1$ Dipartimento di Fisica dell'Universit\`a di Pisa 
and I.N.F.N., I-56126 Pisa, Italy}
\address{${}^2$ Institut f\"ur Physik, Humboldt-Universit\"at
zu Berlin, Invalidenstr.\ 110, D-10115 Berlin, Germany}
\address{${}^3$ Dipartimento di Fisica dell'Universit\`a di Roma I
and I.N.F.N., I-00185 Roma, Italy \\
{{\bf e-mail: }
{\tt Massimo.Campostrini@df.unipi.it}, 
{\tt hasenbus@physik.hu-berlin.de},
{\tt Andrea.Pelissetto@roma1.infn.it},
{\tt Paolo.Rossi@df.unipi.it},
{\tt Ettore.Vicari@df.unipi.it}
}}
\date{\today}

\preprint{IFUP-TH/2000-31, \ HU-EP-00/42, \ Roma1 1303/00}

\maketitle

\begin{abstract}
We improve the theoretical estimates of the critical exponents for the
three-dimensional XY universality class.  We find $\alpha=-0.0146(8)$,
$\gamma=1.3177(5)$, $\nu=0.67155(27)$, $\eta=0.0380(4)$,
$\beta=0.3485(2)$, and $\delta=4.780(2)$.  We observe a discrepancy
with the most recent experimental estimate of $\alpha$; this
discrepancy calls for further theoretical and experimental
investigations.  Our results are obtained by combining Monte Carlo
simulations based on finite-size scaling methods, and high-temperature
expansions.  Two improved models (with suppressed leading scaling
corrections) are selected by Monte Carlo computation.  The critical
exponents are computed from high-temperature expansions specialized to
these improved models.  By the same technique we determine the
coefficients of the small-magnetization expansion of the equation of
state.  This expansion is extended analytically by means of
approximate parametric representations, obtaining the equation of
state in the whole critical region.  We also determine the
specific-heat amplitude ratio.
\end{abstract}


\date{October 24, 2000}

\section{Introduction}
\label{introduction}

In the theory of critical phenomena continuous phase transitions can
be classified into universality classes determined only by a few
properties characterizing the system, such as the space
dimensionality, the range of interaction, the number of components of
the order parameter, and the symmetry.  Renormalization-group (RG)
theory predicts that, within a given universality class, critical
exponents and scaling functions are identical for all systems.  Here
we consider the three-dimensional XY universality class, which is
characterized by a two-component order parameter, ${\rm O}(2)$
symmetry, and short-range interactions.

The superfluid transition of $^4$He, whose order parameter is related
to the complex quantum amplitude of the helium atoms, belongs to the
three-dimensional XY universality class.  It provides an exceptional
opportunity for an experimental test of the RG predictions,
essentially because of the weakness of the singularity in the
compressibility of the fluid, of the purity of the samples, and of the
possibility of performing the experiments, such as the Space Shuttle
experiment reported in \cite{LSNCI-96}, in a microgravity environment,
thereby reducing the gravity-induced broadening of the transition.
Because of these favorable conditions, the specific heat of liquid
helium was accurately measured to within a few nK from the $\lambda$
transition, i.e., very deep in the critical region,
where the scaling corrections to the expected power-law behavior are
small.  The experimental low-temperature data for the specific heat
were analyzed assuming the behavior for $t\equiv (T-T_c)/T_c\to 0$
\begin{equation}
C_H(t) = A |t|^{-\alpha}\left(1 + C |t|^\Delta + D t \right) + B
\end{equation}
with $\Delta = 1/2$.%
\footnote{This value of $\Delta$ is close to the best available
theoretical estimates, i.e., $\Delta=0.529(8)$ from field
theory \cite{GZ-98} and $\Delta=0.531(14)$ from Monte Carlo
simulations \cite{HT-99}.}
This provided the estimate \cite{LSNCI-96,Lipa-etal-00}%
\footnote{
\label{foot:exp}
Ref.\ \cite{LSNCI-96} reported $\alpha=-0.01285(38)$ and
$A^+/A^-=1.054(1)$.  But, as mentioned in footnote [15] of Ref.\ 
\cite{Lipa-etal-00}, the original analysis was slightly in error.
Ref.\ \cite{Lipa-etal-00} reports the new estimates $\alpha=-0.01056$
and $A^+/A^-=1.0442$.  J.~A.~Lipa kindly communicated us
\cite{Lipa-private} that the error on $\alpha$ is the same as in Ref.\ 
\cite{LSNCI-96}.}
\begin{equation}
\alpha = -0.01056(38).
\label{alphaexp}
\end{equation}
This result represents a challenge for theorists because its
uncertainty is substantially smaller than those
of the theoretical calculations.  We mention the best available
theoretical estimates of $\alpha$: $\alpha=-0.0150(17)$ obtained using
high-temperature (HT) expansion techniques \cite{CPRV-00},
$\alpha=-0.0169(33)$ from Monte Carlo (MC) simulations using
finite-size scaling (FSS) techniques \cite{HT-99}, and
$\alpha=-0.011(4)$ from field theory \cite{GZ-98}.

The aim of this paper is to substantially improve the precision of the
theoretical estimates of the critical exponents, reaching an accuracy
comparable with the experimental one.  For this purpose, we will
consider what we call ``improved'' models.  They are characterized by
the fact that the leading correction to scaling is absent in the
expansion of any observable near the critical point.  Moreover, we
will combine MC simulations and analyses of HT series.  We exploit the
effectiveness of MC simulations to determine, by using FSS techniques,
the critical temperature and the parameters of the improved
Hamiltonians, and the effectiveness of HT methods to determine the
critical exponents for improved models, especially when a precise
estimate of $\beta_c$ is available.  Such a combination of lattice
techniques allows us to substantially improve earlier theoretical
estimates.  We indeed obtain
\begin{equation}
\alpha = - 0.0146(8),
\label{finalestimate}
\end{equation}
where, as we will show, the error estimate should be rather
conservative.  The theoretical uncertainty has been substantially
reduced.  We observe a disagreement with the experimental value
(\ref{alphaexp}).  The point to be clarified is whether this
disagreement is significant, or it is due to an underestimate of the
errors reported by us and/or in the experimental papers.  We think
that this discrepancy calls for further theoretical and
experimental investigations.  A new-generation
experiment in microgravity environment is currently in preparation
\cite{Nissen-etal-98}; it should clarify the issue from the experimental
side.

In numerical (HT or MC) determinations of critical quantities,
nonanalytic corrections to the leading scaling behavior represent one
of the major sources of systematic errors.  Considering, for instance,
the magnetic susceptibility, we have
\begin{equation}
\chi = C t^{-\gamma} \left( 1 + a_{0,1} t + a_{0,2}t^2 + ... + a_{1,1}
t^\Delta + a_{1,2} t^{2\Delta} + ... + a_{2,1} t^{\Delta_2} + ... \right).
\label{chiwexp}
\end{equation}
The leading exponent $\gamma$ and the correction-to-scaling exponents
$\Delta,\Delta_2,...$, are universal, while the amplitudes $C$ and
$a_{i,j}$ are nonuniversal.  For three-dimensional XY systems, the
value of the leading correction-to-scaling exponent is
$\Delta\approx 0.53$ \cite{HT-99,GZ-98}, and the value of the
subleading exponent  is $\Delta_2\approx2\Delta$ \cite{NR-84}.

The leading nonanalytic correction $t^\Delta$ is the dominant source
of systematic errors in MC and HT studies.  Indeed, in MC simulations
the presence of this slowly-decreasing term requires careful
extrapolations, increasing the errors in the final estimates.  In HT
studies, nonanalytic corrections introduce large and dangerously
undetectable systematic deviations in the results of the analyses.
Integral approximants \cite{int-appr-ref} (see, e.g., Ref.\ 
\cite{Guttrev} for a review) can in principle cope with an asymptotic
behavior of the form (\ref{chiwexp}); however, in practice, they are
not very effective when applied to the series of moderate length
available today.  Analyses meant to effectively allow for the leading
confluent corrections are based on biased approximants, where the
value of $\beta_c$ and the first non-analytic exponent $\Delta$ are
introduced as external inputs (see e.g.\ Refs.\ 
\cite{Roskies-81,AMP-82,BC-97,PV-98-gr,BC-98,CPRV-99}).  Nonetheless,
their precision is still not comparable to that of the experimental
result (\ref{alphaexp}), see e.g. Ref.~\cite{BC-97}.  The use of
improved Hamiltonians, i.e., models for which the leading correction
to scaling vanishes ($a_{1,1} = 0$ in Eq.\ (\ref{chiwexp})%
\footnote{Actually, for improved models, $a_{1,i} = 0$ for all $i$s.}%
), can lead to an additional improvement of the precision, even
without a substantial extension of the HT series.

The use of improved Hamiltonians was first suggested in the early 80s
by Chen, Fisher, and Nickel \cite{CFN-82} who determined improved
Hamiltonians in the Ising universality class.  The crux of the method
is a precise determination of the optimal value of the parameter
appearing in the Hamiltonian.  One can determine it from the analysis
of HT series, but in this case it is obtained with a relatively large
error \cite{CFN-82,GR-84,FC-85,NR-90,CPRV-99} and the final results do
not significantly improve the estimates obtained from standard
analyses using biased approximants.

Recently \cite{HPV-99,BFMM-98,BFMMPR-99,Hasenbusch-99,HT-99,CPRV-99}
it has been realized that FSS MC simulations are very effective in
determining the optimal value of the parameter, obtaining precise
estimates for several models in the Ising and XY universality classes.
The same holds true of models in the ${\rm O}(3)$ and ${\rm O}(4)$
universality classes \cite{Hasenbusch-00}.  Correspondingly, the
analysis of FSS results obtained in these simulations has provided
significantly more precise estimates of critical exponents.  An
additional improvement of the precision of the results has been
obtained by combining improved Hamiltonians and HT methods.  Indeed,
we already showed that the analysis of HT series for improved models
\cite{CPRV-99,CPRV-00,CPRV-00-2} provides estimates that are
substantially more precise than those obtained from the extrapolation
of the MC data alone.

In this paper we consider again the XY case.  The progress with
respect to the studies of Refs.\ \cite{CPRV-00,CPRV-00-2} is
essentially due to the improved knowledge of $\beta_c$ and of the
parameters of the improved Hamiltonians obtained by means of a
large-scale MC simulation.  The use of this information in the
analysis of the improved HT (IHT) series allows us to substantially
increase the precision and the reliability of the results, especially
of the critical exponents.  As we shall see, in order to determine the
critical exponents, the extrapolation to $\beta_c$ of the IHT series,
using biased integral approximants, is more effective than the
extrapolation $L\rightarrow\infty$ of the FSS MC data.  Moreover, we
consider two improved Hamiltonians.  The comparison of the results
from these two models provides a check of the errors we quote.  The
estimates obtained for the two models are in very good agreement,
providing support for our error estimates and thus confirming our
claim that the systematic error due to confluent singularities is
largely reduced when analyzing IHT expansions.

We consider a simple cubic (sc) lattice, two-component vector
fields $\vec{\phi}_x = (\phi_x^{(1)},\phi_x^{(2)})$, and two classes
of models depending on an irrelevant parameter: the $\phi^4$ lattice
model and the dynamically diluted XY (dd-XY) model.

The Hamiltonian of the $\phi^4$ lattice model is given by
\begin{equation}
{\cal H}_{\phi^4} =
 - \beta\sum_{\left<xy\right>} {\vec\phi}_x\cdot{\vec\phi}_y +\, 
   \sum_x \left[ {\vec\phi}_x^2 + \lambda ({\vec\phi}_x^2 - 1)^2\right].
\label{phi4Hamiltonian}
\end{equation}
The dd-XY model is defined by the Hamiltonian
\begin{equation}
{\cal H}_{\rm dd} =  -  \beta \sum_{\left<xy\right>}   \vec{\phi}_x \cdot 
\vec{\phi}_y - D  \sum_x \vec{\phi}_x^2 ,
\end{equation}
by the local measure  
\begin{equation}
d\mu(\phi_x) = \int d \phi_x^{(1)} \, \int d \phi_x^{(2)} \,
\left[
\delta(\phi_x^{(1)}) \, \delta(\phi_x^{(2)})
 + \frac{1}{2 \pi} \, \delta(1-|\vec{\phi}_x|)
\right],
\label{lmeasure}
\end{equation}
and the partition function
\begin{equation}
\int \prod_x d\mu(\phi_x)\, e^{-{\cal H}_{\rm dd}}.
\end{equation}
In the limit $D \rightarrow \infty$ the standard XY lattice model is
recovered.  We expect the phase transition to become of first order
for $D<D_{\rm tri}$.  $D_{\rm tri}$ vanishes in the mean-field
approximation, while an improved mean-field calculation based on the
``star approximation'' of Ref.\ \cite{DaHa} gives $D_{\rm tri}<0$, so
that we expect $D_{\rm tri}<0$.

The parameters $\lambda$ in ${\cal H}_{\phi^4}$ and $D$ in 
${\cal H}_{\rm dd}$ can be tuned to obtain improved Hamiltonians.  We
performed an accurate numerical study, which provided estimates of
$\lambda^*$, $D^*$, of the inverse critical temperature $\beta_c$ for
several values of $\lambda$ and $D$, as well as estimates of the
critical exponents.  Using the linked-cluster expansion technique, we
computed HT expansions of several quantities for the two theories.  We
analyzed them using the MC results for $\lambda^*$, $D^*$ and
$\beta_c$, obtaining very accurate results, e.g., Eq.\ 
(\ref{finalestimate}).

We mention that the $\phi^4$ lattice model ${\cal H}_{\phi^4}$ has
already been considered in MC and HT studies
\cite{HT-99,CPRV-00,CPRV-00-2}.  With respect to those works, we have
performed additional MC simulations to improve the estimate of
$\lambda^*$ and determine the values of $\beta_c$.  Moreover, we
present a new analysis of the IHT series that uses the MC estimates of
$\beta_c$ to bias the approximants, leading to a substantial
improvement of the results.

In Table \ref{table_exponents} we report our results for the critical
exponents, i.e., our best estimates obtained by combining MC and IHT
techniques---they are denoted by MC+IHT---together with the results
obtained from the analysis of the MC data alone.  There, we also
compare them with the most precise experimental and theoretical
estimates that have been obtained in the latest years.  When only
$\nu$ or $\alpha$ is reported, we used the hyperscaling relation
$2-\alpha=3\nu$ to obtain the missing exponent.  Analogously, if only
$\eta$ or $\gamma$ is quoted, the second exponent was obtained using
the scaling relation $\gamma=(2-\eta)\nu$; in this case the
uncertainty was obtained using the independent-error formula.  The
results we quote have been obtained from the analysis of the HT series
of the XY model (HT), by Monte Carlo simulations (MC) or by
field-theory methods (FT).  The HT results of Ref.\ \cite{BC-97} have
been obtained analyzing the 21st-order HT expansions for the standard
XY model on the sc and the bcc lattice, using biased approximants and
taking $\beta_c$ and $\Delta$ from other approaches, such as MC and
FT.  The FT results of Refs.\ \cite{GZ-98,JK-00} have been derived by
resumming the known terms of the fixed-dimension $g$ expansion: the
$\beta$ function is known to six-loop order \cite{BNGM-77}, while the
critical-exponent series are known to seven loops \cite{MN-91}.  The
estimates from the $\epsilon$ expansion have been obtained resumming
the available $O(\epsilon^5)$ series \cite{CGLT-83,KNSCL-93}.

\begin{table}[tp]
\caption{\label{table_exponents}
Estimates of the critical exponents. See text for the explanation of
the symbols in the second column. We indicate with an asterisk (${}^*$)
the estimates that have been obtained using the hyperscaling relation
$2 - \alpha = 3 \nu $ or the scaling relation $\gamma = (2-\eta)\nu$.
}

\begin{tabular}{rcr@{}lr@{}lr@{}lr@{}l}
\multicolumn{1}{c}{Ref.}&
\multicolumn{1}{c}{Method}&
\multicolumn{2}{c}{$\gamma$}&
\multicolumn{2}{c}{$\nu$}&
\multicolumn{2}{c}{$\eta$}&
\multicolumn{2}{c}{$\alpha$}\\
\tableline \hline
\multicolumn{1}{c}{this work} & 
MC+IHT & 1&.3177(5) & 0&.67155(27)& 0&.0380(4) & $-$0&.0146(8)$^*$\\
\multicolumn{1}{c}{this work} & 
MC & 1&.3177(10)$^*$ & 0&.6716(5)& 0&.0380(5) & $-$0&.0148(15)$^*$\\
\hline
\cite{CPRV-00} (2000) & 
IHT  & 1&.3179(11) & 0&.67166(55)& 0&.0381(3) & $-$0&.0150(17)$^*$\\ 
\cite{BC-99} (1999) & HT & &&&& & & $-$0&.014(9), $-$0.022(6) \\
\cite{BC-97} (1997) & HT, sc & 1&.325(3) & 0&.675(2)
    & 0&.037(7)$^*$ & $-$0&.025(6)$^*$ \\
    & HT, bcc & 1&.322(3) & 0&.674(2) & 0&.039(7)$^*$ & $-$0&.022(6)$^*$ \\
\hline
\cite{HT-99} (1999) & MC & 1&.3190(24)$^*$ 
              & 0&.6723(11) & 0&.0381(4) & $-$0&.0169(33)$^*$ \\
\cite{KL-99}  (1999) & MC  & 1&.315(12)$^*$
              & 0&.6693(58)  & 0&.035(5) &  $-$0&.008(17)$^*$ \\
\cite{BFMM-96} (1996)   & MC & 1&.316(3)$^*$ & 0&.6721(13)  
              & 0&.0424(25) &  $-$0&.0163(39)$^*$ \\
\cite{SM-95} (1995) & MC  &  &             & 0&.6724(17)  & & 
              & $-$0&.017(5)$^*$ \\ 
\cite{HG-93} (1993) & MC  & 1&.307(14)$^*$
              & 0&.662(7)    & 0&.026(6) &  $-$0&.014(21)$^*$ \\
\cite{Janke-90} (1990) & MC        & 1&.316(5) 
              & 0&.670(2)    & 0&.036(14)$^*$ & $-$0&.010(6)$^*$ \\
\hline
\cite{JK-00} (1999) & FT $d=3$ exp 
                 & 1&.3164(8) & 0&.6704(7) & 0&.0349(8) & $-$0&.0112(21) \\
\cite{GZ-98} (1998) & FT $d=3$ exp 
                 & 1&.3169(20) & 0&.6703(15) & 0&.0354(25) & $-$0&.011(4)\\
\cite{GZ-98} (1998) & FT $\epsilon$-exp 
                 & 1&.3110(70) & 0&.6680(35) & 0&.0380(50) &
                 $-$0&.004(11) \\ \hline
\cite{LSNCI-96,Lipa-etal-00}  (1996)  & ${}^4$He  & & & 0&.67019(13)$^*$ & & 
                 & $-$0&.01056(38) \\
\cite{GMA-93} (1993) & ${}^4$He  & && 0&.6705(6)     & &
                 & $-$0&.0115(18)$^*$ \\
\cite{Swanson-etal_92} (1992) & ${}^4$He & & & 0&.6708(4)    & & 
                 & $-$0&.0124(12)$^*$ \\
\cite{SA-84} (1984) & ${}^4$He & & & 0&.6717(4)    & & 
                 & $-$0&.0151(12)$^*$ \\
\cite{LC-83} (1983) & ${}^4$He & & & 0&.6709(9)$^*$ & & 
                 & $-$0&.0127(26) \\
\end{tabular}
\end{table}

We also present a detailed study of the equation of state.  We first
consider its expansion in terms of the magnetization in the
high-temperature phase.  The coefficients of this expansion are
directly related to the zero-momentum $n$-point renormalized
couplings, which were determined by analyzing their IHT expansion.
These results are used to construct parametric representations of the
critical equation of state which are valid in the whole critical
region, satisfy the correct analytic properties (Griffiths'
analyticity), and take into account the Goldstone singularities at the
coexistence curve.  From our approximate representations of the
equation of state we derive estimates of several universal amplitude
ratios.  The specific-heat amplitude ratio is particularly interesting
since it can be compared with experimental results.  We obtain
$A^+/A^-=1.062(4)$, which is not in agreement with the experimental
result $A^+/A^-=1.0442$ of Refs.\ \cite{LSNCI-96,Lipa-etal-00}.  It is
easy to trace the origin of the discrepancy.  In our method as well as
in the analysis of the experimental data, the estimate of $A^+/A^-$ is
strongly correlated with the estimate of $\alpha$.  Therefore, the
discrepancy we observe for this ratio is a direct consequence of the
difference in the estimates of $\alpha$.

Finally, we also discuss the two-point function of the order
parameter, i.e., the structure factor, which is relevant in scattering
experiments with magnetic materials.

The paper is organized as follows.  In Sec.\ \ref{Monte Carlo} we
present our Monte Carlo results.  After reviewing the basic RG ideas
behind our methods, we present a determination of the improved
Hamiltonians and of the critical exponents.  We discuss the several
possible sources of systematic errors, and show that the approximate
improved models we use have significantly smaller corrections than the
standard XY model.  A careful analysis shows that the leading scaling
corrections are reduced at least by a factor of 20.  We also compute
$\beta_c$ to high precision for several values of $\lambda$ and $D$;
this is an important ingredient in our IHT analyses.  Details on the
algorithm appear in App.\ \ref{montecarloupdate}.

In Sec.\ \ref{HTanalysis} we present our results for the critical
exponents obtained from the analysis of the IHT series.  The equation
of state is discussed in Sec.\ \ref{CES}.  After reviewing the basic
definitions and properties, we present the coefficients of the
small-magnetization expansion, again computed from IHT series.  We
discuss parametric representations that provide approximations of the
equation of state in the whole critical region and compute several
universal amplitude ratios.  
In Sec.\ \ref{Gx} we analyze the two-point function of the order parameter.
Details of the IHT analyses are reported
in App.\ \ref{seriesanalysis}.  The definitions of the amplitude
ratios we compute can be found in App.\ \ref{univra}.

\section{Monte Carlo Simulations}
\label{Monte Carlo}

\subsection{The lattice and the quantities that were measured}

We simulated sc lattices of size $V=L^3$, with periodic boundary
conditions in all three directions.  In addition to elementary
quantities like the energy, the magnetization, the specific heat or
the magnetic susceptibility we computed so-called phenomenological
couplings, i.e., quantities that, in the critical limit, are invariant
under RG transformations.  They are well suited to locate the inverse
critical temperature $\beta_c$.  They also play a crucial role in the
determination of the improved Hamiltonians.

In the present study we consider four phenomenological couplings.  We
use the Binder cumulant
\begin{equation}
U_4 \equiv \frac{\langle(\vec{m}^2)^2\rangle}{\langle\vec{m}^2\rangle^2},
\end{equation}
and the analogous quantity with the 6th power of the magnetization
\begin{equation}
U_6 \equiv \frac{\langle(\vec{m}^2)^3\rangle}{\langle\vec{m}^2\rangle^3} 
,
\end{equation}
where $\vec{m} = \frac{1}{V} \, \sum_x \vec{\phi}_x$ is the
magnetization of the system.

We also consider the second-moment correlation length divided by the
linear extension of the lattice $\xi_{\rm 2nd}/L$.  The second-moment
correlation length is defined by
\begin{equation}
\xi_{\rm 2nd}  \equiv  \sqrt{\frac{\chi/F-1}{4 \sin(\pi/L)^2}},
\end{equation}
where
\begin{equation}
\chi  \equiv  \frac{1}{V} 
\left\langle \left(\sum_x \vec{\phi}_x \right)^2 \right\rangle
\end{equation}
is the magnetic susceptibility and
\begin{equation}
F  \equiv  \frac{1}{V} \left \langle
\left|\sum_x \exp\left(i \frac{2 \pi x_1}{L} \right) 
        \vec{\phi}_x \right|^2 
\right \rangle
\end{equation}
is the Fourier transform of the correlation function at the lowest
non-zero momentum.

The list is completed by the ratio $Z_a/Z_p$ of the partition function
$Z_a$ of a system with anti-periodic boundary conditions in one of the
three directions and the partition function $Z_p$ of a system with
periodic boundary conditions in all directions.  Anti-periodic
boundary conditions in the first direction are obtained by changing
sign to the term $\vec{\phi}_x \vec{\phi}_y$ of the Hamiltonian for
links $\left<xy\right>$ that connect the boundaries, i.e., for
$x=(L,x_2,x_3)$ and $y=(1,x_2,x_3)$.  The ratio $Z_a/Z_p$ can be
measured by using the boundary-flip algorithm, which was applied to
the three-dimensional Ising model in Ref.\ \cite{Ha-93} and
generalized to the XY model in Ref.\ \cite{GH-94}.  As in Ref.\ 
\cite{Hasenbusch-99}, in the present work we used a version of the
algorithm that avoids the flip to anti-periodic boundary conditions.
For a detailed discussion see App.\ \ref{zazpapp}.

\subsection{Summary of finite-size methods}
\label{FSStheory}

In this subsection we discuss the FSS methods we used to compute the
inverse critical temperature, the couplings $\lambda^*$ and $D^*$ at
which leading corrections to scaling vanish, and the critical
exponents $\nu$ and $\eta$.

\subsubsection{Summary of basic RG results}

The following discussion of FSS is based on the RG theory of critical
phenomena.  We first summarize some basic results.  In the
three-dimensional XY universality class there exist two relevant
scaling fields $u_t$ and $u_h$, associated to the temperature and the
applied field respectively, with RG exponents $y_t$ and $y_h$.
Moreover, there are several irrelevant scaling fields that we denote
by $u_i$, $i\ge 3$, with RG exponents $0 > y_3 > y_4 > y_5 > ...\,$.

The RG exponent $y_3\equiv-\omega$ of the leading irrelevant scaling
field $u_3$ has been computed by various methods.  The analysis of
field-theoretical perturbative expansions \cite{GZ-98} gives
$\omega=0.802(18)$ ($\epsilon$ expansion) and $\omega=0.789(11)$
($d=3$ expansion).  In the present work we find a result for $\omega$
that is consistent with, although less accurate than, the
field-theoretical predictions.  We also mention the estimate
$\omega=0.85(7)$ that was obtained \cite{NR-84} by the
``scaling-field'' method, a particular implementation of Wilson's
``exact'' renormalization group.  Although it provides an estimate for
$\omega$ that is less precise than those obtained from perturbative
field-theoretic methods, it has the advantage of giving predictions
for the irrelevant RG exponents beyond $y_3$.  Ref.\ \cite{NR-84}
predicts $y_4 =-1.77(7)$ and $y_5=-1.79(7)$ ($y_{421}$ and $y_{422}$
in their notation) for the XY universality class.  Note that, at
present, there is no independent check of these results.  Certainly it
would be worthwhile to perform a Monte Carlo renormalization group
study.  With the computational power available today, it might be
feasible to resolve subleading correction exponents with a
high-statistics simulation.

In the case of $U_4$, $U_6$, and $\xi_{\rm 2nd}/L$ we expect a
correction caused by the analytic background of the magnetic
susceptibility.  This should lead to corrections with $y_6=-(2-\eta)
\approx -1.962$.  We also expect corrections due to the violation of
rotational invariance by the lattice.  For the XY universality class,
Ref.\ \cite{CPRV-98} predicts $y_7=-2.02(1)$.  Note that the numerical
values of $y_6$ and $y_7$ are virtually identical and should hence be
indistinguishable in the analysis of our numerical data.

We wish now to discuss the FSS behavior of a phenomenological coupling
$R$; in the standard RG framework, we can write it as a function of
the thermal scaling field $u_t$ and of the irrelevant scaling fields
$u_i$.  For $L\to\infty$ and $\beta\to\beta_c(\lambda)$, we have
\begin{equation}
R(L,\beta,\lambda) = 
r_0(u_t L^{y_t}) + \sum_{i\ge3} r_i(u_t L^{y_t}) \, u_i L^{y_i} + ... \, ,
\label{Rexp_1}
\end{equation}
where we have neglected terms that are quadratic in the scaling fields
of the irrelevant operators, i.e., corrections of order
$L^{2y_3}\approx L^{-1.6}$.  Note that we include here the corrections
due to the analytic background (with exponent 
$L^{-y_6} \approx L^{\eta - 2}$). In the case of $U_4$, $U_6$, and
$\xi_{\rm 2nd}/L$ (but not $Z_a/Z_p$), in Eq.\ (\ref{Rexp_1}) we have
also discarded terms of order $L^{y_t - 2 y_h} \approx L^{-3.5}$.

The functions $r_0(z)$ and $r_i(z)$ are smooth and finite for $z\to0$,
while $u_t(\beta,\lambda)$ and $u_i(\beta,\lambda)$ are smooth
functions of $\beta$ and $\lambda$.  Note that, by definition,
$u_t(\beta,\lambda)\sim\beta-\beta_c(\lambda)$.  In the limit $t\to 0$
and $u_t L^{y_t}\sim (\beta-\beta_c) L^{1/\nu}\to 0$, we can further
expand Eq.\ (\ref{Rexp_1}), obtaining
\begin{equation}
\label{expandR}
R(L,\beta,\lambda) = 
R^* + c_t(\beta,\lambda) \, L^{y_t}
  + \sum_i  c_i(\beta,\lambda) \, L^{y_i} 
  + O((\beta-\beta_c)^2 L^{2 y_t}, L^{2 y_3}, t L^{y_t+y_3}),
\end{equation}
where $R^*=r_0(0)$ is the value at the critical point of the
phenomenological coupling.

\subsubsection{Locating $\beta_c$}
\label{betacrit}

We locate the inverse critical temperature $\beta_c$ by using Binder's
cumulant crossing method.  This method can be applied in conjunction
with any of the four phenomenological couplings that we computed.

In its simplest version, one considers a phenomenological coupling
$R(\beta,L)$ for two lattice sizes $L$ and $L'=b L$.  The intersection
$\beta_{\rm cross}$ of the two curves $R(\beta,L)$ and $R(\beta,L')$
provides an estimate of $\beta_c$.  The convergence rate of this
estimate $\beta_{\rm cross}$ towards the true value can be computed in
the RG framework.  

By definition, $\beta_{\rm cross}$ at
fixed $b$, $L$, and $\lambda$ is given by the solution of the equation
\begin{equation}
\label{crossing}
R(L,\beta,\lambda) = R(b L,\beta,\lambda) .
\end{equation}
Using Eq.\ (\ref{Rexp_1}), one immediately verifies that 
$\beta_{\rm cross}$ converges to $\beta_c$ faster than $L^{-y_t}$.
Thus, for $L\to\infty$, we can use Eq.\ (\ref{expandR}) and rewrite
Eq.\ (\ref{crossing}) as
\begin{equation}
c_t(\beta,\lambda) \, L^{y_t} 
+  c_3(\beta,\lambda) \, L^{y_3} 
\approx
c_t(\beta,\lambda) \, (b L)^{y_t} 
+ c_3(\beta,\lambda) \, (b L)^{y_3} .
\label{eq17}
\end{equation}
Then, we approximate 
$c_t(\beta,\lambda) \approx c_t' \, (\beta-\beta_c)$ and 
$c_i(\beta,\lambda) \approx c_i(\beta_c,\lambda)=c_i$.
Remember that $c_t(\beta_c,\lambda)=0$ by definition.  Using these
approximations we can explicitly solve Eq.\ (\ref{eq17}) with respect
to $\beta$, obtaining
\begin{equation}
\beta_{\rm cross} = \beta_c +
\frac{c_3 \, (1-b^{y_3}) \, L^{y_3}    }
     {c_t' \, (b^{y_t} - 1) \, L^{y_t}} + ... \, .
\end{equation}
The leading corrections vanish like $L^{-y_t+y_3} \approx L^{-2.3}$.
Inserting $\beta_{\rm cross}$ into Eq.\ (\ref{crossing}), we obtain
\begin{equation}
 R_{\rm cross}  =  R^* +
\frac{b^{y_t} - b^{y_3}}{b^{y_t}-1} \, L^{y_3} + ... \, , 
\end{equation}
which shows that the leading corrections vanish like $L^{y_3}$.

Given a precise estimate of $R^*$, one can locate $\beta_c$ from
simulations of a single lattice size, solving
\begin{equation}
\label{betasimple}
R(L,\beta) = R^*,
\end{equation}
where the corrections vanish like $L^{-y_t+y_3}$.

\subsubsection{Locating $\lambda^*$ and $D^*$}
\label{eliminate}

In order to compute the value $\lambda^*$ for which the leading
corrections to scaling vanish, we use two phenomenological couplings
$R_1$ and $R_2$.  First, we define $\beta_f(L,\lambda)$ by
\begin{equation}
\label{betafix}
R_1(L,\beta_f,\lambda)  =  R_{1,f} \, ,
\end{equation}
where $R_{1,f}$ is a fixed value, which we can choose freely within
the appropriate range.  It is easy to see that 
$\beta_f(L,\lambda)\to \beta_c(\lambda)$ as $L\to\infty$.  Indeed,
using Eq.\ (\ref{Rexp_1}), we have
\begin{equation}
\beta_f(L,\lambda) - \beta_c(\lambda) = z_f L^{-y_t} - 
   {r_{1,3}(z_f)\over r'_{1,0}(z_f)}u_3(\beta_c) L^{y_3-y_t} + ... \, ,
\end{equation}
where we have used $y_t > |y_3|$ and $z_f$ is defined as the solution
of $r_{1,0}(z_f) = R_{1,f}$.  We have added a subscript 1 to make
explicit that all scaling functions refer to $R_1$.  If 
$R_{1,f} \approx R^*$, we can expand the previous formula, obtaining
\begin{equation}
\beta_f(L,\lambda) \approx \beta_c(\lambda) + 
{R_{1,f} - R_1^* \over {c_{1,t}'} } L^{-y_t} - 
   {c_{1,i} \over {c_{1,t}'}} L^{-y_t+y_3} .
\end{equation}
Notice that for $R_{1,f}= R_1^*$ the convergence is faster, and thus
we will always take $R_{1,f}\approx R_1^*$.  Next we define
\begin{equation}
\label{barR}
\bar{R}(L,\lambda)  \equiv  R_2(L,\beta_f,\lambda) .
\end{equation}
For $L\to\infty$ and $R_{1,f} \approx R^*_1$, we have
\begin{eqnarray}
\bar{R}(L,\lambda) &=& 
R_2^* + \frac{c_{2,t}'}{c_{1,t}'} (R_{1,f} - R_1^*) + \sum_i 
\left(c_{2,i} - \frac{c_{2,t}'}{c_{1,t}'} \, c_{1,i} \right) L^{y_i}
\nonumber \\
&=&  \bar{R}^* + \sum_i \bar{c}_{i}(\lambda) \, L^{y_i} ,
\label{barRexp}
\end{eqnarray}
which shows that the rate of convergence is determined by $L^{y_3}$.

In order to find $\lambda^*$, we need to compute the value of
$\lambda$ for which $\bar{c}_{i}(\lambda) = 0$.  We can obtain
approximate estimates of $\lambda^*$ by solving the equation
\begin{equation}
\bar{R}(L,\lambda)  =  \bar{R}(b L,\lambda) .
\end{equation}
Using the approximation (\ref{barRexp}) one finds
\begin{equation}
\lambda_{\rm cross}  = 
  \lambda^* - \frac{\bar{c}_{4}}{\bar{c}^{\,\prime}_{3}}
  \frac{1-b^y_4}{1-b^y_3} \, L^{y_4-y_3} + ... \, ,
\end{equation} 
where $\bar{c}^{\,\prime}_{3}$ is the derivative of $\bar{c}_{3}$ with
respect to $\lambda$, and
\begin{equation} 
\bar{R}_{\rm cross}   = \bar{R}^*
- \bar{c}_{4} \frac{b^y_3 - b^y_4}{1 -b^y_3} \, L^{y_4} . 
\end{equation}  

In principle, any pair $R_1$, $R_2$ of phenomenological couplings can
be used in this analysis.  However, in practice we wish to see a good
signal for the corrections.  This means, in particular, that in
$c_{1,t}' \, c_{2,3} - c_{2,t}' \, c_{1,3}$ the two terms should add
up rather than cancel.  Of course, also the corrections due to the
subleading scaling fields should be small.

\subsubsection{The critical exponents}
\label{rg-exponents}

Typically, the thermal RG exponent $y_t=1/\nu$ is computed from the
FSS of the derivative of a phenomenological coupling $R$ with respect
to $\beta$ at $\beta_c$.  Using Eq.\ (\ref{Rexp_1}) one obtains
\begin{equation}
\label{numethod}
\left. \frac{\partial R}{\partial\beta} \right|_{\beta_c} = 
r_0'(0)\,L^{y_t} + \sum_{i=3} r'_i(0)\,u_i(\beta_c)\,L^{y_i + y_t} + 
   \sum_{i=3} r_i(0)\,u'_i(\beta_c)\,L^{y_i} + ... \, .
\label{scaling-derivative-R}
\end{equation}
Hence, the leading corrections scale with $L^{y_3}$.  However, in
improved models in which $u_3(\beta_c) = 0$, the leading correction is
of order $L^{y_4}$.  Note that corrections proportional to
$L^{y_3-y_t} \approx L^{-2.3}$ are still present even if the model is
improved.  In Ref.\ \cite{HPV-99}, for the spin-1 Ising model, an
effort was made to eliminate also this correction by taking the
derivative with respect to an optimal linear combination of $\beta$
and $D$ instead of $\beta$.  Here we make no attempt in this
direction, since corrections of order $L^{-2.3}$ are subleading with
respect to those of order $L^{y_4} \approx L^{-1.8}$.

In practice it is difficult to compute the derivative at $\beta_c$,
since $\beta_c$ is only known numerically, and therefore, it is more
convenient to evaluate $\partial R /\partial \beta$ at $\beta_f$ (see
Eq.\ (\ref{betafix})).  This procedure has been used before, e.g., in
Ref.\ \cite{BFMM-96}.  In this case, Eq.\ (\ref{scaling-derivative-R})
still holds, although with different amplitudes that depend on the
particular choice of the value of $R_f$.

The exponent $\eta$ is computed from the finite-size behavior of the
magnetic susceptibility, i.e.,
\begin{equation}
\left. \chi \right |_{\beta_f} \propto L^{2-\eta} .
\end{equation}
Also here the corrections are of order $L^{y_3}$ for generic models,
and of order $L^{y_4}$ for improved ones.

\subsubsection{Estimating errors caused by residual
               leading scaling corrections}
\label{est-residual-err}

In Ref.\ \cite{BFMMPR-99}, the authors pointed out that with the
method discussed in Sec.\ \ref{eliminate}, the leading corrections are
only approximately eliminated, so that there is still a small leading
scaling correction which causes a systematic error in the estimates
of, e.g., the critical exponents.  The most naive solution to this
problem consists in adding a term $L^{-\omega}$ to the fit ansatz,
i.e., in considering
\begin{equation}
\left . \frac{\partial R }{\partial \beta} \right|_{\beta_f} = 
A \, L^{1/\nu} \left(1 + B \, L^{-\omega}\right) .
\end{equation}
However, by adding such a correction term, the precision of the result
decreases, so that there is little advantage in using (approximately)
improved models.  A more sophisticated approach is based on the fact
that ratios of leading correction amplitudes are universal
\cite{AHP-91}.

Let us consider a second phenomenological coupling 
$\bar{R} = R(L,\beta_f,\lambda)$, which, for $L\to\infty$, behaves as
\begin{equation}
\bar{R} = \bar{R}^* +  \bar{c}_3 \, L^{-\omega} .
\end{equation}
The universality of the correction amplitudes implies that the ratio
$B/\bar{c}_3$ is the same for the $\phi^4$ model and the dd-XY model
and is independent of $\lambda$ and $D$.  Therefore, this ratio can be
computed in models that have large corrections to scaling, e.g., in
the standard XY model.  Then, we can compute a bound on $B$ for the
(approximately) improved model from the known ratio $B/\bar{c}_3$ and
a bound for $\bar{c}_3$.  This procedure was proposed in Ref.\ 
\cite{HPV-99}.

\subsection{The simulations}

We simulated the $\phi^4$ and the dd-XY model using the wall-cluster
update algorithm of Ref.\ \cite{HPV-99} combined with a local update
scheme.  The update algorithm is discussed in detail in App.\ 
\ref{montecarloupdate}, where we also report an analysis of its
performance and the checks we have done.

Most of the analyses need the quantities as functions of $\beta$.
Given the large statistics, we could not store all individual
measurements of the observables.  Therefore, we did not use the
reweighting method.  Instead, we determined the Taylor coefficients of
all quantities of interest up to the third order in $(\beta-\beta_s)$,
where $\beta_s$ is the value of $\beta$ at which the simulation was
performed.  We checked carefully that this is sufficient for our
purpose.  For details, see App.\ \ref{montecarloupdate}.

Most of our simulations were performed at $\lambda=2.1$ in the
case of the $\phi^4$ model and at $D=1.03$ in the case of the dd-XY
model.  $\lambda=2.1$ is the estimate of $\lambda^*$ of Ref.\ 
\cite{HT-99}, and $D=1.03$ is the result for $D^*$ of a preliminary
analysis of MC data obtained on small lattices.

In addition, we performed simulations at $\lambda=2.0$ and $2.2$ for
the $\phi^4$ model and $D=0.9$ and $1.2$ for the dd-XY model in order
to obtain an estimate of the derivative of the amplitude of the
leading corrections to scaling with respect to $\lambda$ and $D$,
respectively.

We also performed simulations of the standard XY model in order
to estimate the effect of the leading corrections to scaling on the
estimates of the critical exponents obtained from the FSS analysis.

\subsection{$\beta_c$ and the critical value 
            of phenomenological couplings}

In a first step of the analysis we computed $R^*$ and the inverse
critical temperature $\beta_c$ at $\lambda=2.1$ and $D=1.03$
respectively.

For $\lambda=2.1$ and $D=1.03$ we simulated sc lattices of linear size
$L$ from $4$ to $16$ and $L=18$, $20$, $22$, $24$, $26$, $28$, $32$,
$36$, $40$, $48$, $56$, $64$, and $80$.  For all lattice sizes smaller
than $L=24$ we performed $10^8$ measurements, except for the dd-XY
model at $L=16$ where approximately $1.5 \times 10^8$ measurements
were performed.  The statistics for the larger lattices is given in
Table \ref{statist}.  A measurement was performed after an update
cycle as discussed in App.\ \ref{update-cycle}.

\begin{table}[tp]
\caption{\label{statist}
The number of measurements/1000 for the $\phi^4$ model at
$\lambda=2.1$ and for the dd-XY model at $D=1.03$ for linear lattice
sizes $L\ge 24$.
}
\begin{tabular}{crr}
\multicolumn{1}{c}{$L$} &  \multicolumn{1}{c}{$\phi^4$} & 
\multicolumn{1}{c}{dd-XY} \\
\hline
24  &  100,000  &   112,740   \\
26  &   28,475  &    60,900   \\
28  &  101,235  &    80,645   \\
32  &   48,005  &    55,560   \\
36  &   18,220  &    22,880   \\
40  &   21,120  &    27,115   \\
48  &    9,735  &    16,005   \\
56  &    6,550  &     7,495   \\
64  &    4,725  &     5,240   \\
80  &      585  &       780   \\
\end{tabular}
\end{table}

Instead of computing $R^*$ and $\beta_c$ from two lattice sizes as
discussed in Sec.\ \ref{betacrit}, we perform a fit with the ansatz
\begin{equation}
R^* = R(L,\beta_c) ,
\end{equation}
where $R^*$ and $\beta_c$ are free parameters.  We compute
$R(L,\beta)$ using its third-order Taylor expansion
\begin{equation}
\label{bindercross}
R^* = R(L,\beta_s) + d_1(L,\beta_s) (\beta_c-\beta_s)
      + \frac{1}{2} \, d_2(L,\beta_s) (\beta_c-\beta_s)^2
      + \frac{1}{6} \, d_3(L,\beta_s) (\beta_c-\beta_s)^3,
\end{equation}
where $\beta_s$ is the $\beta$ at which the simulation was performed,
and $R$, $d_1$, $d_2$, and $d_3$ are determined in the MC simulation.

First, we perform fits for the two models separately.  We obtain
consistent results for $R^*$ for all four choices of phenomenological
couplings.  In order to obtain more precise results for $\beta_c$ and
$R^*$, we perform joint fits of both models.  Here, we exploit
universality by requiring that $R^*$ takes the same value in both
models.  Hence, such fits have three free parameters: $R^*$ and the
two values of $\beta_c$.  In the following we shall only report the
results of such joint fits.

Let us discuss in some detail the results for $R=Z_a/Z_p$ that are
summarized in Table \ref{joint1}.  In each fit, we take all data
with $L_{\rm min} \le L \le L_{\rm max}$ into account.  For 
$L_{\rm max}=80$, the $\chi^2/$d.o.f.\ becomes approximately one
starting from $L_{\rm min}=15$.  However, we should note that a
$\chi^2/$d.o.f.\ close to one does not imply that the systematic
errors due to corrections that are not taken into account in the
ansatz are negligible.

Our final result is obtained from the fit with $L_{\rm min}=28$ and
$L_{\rm max}=80$.  The systematic error is estimated by comparing this
result with that obtained using $L_{\rm min}=10$ and $L_{\rm max}=28$.
The systematic error on $\beta_c$ is estimated by the difference of
the results from the two fits divided by $2.8^{2.3}-1$, where $2.8$ is
the scale factor between the two intervals and $y_t-y_3 \approx 2.3$.
Estimating the systematic error by comparison with the interval
$L_{\rm min}=14$ and $L_{\rm max}=40$ leads to a similar result.  We
obtain $\beta_c=0.5091507(6)[7]$ for the $\phi^4$ model at
$\lambda=2.1$ and $\beta_c=0.5627975(7)[7]$ for the dd-XY model at
$D=1.03$.  In parentheses we give the statistical error and in the
brackets the systematic one.  Our final result for the critical ratio
of partition functions is $(Z_a/Z_p)^* = 0.3202(1)[5]$.  Here the
systematic error is computed by dividing the difference of the results
of the two fits by $2.8^{0.8}-1$.

\begin{table}[tp]
\caption{\label{joint1}
Joint fits  of the $Z_a/Z_p$ data of the $\phi^4$ model and the dd-XY
model with the ansatz (\ref{bindercross}). All lattice sizes $L_{\rm
min} \le L \le L_{\rm max}$ are used in the fit.  In column four we
give the results of the fits for $\beta_c$ of the $\phi^4$ model at
$\lambda=2.1$ and in  column five the results for  $\beta_c$ of the
dd-XY model at $D=1.03$. Finally, in column six we give the results
for the fixed-point value $(Z_a/Z_p)^*$.  The final results and an
estimate of the systematic errors are given in the text.
}
\begin{tabular}{ccclll}
\multicolumn{1}{c}{$L_{\rm min}$}& 
\multicolumn{1}{c}{$L_{\rm max}$}& 
\multicolumn{1}{c}{$\chi^2$/d.o.f.} &
\multicolumn{1}{c}{$\beta_c$, $\lambda=2.1$} &
\multicolumn{1}{c}{$\beta_c$, $D=1.03$}  
& \multicolumn{1}{c}{$(Z_a/Z_p)^*$}  \\
\hline
11 &  80 &  3.25 & 0.50915354(33) & 0.56280014(35) & 0.319794(25) \\
13 &  80 &  2.48 & 0.50915287(35) & 0.56279938(38) & 0.319883(29) \\
15 &  80 &  1.06 & 0.50915192(38) & 0.56279834(41) & 0.320019(35) \\
20 &  80 &  0.91 & 0.50915142(46) & 0.56279784(49) & 0.320093(52) \\
24 &  80 &  0.89 & 0.50915109(53) & 0.56279740(56) & 0.320162(72) \\
28 &  80 &  0.73 & 0.50915074(63) & 0.56279747(66) & 0.320195(102) \\
32 &  80 &  0.85 & 0.50915065(75) & 0.56279746(82) & 0.320208(149) \\
10 &  28 &  3.47 & 0.50915783(53) & 0.56280463(59) & 0.319524(32) \\
14 &  40 &  1.78 & 0.50915337(48) & 0.56279981(53) & 0.319877(39) \\
\end{tabular}
\end{table}

We repeat this analysis for $\xi_{\rm 2nd}$, $U_4$ and $U_6$.  The
final results are summarized in Table \ref{summary1}.

\begin{table}[tp]
\caption{\label{summary1}
Summary of the final results for $\beta_c$ and $R^*$.  In column one
the choice of the phenomenological coupling $R$ is listed. In columns
two and three we report our estimates of $\beta_c$ for the $\phi^4$
model at $\lambda=2.1$ and the dd-XY model at $D=1.03$, and finally in
column four the result for the fixed-point value of the
phenomenological coupling.  Note that the estimates of $\beta_c$,
based on the four different choices of $R$, are consistent within
error bars.
}
\begin{tabular}{clll}
$R$ &  
\multicolumn{1}{c}{$\beta_c$, $\lambda=2.1$} &
\multicolumn{1}{c}{$\beta_c$, $D=1.03$}
& \multicolumn{1}{c}{$R^*$} \\
\hline
$Z_a/Z_p$ & 0.5091507(6)[7] & 0.5627975(7)[7]  &0.3202(1)[5]\\
$\xi_{\rm 2nd}/L$ & 0.5091507(7)[3]  &  0.5627971(7)[2] & 0.5925(1)[2] \\
$ U_4$ & 0.5091495(9)[10] & 0.5627972(10)[11] & 1.2430(1)[5] \\
$ U_6$ & 0.5091498(9)[15] &  0.5627976(10)[15] & 1.7505(3)[25] \\
\end{tabular}
\end{table}

Next we compute $\beta_c$ at additional values of $\lambda$ and $D$.
For this purpose we simulated lattices of size $L=96$ and compute
$\beta_c$ using Eq.\ (\ref{betasimple}).  We use only $R=Z_a/Z_p$ with
the above-reported estimate $(Z_a/Z_p)^*=0.3202(6)$.  The results are
summarized in Table \ref{otherbetac}.

\begin{table}[tp]
\caption{\label{otherbetac}
Estimates of $\beta_c$ from simulations of $96^3$ lattices.  $\beta_c$
is obtained from Eq.\ (\ref{betasimple}) using $Z_a/Z_p$ as
phenomenological coupling. In parentheses we give the statistical
error and in brackets the error due to the error on $(Z_a/Z_p)^*$.
``stat'' gives (the number of measurements)/1000.
}
\begin{tabular}{clrl}
model   &  \multicolumn{1}{c}{$\lambda$ ; $D$} & 
\multicolumn{1}{r}{stat} & \multicolumn{1}{c}{$\beta_c$}  \\
\hline
$\phi^4$  & 2.07  &  545  &  0.5093853(16)[8]  \\
$\phi^4$  & 2.2   &  510  &  0.5083366(16)[8]  \\
\hline
dd-XY & 0.9   &    720  &  0.5764582(15)[9] \\
dd-XY & 1.02  &  1,215  &  0.5637972(12)[9] \\
dd-XY & 1.2   &    665  &  0.5470377(17)[9] \\
\end{tabular}
\end{table}

\subsection{Eliminating leading corrections to scaling}

In this subsection we determine $\lambda^*$ and $D^*$.  For this
purpose, we compute the correction amplitude $\bar{c}_3$ for various
choices of $R_1$ and $R_2$ for the $\phi^4$ model at $\lambda=2.1$ and
the dd-XY model at $D=1.03$.  In order to convert these results into
estimates of $\lambda^*$ and $D^*$, we determine the derivative of the
correction amplitude $\bar{c}_3$ with respect to $\lambda$ (resp.\ 
$D$) at $\lambda=2.1$ (resp.\ $D=1.03$).  We also simulated the XY
model in order to obtain estimates of the residual systematic error
due to the leading corrections to scaling.  Note that, in the
following, we always use as the value of $R_{1,f}$ in Eq.\ 
(\ref{betafix}) the estimates of $R^*$ given in Table \ref{summary1}.

\subsubsection{Derivative of the correction amplitude 
               with respect to $\lambda$ or $D$}

For this purpose we simulated the dd-XY model at $D=0.9$ and $D=1.2$
on lattices of size $L=5$, $6$, $7$, $8$, $9$, $10$, $12$ and $16$.
The $\phi^4$ model was simulated at $\lambda=2.0$ and $\lambda=2.2$ on
lattices of size $L=3$, $4$, $5$, $6$, $7$, $8$ and $9$.  In the case
of the dd-XY model we performed $100 \times 10^6$ measurements for
each parameter set.  In the case of the $\phi^4$ model 
$250 \times 10^6$ measurements were performed.

In the following we discuss only the dd-XY model, since the analysis
of the $\phi^4$ data is performed analogously.

In Refs.\ \cite{Hasenbusch-99,HT-99} it was observed that subleading
corrections to scaling cancel to a large extent when one considers the
difference of $\bar{R}$ at close-by values of $\lambda$.  In order to
get an idea of the size of the corrections, we report in Table
\ref{diffanders}
\begin{equation}
\Delta \bar{R} \, L^{0.8} =
(\bar{R}|_{D=0.9} - \bar{R}|_{D=1.2}) \, L^{0.8} 
\end{equation}
for various choices of $R_1$ and $R_2$.  We see that this quantity
varies little with $L$ in all cases.  In the case of $R_1=Z_a/Z_p$ and
$R_2=U_4$, $\Delta \bar{R} \, L^{0.8}$ is already constant within
error bars starting from $L=5$.

In order to compute $\overline{c}_3$, see Eq.\ (\ref{barRexp}), we
need $\Delta\bar{R}\,L^{0.8}$ to be as flat as possible and especially
$\Delta\bar{R}$ large compared to the statistical errors.  Looking at
Table \ref{diffanders}, we see that the two combinations
$R_1=\xi_{\rm 2nd}/L$, $R_2=Z_a/Z_p$ and $R_1=U_4$, $R_2=U_6$ are
unfavorable compared with the other four combinations.

\begin{table}[tp]
\caption{\label{diffanders}
The quantity $(\bar{R}|_{D=0.9} - \bar{R}|_{D=1.2}) L^{0.8}$ for the
dd-XY model.  In the top row we give the choice of $R_1$ and $R_2$.
For instance, $U_4$ at $(Z_a/Z_{p})_f$ means that $R_1 = Z_a/Z_{p}$
and $R_2=U_4$.
}
\begin{tabular}{rllllll}
$L$ & \multicolumn{1}{c}{$U_4$ at $(Z_a/Z_{p})_f$} &
\multicolumn{1}{c}{$U_6$ at $(Z_a/Z_{p})_f$}&
\multicolumn{1}{c}{$U_4$ at $(\xi_{\rm 2nd}/L)_f$}&
\multicolumn{1}{c}{$U_6$ at $(\xi_{\rm 2nd}/L)_f$} &
\multicolumn{1}{c}{$Z_a/Z_p$ at $(\xi_{\rm 2nd}/L)_f$} &
\multicolumn{1}{c}{$U_6$ at $U_{4,f}$} \\
\hline
 5 & 0.0366(2)& 0.1294(7) & 0.0400(2)& 0.1404(7) & 0.0057(2)& 0.0039(1) \\
 6 & 0.0365(2)& 0.1297(8) & 0.0409(3)& 0.1442(9) & 0.0069(2)& 0.0042(1) \\
 7 & 0.0368(3)& 0.1312(9) & 0.0415(3)& 0.1469(10)& 0.0073(3)& 0.0046(2) \\
 8 & 0.0369(3)& 0.1312(10)& 0.0421(3)& 0.1489(11)& 0.0081(3)& 0.0046(2) \\
 9 & 0.0366(4)& 0.1301(12)& 0.0415(3)& 0.1468(13)& 0.0076(3)& 0.0044(2) \\
10 & 0.0368(4)& 0.1311(13)& 0.0419(4)& 0.1483(14)& 0.0078(4)& 0.0045(2) \\
12 & 0.0372(4)& 0.1324(15)& 0.0427(5)& 0.1511(17)& 0.0084(4)& 0.0045(3) \\
16 & 0.0360(6)& 0.1286(20)& 0.0411(7)& 0.1460(22)& 0.0078(5)& 0.0046(4) \\
\end{tabular}
\end{table}

In order to see whether we can predict the exponent $\omega$, we
perform a fit with the ansatz
\begin{equation}
\Delta\bar{R} = k \, L^{-\omega},
\end{equation}
with $k$ and $\omega$ as free parameters.  From $U_4$ at
$Z_a/Z_p=0.3202$ we get $\omega=0.795(9)$ with
$\chi^2/\mbox{d.o.f.}=0.66$, using all available data.  This value is
certainly consistent with field-theoretical results.  Note however,
that we would like to vary the range of the fit in order to estimate
systematic errors.  For this purpose more data at larger values of $L$
are needed.

In the following we need estimates of
\begin{equation}
\left . \frac{ d \bar{c}_3 } { d D } \right |_{D=1.03}
\end{equation}
and of the corresponding quantity for the $\phi^4$ model, in order to
determine $D^*$ and $\lambda^*$.  We approximated this derivative
by a finite difference between $D=0.9$ and $D=1.2$.  The coefficient
$\bar{c}_3$ is determined by fixing $\omega=0.8$.  Our final result
is the average of the estimates for $L=10,12$ and 16 in Table
\ref{diffanders}.  In a similar way we proceed in the case of the
$\phi^4$ model, averaging the $L=8,9$ results.  The results are
summarized in Table \ref{derivatives}.  We make no attempt to estimate
error bars.  Sources of error are the finite difference in $D$,
subleading corrections, the error on $\omega$ and the statistical
errors.  Note however that these errors are small enough to be
neglected in the following.

\begin{table}[tp]
\caption{\label{derivatives}
Estimates for $d\bar{c}_3/d D$ at $D=1.03$ (dd-XY) and
$d\bar{c}_3/d\lambda$ at $\lambda=2.1$ ($\phi^4$). In the first row
we give the combination of $R_1$ and $R_2$.
}
\begin{tabular}{lllll}
Model   & $U_4$ at $(Z_a/Z_{p})_f$ & $U_6$ at $(Z_a/Z_{p})_f$ & 
$U_4$ at $(\xi_{\rm 2nd}/L)_f$ & $U_6$ at $(\xi_{\rm 2nd}/L)_f$ \\
\hline
dd-XY    & $-$0.122  & $-$0.435 & $-$0.140  & $-$0.495  \\
$\phi^4$ & $-$0.0490 & $-$0.175 & $-$0.0546 & $-$0.194  \\
\end{tabular}
\end{table}

\subsubsection{Finding $\bar{R}^*$, $\lambda^*$ and $D^*$}

For this purpose we fit our results at $D=1.03$ and $\lambda=2.1$
with the ansatz
\begin{equation}
\label{fitbar}
\bar{R}  =  \bar{R}^* +\bar{c}_3 \, L^{-\omega} ,
\end{equation}
where we fix $\omega=0.8$.  We convinced ourselves that setting
$\omega$ = $0.75$ or $0.85$ changes the final results very little
compared with statistical errors and errors caused by subleading
corrections.  We perform joint fits, by requiring $\bar{R}^*$ to be
equal in both models.

The results of the fits for four different combinations of $R_1$ and
$R_2$ are given in Table \ref{binderfixzazp}, where we have already
translated the results for $\bar{c}_3$ into an estimate of $\lambda^*$
and $D^*$, by using
\begin{equation}
\label{translate}
\lambda^*  =  2.1 - \bar{c}_3 
     \left(\frac{ d \bar{c}_3}{ d \lambda}\right)^{-1}
\end{equation}
for the $\phi^4$ model and the analogous formula for the dd-XY model,
and the results of Table \ref{derivatives}.

\begin{table}[tp]
\caption{\label{binderfixzazp}
Results of fits with the ansatz (\ref{fitbar}).  The coefficients
$\bar{c}_3$ are converted into $D^*$ and $\lambda^*$ using Eq.\
(\ref{translate}). All data with $L_{\rm min} \le L \le L_{\rm max}$
are fitted.
}
\begin{tabular}{rrllll}
\rule[0mm]{0mm}{4mm}
$L_{\rm min}$ & $L_{\rm max}$ & $\chi^2$/d.o.f. & 
  $\bar{R}^*$ & $\lambda^*$ & $D^*$ \\
\hline
\multicolumn{6}{c}{$R_{1,f}=(Z_a/Z_{p})_f=0.3202$  and $R_2=U_4$.} \\
\hline
 8& 80& 1.55&  1.24303(2)&  2.077(4) & 1.020(2) \\
12& 80& 1.01&  1.24304(4)&  2.071(8) & 1.022(3) \\
16& 80& 1.07&  1.24308(6)&  2.057(14)& 1.019(5) \\
20& 80& 1.12&  1.24301(8)&  2.073(22)& 1.028(9) \\
 8& 40& 1.62&  1.24304(3)&  2.077(4) & 1.020(2) \\
10& 40& 1.02&  1.24305(3)&  2.070(6) & 1.020(2) \\
\hline
\multicolumn{6}{c}{$R_{1,f}=(Z_a/Z_{p})_f=0.3202$  and $R_2=U_6$.} \\
\hline
  8& 80& 2.15&  1.75156(8) &  2.006(4) &   0.990(2) \\
 12& 80& 1.15&  1.75126(13)&  2.018(7) &   1.000(3) \\
 16& 80& 1.22&  1.75120(19)&  2.017(13)&   1.003(5) \\
 20& 80& 1.19&  1.75085(27)&  2.043(21)&   1.015(8) \\
  8& 40& 2.21&  1.75160(9) &  2.004(4) &   0.989(2) \\
 10& 40& 1.24&  1.75143(11)&  2.010(6) &   0.994(2) \\
\hline
\multicolumn{6}{c}{$R_{1,f}=(\xi_{\rm 2nd}/L)_f=0.5925 $  and $R_2=U_4$.} \\
\hline
  8& 80& 4.01&  1.24352(3)&   1.977(4) &   0.987(2) \\
 12& 80& 1.19&  1.24322(4)&   2.031(8) &   1.010(3) \\
 16& 80& 1.29&  1.24314(6)&   2.049(14)&   1.019(5) \\
 20& 80& 1.13&  1.24299(9)&   2.083(23)&   1.035(9) \\
  8& 40& 4.19&  1.24355(3)&   1.973(4) &   0.985(2) \\
 10& 40& 1.49&  1.24335(4)&   2.006(6) &   1.000(2) \\
\hline
\multicolumn{6}{c}{$R_{1,f}=(\xi_{\rm 2nd}/L)_f=0.5925$  and $R_2=U_6$.} \\
\hline
 8& 80& 6.62&  1.75323(10)&   1.915(4) &   0.961(2) \\
12& 80& 1.55&  1.75189(14)&   1.985(7) &   0.991(3) \\
16& 80& 1.51&  1.75142(22)&   2.013(13)&   1.004(5) \\
20& 80& 1.24&  1.75078(32)&   2.055(22)&   1.024(8) \\
 8& 40& 7.02&  1.75336(10)&   1.911(4) &   0.959(2) \\
10& 40& 2.21&  1.75248(12)&   1.953(6) &   0.978(2) \\
\end{tabular}
\end{table}

A $\chi^2$/d.o.f.\ close to $1$ is reached for $L_{\rm min}=10$ and
$L_{\rm max}=80$ in the case of $U_4$ at $(Z_a/Z_{p})_f=0.3202$.  This
has to be compared with $L_{\rm min}=11$, $11$, and $14$ in the case
of $U_6$ at $(Z_a/Z_{p})_f=0.3202$, $U_4$ at 
$(\xi_{\rm 2nd}/L)_f=0.5925$ and $U_6$ at 
$(\xi_{\rm 2nd}/L)_f=0.5925$.

This indicates that $U_4$ at $(Z_a/Z_{p})_f=0.3202$ has the least bias
due to subleading corrections to scaling.  Therefore we take as our
final result $\lambda^*=2.07$ and $D^*=1.02$ which is the result of
$L_{\rm min}=12$ and $L_{\rm max}=80$ in Table \ref{binderfixzazp}.
Starting from $L_{\rm min}=20$ all results for $\lambda^*$ and $D^*$
are within $2\sigma$ of our final result quoted above.

Our final results are $\lambda^*=2.07(5)$ and $D^*=1.02(3)$.  The
error bars are such to include all results in Table
\ref{binderfixzazp} with $L_{\rm min}=20$ and $L_{\rm max}=80$,
including the statistical error, and therefore should take into proper
account all systematic errors.

From these results, it is also possible to obtain a conservative 
upper bound on the coefficient $\bar{c}_3$ for $\lambda= 2.1$ and 
$D = 1.03$.  Indeed, using the estimates of $\lambda^*$ and $D^*$ 
and their errors, we can obtain the upper bounds
$|2.1 - \lambda^*| < \Delta \lambda = 0.08$ and 
$|1.03 - D^*| < \Delta D = 0.04$.  Then, we can estimate
$|\bar{c}_3(\lambda = 2.1)| < \Delta \lambda (d \bar{c}_3/d\lambda)$,
and analogously $|\bar{c}_3(D = 1.03)| < \Delta D (d \bar{c}_3/dD)$.
For $U_4$ at $(Z_a/Z_p)_f = 0.3202$, using the results of Table 
\ref{derivatives}, we have 
\begin{equation}
|\bar{c}_3(\lambda = 2.1)| < 0.004, \qquad\qquad
|\bar{c}_3(D = 1.03)| < 0.005.
\label{c3-improvedmodel}
\end{equation}

\subsubsection{Corrections to scaling in the standard XY model}

We simulated the standard XY model on lattices with linear sizes
$L=6$, $8$, $10$, $12$, $16$, $18$, $20$, $22$, $24$, $28$, $32$,
$48$, $56$, and $64$ at $\beta_s=0.454165$, which is the estimate of
$\beta_c$ of Ref.\ \cite{BFMM-96}.  Here, we used only the
wall-cluster algorithm for the update.  In one cycle we performed 12
wall-cluster updates.  For $L\le 16$ we performed $10^8$ cycles.  For
lattice sizes $16 \le L \le 64$, we spent roughly the same amount of
CPU time for each lattice size.  For $L=64$ the statistics is
$2.73\times10^6$ measurements.

We determine the amplitude of the corrections to scaling for
$\bar{R}$ with $R_{1,f}=(Z_a/Z_{p})_f=0.3202$ and $R_2=U_4$.  Other
choices lead to similar results.  We fit our numerical results
with the ansatz (\ref{fitbar}), where we fix $\omega=0.8$.  The
results are given in Table \ref{XYcorrections}.
\begin{table}[tp]
\caption{\label{XYcorrections}
Corrections to scaling in the standard XY model for 
$\bar{R}$ with $R_2=U_4$ and $R_{1,f}=(Z_a/Z_{p})_f=0.3202$.
We use the ansatz (\ref{fitbar}) with $\omega=0.8$ fixed.
}
\begin{tabular}{ccccc}
$L_{\rm min}$ & $L_{\rm max}$ & $\chi^2$/d.o.f. & 
  $\bar{R}^*$ & $\bar{c}_3$ \\
\hline
 12  &  64  &   1.78  & 1.2432(1)   &  $-$0.1120(7)\0  \\
 16  &  64  &   0.73  & 1.2430(1)   &  $-$0.1087(13)\\
 20  &  64  &   0.38  & 1.2427(2)   &  $-$0.1048(22)\\
 24  &  64  &   0.24  & 1.2429(2)   &  $-$0.1083(34)\\
 12  &  32  &   2.32  & 1.2433(1)   &  $-$0.1124(8)\0  \\
\end{tabular}
\end{table}
Note that the results for $\bar{R}^*$ are consistent with the result
obtained from the joint fit of the two improved models.  In Table
\ref{binderfixzazp} we obtained, e.g., $\bar{R}^*=1.24301(8)$ with
$L_{\rm min}=20$ and $L_{\rm max}=80$.

\begin{figure}[tp]
\null\vskip 5mm
\centerline{\psfig{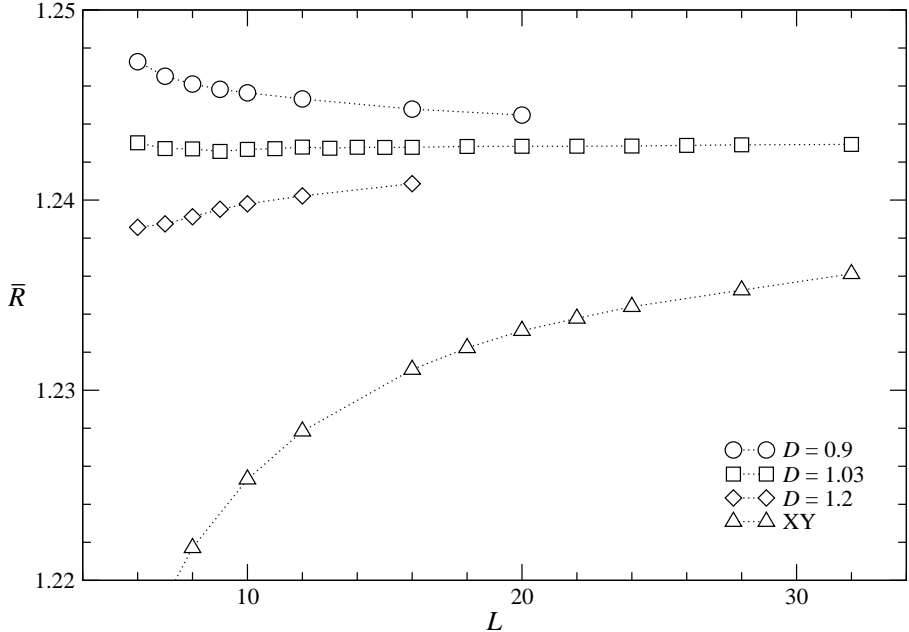}}
\vskip 2mm
\caption{\label{figcor}
Corrections to scaling for the dd-XY model at $D=0.9$, $1.03$, and
$1.2$, and for the standard XY model. We plot $\bar{R}$ with
$R_{1,f}=(Z_a/Z_{p})_f = 0.3202$ and $R_2=U_4$ as a function of the
lattice size.  The dotted lines should only guide the eye.
}
\end{figure}

Corrections to scaling are clearly visible, see Fig.\ \ref{figcor}.
From the fit with $L_{\rm min}=20$ and $L_{\rm max}=64$ we obtain
$\bar{c}_3=-0.1048(22)$.  For the following discussion no estimate of
the possible systematic errors of $\bar{c}_3$ is needed.  Comparing
with Eq.\ (\ref{c3-improvedmodel}), we see that in the (approximately)
improved models the amplitude of the leading correction to scaling is
at least reduced by a factor of 20.  Note, that even if this result
was obtained by considering a specific observable, $U_4$ at fixed
$Z_a/Z_p$, the universality of the ratios of the subleading
corrections implies the same reduction for any quantity.  In the
following section we will use this result to estimate the systematic
error on our results for the critical exponents.

\subsection{Critical exponents from finite-size scaling}

As discussed in Sec.\ \ref{FSStheory}, we may use the derivative of
phenomenological couplings taken at $\beta_f$ in order to determine
$y_t$.  Given the four phenomenological couplings that we have
implemented, this amounts to 16 possible combinations.  In the
following we will restrict the discussion to two choices: in both
cases we fix $\beta_f$ by $(Z_a/Z_{p})_f=0.3202$.  At $\beta_f$ we
consider the derivative of the Binder cumulant and the derivative of
$Z_a/Z_p$.  In Table \ref{ddfzazpdb} we summarize the results of the
fits with the ansatz
\begin{equation}
\label{nufit}
\left . \frac{\partial R}{\partial \beta} \right|_{\beta_f}
 =  c  L^{1/\nu}
\end{equation}
for the $\phi^4$ model at $\lambda=2.1$, the 
dd-XY model at $D=1.03$, and the standard XY model.  

\begin{table}[tp]
\caption{\label{ddfzazpdb}
Fit results for the critical exponent $\nu$ obtained from the
ansatz~(\ref{nufit}).  In all cases $\beta_f$ is fixed by 
$Z_a/Z_{p,f} =0.3202$.  We analyze the $\phi^4$ model at
$\lambda=2.1$, the dd-XY model at $D=1.03$, and the standard XY model.
We consider the slope of the Binder cumulant $U_4$ and of the ratio of
partition functions $Z_a/Z_p$.  We included all data with 
$L_{\rm min} \le L \le L_{\rm max}$ into the fit.
}
\begin{tabular}{cccl}
$L_{\rm min}$ & $L_{\rm max}$ & $\chi^2/$d.o.f. &
\multicolumn{1}{c}{$\nu$} \\
\hline
\multicolumn{4}{c}{$\phi^4$ model:  derivative of $U_4$}\\
\hline
\07&    80& 1.17& 0.67168(12) \\
\09&    80& 0.79& 0.67188(15) \\
 11&    80& 0.85& 0.67181(19) \\
 16&    80& 0.98& 0.67192(34) \\
\hline
\multicolumn{4}{c}{$\phi^4$ model: derivative of $Z_a/Z_p$}\\
\hline
 12&   80& 3.01&  0.67042(9) \\
 16&   80& 1.61&  0.67104(15) \\
 20&   80& 1.04&  0.67139(22) \\
 24&   80& 0.54&  0.67194(32) \\
\hline
\multicolumn{4}{c}{dd-XY model: derivative of $U_4$}\\
\hline
\07&    80 & 2.06 & 0.67258(12) \\
\09&    80 & 1.13 & 0.67216(15) \\
 11&    80 & 1.19 & 0.67209(19) \\
 16&    80 & 0.97 & 0.67154(31) \\
\hline
\multicolumn{4}{c}{dd-XY model:  derivative of $Z_a/Z_p$}\\
\hline
 12&    80& 1.89&  0.67017(9)  \\
 16&    80& 1.60&  0.67046(14) \\
 20&    80& 0.79&  0.67099(21) \\
 24&    80& 0.80&  0.67113(30) \\
\hline
\multicolumn{4}{c}{XY model:  derivative of $U_4$}\\
\hline
 12 & 64 & 4.48 &  0.66450(28) \\
 16 & 64 & 1.30 &  0.66618(42) \\
 20 & 64 & 0.54 &  0.66740(63) \\
\hline
\multicolumn{4}{c}{XY model: derivative of $Z_a/Z_p$}\\
\hline
 12 & 64 & 1.33 & 0.67263(13)\\
 16 & 64 & 0.69 & 0.67300(19)\\
 20 & 64 & 0.30 & 0.67325(30)\\
 24 & 64 & 0.25 & 0.67327(41)\\
\end{tabular}
\end{table}

We see that for the same $L_{\rm min}$ and $L_{\rm max}$ the
statistical error on the estimate of $\nu$ obtained from the
derivative of $Z_a/Z_p$ is smaller than that obtained from the
derivative of $U_4$.  On the other hand, for the two improved models,
scaling corrections seem to be larger for $Z_a/Z_p$ than for $U_4$.

In the case of $Z_a/Z_p$, for both improved models, the result of the
fit for $\nu$ is increasing with increasing $L_{\rm min}$.  In the
case of the Binder cumulant, it is increasing with $L_{\rm min}$ for
the $\phi^4$ model and decreasing for the dd-XY model.
The fact that scaling corrections affect the two quantities and the
two improved models in a quite different way suggests that systematic
errors in the estimate of $\nu$ can be estimated from the variation of
the fits presented in Table \ref{ddfzazpdb}.

As our final result we quote $\nu=0.6716(5)$ which is consistent with
the two results from $Z_a/Z_p$ at $L_{\rm min}=24$ and with the
results from $U_4$ at $L_{\rm min}=16$.

In the case of the standard XY model, the derivative of $U_4$ requires
a much larger $L_{\rm min}$ to reach a small $\chi^2/$d.o.f.\ than for
the improved models.  For the derivative of $Z_a/Z_p$ instead a
$\chi^2/$d.o.f.\ $\approx1$ is obtained for an $L_{\rm min}$ similar
to that of the improved models.  Note that the result for $\nu$ from
the derivative of $U_4$ for $L_{\rm min}=16$ is by several standard
deviations smaller than our final result from the improved models,
while the result from the derivative of $Z_a/Z_p$ is by several
standard deviations larger!  Again we have a nice example that a
$\chi^2/$d.o.f.\ $\approx 1$ does not imply that systematic errors due
to corrections that have not been taken into account in the fit are
small.

Remember that in improved models the leading corrections to scaling
are suppressed at least by a factor of 20 with respect to the
standard XY model.  Since the range of lattice sizes is roughly the
same for the XY model and for the improved models, we can just divide
the deviation of the XY results from $\nu=0.6716(5)$ by 20 to obtain
an estimate of the possible systematic error due to the residual
leading corrections to scaling.  For the derivative of $Z_a/Z_p$ we
end up with $0.0001$ and for the derivative of $U_4$ with $0.0003$.

We think that these errors are already taken into account by the
spread of the results for $\nu$ from the derivatives of $U_4$ and
$Z_a/Z_p$ and the two improved models.  Therefore, we keep our
estimate $\nu=0.6716(5)$ with its previous error bar.

Next we compute the exponent $\eta$.  For this purpose we study the
finite-size behavior of the magnetic susceptibility at $\beta_f$.  In
the following we fix $\beta_f$ by $R_{1,f}=(Z_a/Z_{p})_f=0.3202$.
Other choices for $R_{1,f}$ give similar results.

In a first attempt we fit the data of the two improved models and
the standard XY model to the simple ansatz
\begin{equation}
\label{etasimple}
\left . \chi \right |_{\beta_f} = c \, L^{2-\eta} .
\end{equation}
The results are summarized in Table \ref{etasimpletable}.

\begin{table}[tp]
\caption{\label{etasimpletable}
Results for the critical exponent $\eta$ from the FSS of the magnetic
susceptibility. Fits with ansatz (\ref{etasimple}).  All data with
$L_{\rm min} \le  L \le L_{\rm max}$ are taken into account.
}
\begin{tabular}{cccc}
$L_{\rm min}$ & $L_{\rm max}$ & $\chi^2/$ d.o.f. & $\eta$ \\
\hline
\multicolumn{4}{c}{$\phi^4$ model}\\
\hline
20   &   80  &   2.44 &  0.0371(1) \\
24   &   80  &   0.73 &  0.0375(1) \\
28   &   80  &   0.94 &  0.0375(2) \\
32   &   80  &   0.41 &  0.0378(3) \\
\hline
\multicolumn{4}{c}{dd-XY model}\\
\hline
20   &   80  &   1.88 &  0.0371(1) \\
24   &   80  &   1.19 &  0.0373(1) \\
28   &   80  &   1.52 &  0.0374(2) \\
32   &   80  &   1.24 &  0.0376(2) \\
\hline
\multicolumn{4}{c}{XY model}\\
\hline
 20  & 64  & 7.92  & 0.0325(2)  \\
 24  & 64  & 1.81  & 0.0344(2)  \\
 28  & 64  & 0.27  & 0.0340(3)  \\
 32  & 64  & 0.06  & 0.0342(4)  \\
\end{tabular}
\end{table}

For all three models rather large values of $L_{\rm min}$ are needed
in order to reach a $\chi^2/$d.o.f.\ close to one.  In all cases the
estimate of $\eta$ is increasing with increasing $L_{\rm min}$.  For
$L_{\rm min}=24$ the result for $\eta$ from the standard XY model is
lower than that of the improved models by an amount of approximately
$0.0030$ .  Therefore, the systematic error due to leading corrections
on the results obtained in the improved models should be smaller than
$0.0030/20=0.00015$.  Given this tiny effect, it seems plausible that,
for the improved models, the increase of the estimate of $\eta$ with
increasing $L_{\rm min}$ is caused by subleading corrections.
Therefore, we consider $0.0375$, which is the result of the fit with
$L_{\rm min}=24$ in the $\phi^4$ model, as a lower bound of $\eta$.

Finally, we perform a fit which takes into account the analytic
background of the magnetic susceptibility.  In Ref.\ \cite{HT-99}, it
was shown that the addition of a constant term to Eq.\ 
(\ref{etasimple}) leads to a small $\chi^2$/d.o.f.\ already for small
$L_{\rm min} < 10$.  Similar results have been found for the Ising
universality class.  This ansatz is not completely correct, since it
does not take into account corrections proportional to $L^{2+y}$ with
$y\approx-1.8$, which formally are more important than the analytic
background.  However, the difference between these exponents is small,
and a four-parameter fit is problematic Therefore, we decided to fit
our data with the ansatz
\begin{equation}
\label{etaback}
\left. \chi \right|_{\beta_f} = c \, L^{2-\eta} + b \, L^\kappa \, ,
\end{equation}
with $\kappa$ fixed to $0.0$ and to $0.2$.  The difference between
the results of the fits with the two values of $\kappa$ will give an
estimate of the systematic error of the procedure.  Results for all
three models are summarized in Table \ref{etabacktable}.

\begin{table}[tp]
\caption{\label{etabacktable}
Results for the critical exponent $\eta$ from the FSS of the magnetic
susceptibility. Fits with ansatz (\ref{etaback}).  All data with
$L_{\rm min} \le  L  \le L_{\rm max}$ are taken into account.
}
\begin{tabular}{llllll}
 & & \multicolumn{2}{c}{fit with $\kappa=0.0$} &
     \multicolumn{2}{c}{fit with $\kappa=0.2$} \\
\multicolumn{1}{c}{$L_{\rm min}$} &
\multicolumn{1}{c}{$L_{\rm max}$} &
\multicolumn{1}{c}{$\chi^2/$ d.o.f.} & 
\multicolumn{1}{c}{$\eta$} &
\multicolumn{1}{c}{$\chi^2/$ d.o.f.} & 
\multicolumn{1}{c}{$\eta$} \\
\hline
\multicolumn{6}{c}{$\phi^4$ model}\\
\hline
\08 & 80 & 0.72 & 0.0386(1) & 1.16 & 0.0391(1) \\
 10 & 80 & 0.68 & 0.0385(1) & 1.27 & 0.0388(1) \\
 12 & 80 & 0.75 & 0.0385(1) & 0.81 & 0.0388(1) \\
 14 & 80 & 0.84 & 0.0386(2) & 0.92 & 0.0388(2) \\
 16 & 80 & 0.72 & 0.0384(2) & 0.73 & 0.0386(2) \\
 20 & 80 & 0.88 & 0.0384(3) & 0.88 & 0.0385(4) \\
\hline
\multicolumn{6}{c}{dd-XY model}\\
\hline
\08 & 80 & 1.85 & 0.0387(1) & 3.06 & 0.0391(1) \\
 10 & 80 & 0.95 & 0.0384(1) & 1.15 & 0.0388(1) \\
 12 & 80 & 0.99 & 0.0384(1) & 1.04 & 0.0386(1) \\
 14 & 80 & 0.94 & 0.0384(2) & 1.03 & 0.0386(2) \\ 
 16 & 80 & 0.85 & 0.0383(2) & 1.14 & 0.0384(2) \\
 20 & 80 & 0.90 & 0.0381(4) & 0.90 & 0.0382(4) \\
\hline
\multicolumn{6}{c}{XY model}\\
\hline
 12 & 64 & 0.64 & 0.0350(2) \\
 16 & 64 & 0.48 & 0.0353(3) \\
 20 & 64 & 0.37 & 0.0358(5) \\
\end{tabular}
\end{table}

The value $\chi^2$/d.o.f.\ is close to one for $L_{\rm min}=8$ for the
$\phi^4$ model and $L_{\rm min}=10$ for the dd-XY model, and it does
not allow to discriminate between the two choices of $\kappa$.  The
values of $\eta$ are rather stable as $L_{\rm min}$ is varied,
although there is a slight trend towards smaller results as $L_{\rm
min}$ increases; the trend seems to be stronger for $\kappa=0.2$.
Moreover, the results from the two models are in good agreement.

The fits for the XY model also give a good $\chi^2$/d.o.f.\ for
$L_{\rm min}\ge12$; the value of $\eta$ is however much too small, and
shows an increasing trend.  We can estimate from the difference between
the XY model and the improved models at $L_{\rm min}=16$ that the
error on the value of $\eta$ obtained from improved models, induced by
residual leading scaling corrections, is smaller than
$0.003/20=0.00015$.

From the results for the improved models reported in Table
\ref{etabacktable}, one would be tempted to take $\eta=0.0384$ as
the final result.  However, as we can see from the results for the XY
model, we should not trust blindly the good $\chi^2/$d.o.f.\ of these
fits.  Taking into account the decreasing trend of the values of
$\eta$ for the improved models, we assign the conservative upper bound
$\eta<0.0385$.  By combining it with the lower bound obtained from
ansatz (\ref{etasimple}), we obtain our final result
\begin{equation}
 0.0375 < \eta < 0.0385, \qquad
 \hbox{i.e., } \qquad \eta = 0.0380(5) .
\end{equation}

\section{High-temperature determination of critical exponents}
\label{HTanalysis}

In this Section we report the results of our analyses of the HT
series.  The details are reported in App.\ \ref{seriesanalysis}.

We compute $\gamma$ and $\nu$ from the analysis of the HT expansion
to $O(\beta^{20})$ of the magnetic susceptibility and of the
second-moment correlation length.  In App.\ \ref{exponents} we report
the details and many intermediate results so that the reader can judge
the quality of our results without the need of performing his own
analysis.  This should give an idea of the reliability of our
estimates and of the meaning of the errors we quote, which depend on
many somewhat arbitrary choices and are therefore partially
subjective.

We analyze the HT series by means of integral approximants (IA's) of
first, second, and third order.  The most precise results are obtained
biasing the value of $\beta_c$, using its MC estimate.  We consider
several sets of biased IA's, and for each of them we obtained
estimates of the critical exponents.  These results are reported in
App.\ \ref{exponents}.  All sets of IA's give substantially consistent
results. Moreover, the results are also stable with respect to the
number of terms of the series, so that there is no need to perform
problematic extrapolations in the number of terms in order to obtain
the final estimates.  The error due to the uncertainty on $\lambda^*$
and $D^*$ is estimated by considering the variation of the results
when changing the values of $\lambda$ and $D$.

Using the intermediate results reported in App.\ \ref{exponents}, we
obtain the estimates of $\gamma$ and $\nu$ shown in Table
\ref{finalres}.  We report on $\gamma$ and $\nu$ three contributions
to the error.  The number within parentheses is basically the spread
of the approximants at the central estimate of $\lambda^*$ ($D^*$)
using the central value of $\beta_c$.  The number within brackets is
related to the uncertainty on the value of $\beta_c$ and is estimated
by varying $\beta_c$ within one error bar at $\lambda=\lambda^*$ or
$D=D^*$ fixed.  The number within braces is related to the uncertainty
on $\lambda^*$ or $D^*$, and is obtained by computing the variation of
the estimates when $\lambda^*$ or $D^*$ vary within one error bar,
changing correspondingly the values of $\beta_c$.  The sum of these
three numbers should be a conservative estimate of the total error.

\begin{table}[tp]
\caption{\label{finalres}
Estimates of the critical exponents obtained from the analysis of the
HT expansion of the improved $\phi^4$ lattice Hamiltonian and dd-XY
model.
}
\begin{tabular}{cr@{}lr@{}lr@{}lr@{}l}
\multicolumn{1}{c}{}&
\multicolumn{2}{c}{$\gamma$}&
\multicolumn{2}{c}{$\nu$}&
\multicolumn{2}{c}{$\eta$}&
\multicolumn{2}{c}{$\alpha$}\\
\tableline \hline
$\phi^4$ Hamiltonian &  1&.31780(10)[27]\{15\} & 0&.67161(5)[12]\{10\}
            & 0&.0380(3)\{1\} & $-$0&.0148(8)  \\
dd-XY model          &  1&.31748(20)[22]\{18\} & 0&.67145(10)[10]\{15\}
            & 0&.0380(6)\{2\} & $-$0&.0144(10) \\
\end{tabular}
\end{table}

We determine our final estimates by combining the results for the two
improved Hamiltonians: we take the weighted average of the two
results, with an uncertainty given by the smallest of the two errors.
We obtain for $\gamma$ and $\nu$
\begin{eqnarray}
\gamma &=& 1.3177(5), \label{resga}\\
\nu    &=& 0.67155(27), \label{resnu}
\end{eqnarray}
and by the hyperscaling relation $\alpha=2-3\nu$
\begin{equation}
\alpha=-0.0146(8).
\end{equation} 
Consistent results, although significantly less precise (approximately
by a factor of two), are obtained from the IHT analysis without
biasing $\beta_c$ (see App.\ \ref{exponents}).

From the results for $\gamma$ and $\nu$, we can obtain $\eta$ by the
scaling relation $\gamma=(2-\eta)\nu$. This gives $\eta=0.0379(10)$,
where the error is estimated by considering the errors on $\gamma$ and
$\nu$ as independent, which is of course not true.  We can obtain an
estimate of $\eta$ with a smaller, yet reliable, error by applying the
so-called critical-point renormalization method (CPRM) (see, e.g.,
Refs.\ \cite{int-appr-ref} and references therein) to the series of
$\chi$ and $\xi^2$. The results are reported in Table \ref{finalres}.
We report two contributions to the error on $\eta$, as discussed for
$\gamma$ and $\nu$; the uncertainty on $\beta_c$ does not contribute
in this case. Our final estimate is
\begin{equation}
\eta=0.0380(4).
\label{reseta}
\end{equation}
Moreover, using the scaling relations we obtain
\begin{eqnarray}
\delta &=& {5- \eta\over 1 +\eta}= 4.780(2), \label{deltaex}\\
\beta &=& {\nu\over 2} \left(1 + \eta\right) = 0.3485(2), \label{betaex}
\end{eqnarray}
where the error on $\beta$ has been estimated by considering the
errors of $\nu$ and $\eta$ as independent.

\section{The critical equation of state}
\label{CES}

\subsection{General properties of the critical equation of state
            of XY models}

We begin by introducing the Gibbs free-energy density
\begin{equation}
G(H) = {1\over V} \log Z(H),
\end{equation}
and the related Helmholtz free-energy density
\begin{equation}
{\cal F}(M) = \vec{M}\cdot\vec{H} - G(H),
\end{equation}
where $V$ is the volume, $\vec{M}$ the magnetization density,
${\vec{H}}$ the magnetic field, and the dependence on the temperature
is understood in the notation. A strictly related quantity is the
equation of state which relates the magnetization to the external field
and the temperature:
\begin{equation}
 \vec{H} = {\partial {\cal F}(M)\over \partial \vec{M}}.
\end{equation}
In the critical limit, the Helmholtz free energy obeys a general
scaling law.  Indeed, for $t\to 0$, $|M|\to 0$, and $t|M|^{-1/\beta}$
fixed, it can be written as
\begin{equation}
\Delta {\cal F} = {\cal F}(M) - {\cal F}_{\rm reg}(M) \sim
    t^{2-\alpha} \widehat{\cal F} (|M| t^{-\beta}),
\label{scaling-calF}
\end{equation}
where ${\cal F}_{\rm reg}(M)$ is a regular background contribution.
The function $\widehat{\cal F}$ is universal apart from trivial
rescalings.

The Helmholtz free energy is analytic outside the critical point and
the coexistence curve (Griffiths' analyticity).  Therefore, it has a
regular expansion in powers of $|M|$ for $t>0$ fixed, which we write
in the form
\begin{equation}
\Delta {\cal F} =\, 
{1\over 2} m^2\varphi^2 +
  \sum_{j=2} m^{3-j} {1\over (2j)!} g_{2j} \varphi^{2j},
\end{equation}
where $m=1/\xi$, $\xi$ is the second-moment correlation length,
$\varphi$ is a renormalized magnetization, and $g_{2j}$ are the
zero-momentum $2j$-point couplings.  By performing a further rescaling
$\varphi = m^{1/2}z/\sqrt{g_4} $, the free energy can be written as
\cite{GZ-97}
\begin{equation}
\Delta {\cal F} = {m^3\over g_4}A(z),
\label{dAZ}
\end{equation}
where
\begin{equation}
A(z) = {1\over 2} z^2 + {1\over 4!} z^4
       + \sum_{j=3} {1\over (2j)!} r_{2j} z^{2j}.
\label{AZ}
\end{equation}
Note that $z\propto |M| t^{-\beta}$ for $t\to0$, so that Eq.\ 
(\ref{AZ}) is nothing but the expansion of 
$\widehat{\cal F}(|M| t^{-\beta})$ for $|M| t^{-\beta}\to 0$.
Correspondingly, by using the scaling relation $\beta(1 + \delta) = 2
- \alpha$, we obtain for the equation of state
\begin{equation}
H\propto t^{\beta\delta} F(z),
\label{eqa}
\end{equation}
with
\begin{equation}
F(z) \equiv {\partial A(z)\over \partial z} =
z + \case{1}{6}z^3 + \sum_{j=3} {r_{2j}\over(2j-1)!} z^{2j-1}.
\label{Fzdef}
\end{equation}
Because of Griffiths' analyticity, ${\cal F}(M)$ has also a regular
expansion in powers of $t$ for $|M|$ fixed. Therefore,
\begin{equation}
\Delta {\cal F}(M) = 
   \sum_{k=0}^\infty {\cal F}_k(M)\, t^k = \,
   t^{2-\alpha} \sum_{k=0}^\infty 
   \left[ {\cal F}_k(M)\,  |M|^{(k+\alpha-2)/\beta} \right] 
   \left( t^{-\beta} |M|\right )^{(2 - \alpha - k)/\beta},
\label{Fisoterma}
\end{equation}
where, because of Eq.\ (\ref{scaling-calF}), the coefficients 
${\cal F}_k(M)$ scale as $|M|^{-(k+\alpha-2)/\beta}$ for $|M|\to 0$.
From this expression we immediately obtain the large-$z$ expansion of
$F(z)$,
\begin{equation}
F(z) = z^\delta \sum_{k=0} F^{\infty}_k z^{-k/\beta},
\label{asyFz}
\end{equation}
where we have used again $\beta(1 + \delta) = 2 - \alpha$.  The
function $F(z)$ is defined only for $t > 0$, and thus, in order to
describe the low-temperature region $t<0$, one should perform an
analytic continuation in the complex $t$ plane \cite{ZJbook,GZ-97}.
The coexistence curve corresponds to a complex 
$z_0 = |z_0| e^{-i\pi\beta}$ such that $F(z_0) = 0$.  Therefore, the
behavior near the coexistence curve is related to the behavior of
$F(z)$ in the neighborhood of $z_0$.  The constants $F_0^\infty$ and
$|z_0|$ can be expressed in terms of universal amplitude ratios, by
using the asymptotic behavior of the magnetization along the critical
isotherm and at the coexistence curve:
\begin{eqnarray}
F_0^\infty &=& { (C^+)^{(3\delta-1)/2} \over (\delta C^c)^\delta
(-C_4^+)^{(\delta-1)/2}},
\label{f0inf} \\
|z_0|^2 &=& R_4^+\equiv - C_4^+B^2 / (C^+)^{3},
\label{z0}
\end{eqnarray}
where the critical amplitudes are defined in App.\ \ref{univra}.

The function $F(z)$ provides in principle the full equation of state.
However, it has the shortcoming that temperatures $t<0$ correspond to
imaginary values of the argument. It is thus more convenient to use a
variable proportional to $t|M|^{-1/\beta}$, which is real for all
values of $t$. Therefore, it is convenient to rewrite the equation of
state in a different form,
\begin{equation}
\vec{H} = \vec{M} |M|^{\delta-1} f(x), \qquad\qquad 
          x \propto t|M|^{-1/\beta},
\label{eqstfx}
\end{equation}
where $f(x)$ is a universal scaling function normalized in such a way
that $f(-1)=0$ and $f(0)=1$.  The two functions $f(x)$ and $F(z)$ are
clearly related:
\begin{equation}
z^{-\delta} F(z) = F_0^\infty f(x), \qquad\qquad z = |z_0| x^{-\beta}.
\end{equation}
It is easy to reexpress the results we have obtained for $F(z)$ in
terms of $x$. The regularity of $F(z)$ for $z\to 0$ implies a
large-$x$ expansion of the form
\begin{equation}
f(x) = x^\gamma \sum_{n=0}^\infty f_n^\infty x^{-2n\beta}.
\label{largexfx}
\end{equation}
The coefficients $f_n^\infty$ can be expressed in terms of $r_{2n}$
using Eq.\ (\ref{Fzdef}). We have
\begin{equation}
f_n^\infty = |z_0|^{2n+1-\delta} {r_{2n+2}\over F_0^\infty (2 n + 1)!},
\end{equation}
where we set $r_2 = r_4 = 1$. In particular, using Eqs.\ (\ref{f0inf})
and (\ref{z0}),
\begin{equation} 
f_0^\infty = R_\chi^{-1},
\end{equation}
where $R_\chi$ is defined in App.\ \ref{univra}.  Finally, we notice
that Griffiths' analyticity implies that $f(x)$ is regular for $x>-1$.
In particular, it has a regular expansion in powers of $x$. The
corresponding coefficients are easily related to those appearing in
Eq.\ (\ref{asyFz}).

We want now to discuss the behavior of $f(x)$ for $x\to -1$, i.e., at
the coexistence curve.  General arguments predict that at the
coexistence curve the transverse and longitudinal magnetic
susceptibilities behave respectively as
\begin{equation}
\chi_T = {M\over H}  ,\qquad\qquad
\chi_L = {\partial M\over \partial H} \propto H^{-1/2}.
\label{chitl}
\end{equation}
In particular the singularity of $\chi_L$ for $t<0$ and $H\to0$ is
governed by the zero-temperature infrared-stable Gaussian fixed point
\cite{BW-73,BZ-76,Lawrie-81}, leading to the prediction
\begin{equation}
f(x) \sim  c_f \,(1+x)^2 \qquad\qquad {\rm for}
\qquad x\rightarrow -1.
\label{fxcc} 
\end{equation}
The nature of the corrections to the behavior (\ref{fxcc}) is less
clear.  It has been conjectured \cite{WZ-75,SH-78,Lawrie-81}, using
essentially $\epsilon$-expansion arguments, that, for $y\to 0$, i.e.,
near the coexistence curve, $v\equiv 1+x$ has a double expansion in
powers of $y\equiv H M^{-\delta}$ and $y^{(d-2)/2}$.  This implies
that in three dimensions $f(x)$ can be expanded in powers of $v$ at
the coexistence curve.  On the other hand, an explicit calculation
\cite{PV-99} to next-to-leading order in the $1/N$ expansion shows the
presence of logarithms in the asymptotic expansion of $f(x)$ for
$x\rightarrow -1$.  However, they are suppressed by an additional
factor of $v^2 \log v$ compared to the leading behavior (\ref{fxcc}).

It should be noted that for the $\lambda$ transition in ${}^4$He the
order parameter is related to the complex quantum amplitude of helium
atoms.  Therefore, the ``magnetic'' field is not experimentally
accessible, and the function appearing in Eq.\ (\ref{eqstfx}) cannot
be measured directly in experiments.  The physically interesting
quantities are universal amplitude ratios of quantities formally
defined at zero external field, such as $U_0\equiv A^+/A^-$, for which
accurate experimental estimates have been obtained.  On the other
hand, the scaling equation of state (\ref{eqstfx}) is physically
relevant for planar ferromagnetic systems.

\subsection{Small-$M$ expansion of the equation of state
            in the high-temperature phase}
\label{EP}

Using HT methods, it is possible to compute the first coefficients
$g_{2j}$ and $r_{2j}$ appearing in the expansion of the Helmholtz free
energy and of the equation of state, see Eqs.\ (\ref{AZ}) and
(\ref{Fzdef}).  Indeed, these quantities can be expressed in terms of
zero-momentum $2j$-correlation functions and of the correlation
length.

Details of the analysis of the HT series of $g_4$, $r_6$, $r_8$, and
$r_{10}$ are reported in App.\ \ref{ratioofamp}.  We obtained the
results shown in Table \ref{finalres2}.  In Table \ref{summarygj} we
report our final estimates (denoted by IHT), obtained by combining the
results of the two models; we also compare them with the estimates
obtained using other approaches.  Note that our final estimate of
$g_4$ is slightly larger than the result reported in Ref.\ 
\cite{CPRV-00-2} (see Table \ref{summarygj}).  The difference is
essentially due to the different analysis employed here, which should
be more reliable. This point is further discussed in App.\ 
\ref{ratioofamp}.

\begin{table}[tp]
\caption{\label{finalres2}
Estimates of $g_4$, $r_6$, $r_8$, and $r_{10}$ obtained from the
analysis of the HT series for the two improved Hamiltonians.
Final results will be reported in Table \protect\ref{summarygj}.
}
\begin{tabular}{cr@{}lr@{}lr@{}lr@{}l}
\multicolumn{1}{c}{}&
\multicolumn{2}{c}{$g_4$}&
\multicolumn{2}{c}{$r_6$}&
\multicolumn{2}{c}{$r_8$}&
\multicolumn{2}{c}{$r_{10}$}\\
\tableline \hline
$\phi^4$ Hamiltonian & 21&.15(6) & 1&.955(20) & 1&.37(15) & $-$13&(7) \\
dd-XY model          & 21&.13(7) & 1&.948(15) & 1&.47(10) & $-$11&(14)\\
\end{tabular}
\end{table}

\begin{table}[tp]
\caption{\label{summarygj}
Estimates of $g_4$, $r_{6}$, $r_8$, and $r_{10}$ obtained using the
following methods: analyses of improved HT expansions (IHT), of HT
expansions for the standard XY model (HT), of fixed-dimension
perturbative expansions ($d=3$ $g$-exp.), and of $\epsilon$ expansions
($\epsilon$-exp.).  A more precise determination of $r_{10}$ will be
reported in Table \protect\ref{eqstresAB}.
}
\begin{tabular}{cllll}
\multicolumn{1}{c}{}&
\multicolumn{1}{c}{IHT}&
\multicolumn{1}{c}{HT}&
\multicolumn{1}{c}{$d=3$ $g$-exp.}&
\multicolumn{1}{c}{$\epsilon$-exp.}\\
\tableline \hline
$g_4$   & 21.14(6)  & 21.28(9) \cite{BC-98}  & 21.16(5) \cite{GZ-98} &
          21.5(4) \cite{PV-00,PV-98-gr}\\
        & 21.05(6)\cite{CPRV-00-2}           & 21.34(17)\cite{PV-98-gr} &
          21.11 \cite{MN-91} & \\
\hline
$r_6$   & 1.950(15) & 2.2(6) \cite{Reisz-95} & 1.967 \cite{SOUK-99} &
          1.969(12) \cite{PV-00,PV-98-ef} \\
        & 1.951(14) \cite{CPRV-00-2} &  &  & \\
\hline
$r_8$   & 1.44(10)  &                        &  1.641 \cite{SOUK-99} &
          2.1(9) \cite{PV-00,PV-98-ef} \\
        & 1.36(9)\cite{CPRV-00-2}   &   &  &  \\ \hline
$r_{10}$& $-$13(7)  &  &  &  \\ 
\end{tabular}
\end{table}

\subsection{Parametric representations of the equation of state}
\label{preq}

In order to obtain a representation of the equation of state that is
valid in the whole critical region, we need to extend analytically the
expansion (\ref{Fzdef}) to the low-temperature region $t<0$. For this
purpose, we use parametric representations that implement the expected
scaling and analytic properties.  They can be obtained by writing
\cite{Schofield-69,SLH-69,Josephson-69}
\begin{eqnarray}
M &=& m_0 R^\beta m(\theta) ,\nonumber \\
t &=& R(1-\theta^2), \nonumber \\
H &=& h_0 R^{\beta\delta}h(\theta), \label{parrep}
\end{eqnarray}
where $h_0$ and $m_0$ are normalization constants.  The variable $R$
is nonnegative and measures the distance from the critical point in
the $(t,H)$ plane, while the variable $\theta$ parametrizes the
displacement along the lines of constant $R$. The functions
$m(\theta)$ and $h(\theta)$ are odd and regular at $\theta=0$ and at
$\theta=1$.  The constants $m_0$ and $h_0$ can be chosen so that
$m(\theta)=\theta+O(\theta^3)$ and $h(\theta)=\theta+O(\theta^3)$.
The smallest positive zero of $h(\theta)$, which should satisfy
$\theta_0>1$, corresponds to the coexistence curve, i.e., to $T<T_c$
and $H\to 0$. The singular part of the free energy is then given by
\begin{equation}
\Delta {\cal F} = h_0 m_0 R^{2-\alpha} g(\theta),
\end{equation}
where $g(\theta)$ is the solution of the first-order differential
equation
\begin{equation}
(1-\theta^2) g'(\theta) + 2(2-\alpha)\theta g(\theta) = 
\left[(1-\theta^2)m'(\theta) + 2\beta\theta m(\theta)\right] h(\theta)
\label{pp1}
\end{equation}
that is regular at $\theta=1$.  In particular, the ratio $A^+/A^-$ of
the specific-heat amplitudes in the two phases can be derived by using
the relation
\begin{equation}
A^+/A^- = (\theta_0^2 - 1 )^{2-\alpha} {g(0)\over g(\theta_0)}.
\label{APAM}
\end{equation}
The parametric representation satisfies the requirements of regularity
of the equation of state. Singularities can appear only at the
coexistence curve (due, for example, to the logarithms discussed in
Ref.\ \cite{PV-99}), i.e., for $\theta=\theta_0$.  Notice that the
mapping (\ref{parrep}) is not invertible when its Jacobian vanishes,
which occurs when
\begin{equation}
Y(\theta) \equiv (1-\theta^2)m'(\theta) + 2\beta\theta m(\theta)=0.
\label{Yfunc}
\end{equation}
Thus, parametric representations based on the mapping (\ref{parrep})
are acceptable only if $\theta_0<\theta_l$ where $\theta_l$ is the
smallest positive zero of the function $Y(\theta)$.  One may easily
verify that the asymptotic behavior (\ref{fxcc}) is reproduced simply
by requiring that
\begin{equation}
h(\theta)\sim \left( \theta_0 - \theta\right)^2 
        \qquad\qquad{\rm for}\qquad \theta \rightarrow \theta_0.
\label{hcoex}
\end{equation}
The functions $m(\theta)$ and $h(\theta)$ are related to $F(z)$ by
\begin{eqnarray}
&&z = \rho \,m(\theta) \left( 1 - \theta^2\right)^{-\beta},
\label{thzrel} \\
&&F(z(\theta)) = \rho \left( 1 - \theta^2 \right)^{-\beta\delta} h(\theta),
\label{hFrel}
\end{eqnarray}
where $\rho$ is a free parameter \cite{GZ-97,CPRV-99}.  In the exact
parametric equation the value of $\rho$ may be chosen arbitrarily but,
as we shall see, when adopting an approximation procedure the
dependence on $\rho$ is not eliminated.  In our approximation scheme
we will fix $\rho$ to ensure the presence of the Goldstone
singularities at the coexistence curve, i.e., the asymptotic behavior
(\ref{hcoex}).  Since $z=\rho\,\theta+O(\theta^3)$, expanding
$m(\theta)$ and $h(\theta)$ in (odd) powers of $\theta$,
\begin{eqnarray}
m(\theta) &=& \theta  + \sum_{n=1} m_{2n+1}\theta^{2n+1} , \nonumber \\
h(\theta) &=& \theta  + \sum_{n=1} h_{2n+1}\theta^{2n+1} , \label{mhexp}
\end{eqnarray}
and using Eqs.\ (\ref{thzrel}) and (\ref{hFrel}), one can find the
relations among $\rho$, $m_{2n+1}$, $h_{2n+1}$ and the coefficients
$r_{2n}$ of the expansion of $F(z)$.

Following Ref.\ \cite{CPRV-00-2}, we construct approximate polynomial
parametric representations that have the expected singular behavior at
the coexistence curve \cite{BW-73,BZ-76,Lawrie-81,PV-99} (Goldstone
singularity) and match the known small-$z$ expansion (\ref{Fzdef}).
We will not repeat here in full the discussion of Ref.\ 
\cite{CPRV-00-2}, which should be consulted for more details.  We
consider two distinct schemes of approximation.  In the first one,
which we denote by (A), $h(\theta)$ is a polynomial of fifth order
with a double zero at $\theta_0$, and $m(\theta)$ a polynomial of
order $(1+2n)$:
\begin{eqnarray}
{\rm scheme}\quad({\rm A}):\qquad\qquad 
&&m(\theta) = \theta 
    \left(1 + \sum_{i=1}^n c_{i}\theta^{2i}\right), \nonumber \\
&&h(\theta) = \theta \left( 1 - \theta^2/\theta_0^2 \right)^2. 
\label{scheme1}
\end{eqnarray}
In the second scheme, denoted by (B), we set 
\begin{eqnarray}
{\rm scheme}\quad({\rm B}):\qquad\qquad 
&&m(\theta) = \theta, \nonumber \\
&&h(\theta) = \theta 
    \left(1 - \theta^2/\theta_0^2 \right)^2
    \left( 1 + \sum_{i=1}^n c_{i}\theta^{2i}\right).
\label{scheme2}
\end{eqnarray}
Here $h(\theta)$ is a polynomial of order $5+2n$ with a double zero at
$\theta_0$.  Note that for scheme (B)
\begin{equation}
Y(\theta) = 1 - \theta^2 + 2\beta\theta^2,
\label{Ytheta_def}
\end{equation}
independently of $n$, so that $\theta_l = (1-2\beta)^{-1}$.
Concerning scheme (A), we note that the analyticity of the
thermodynamic quantities for $|\theta|<\theta_0$ requires the
polynomial function $Y(\theta)$ not to have complex zeroes closer to
the origin than $\theta_0$.

In both schemes the parameter $\rho$ is fixed by the requirement
(\ref{hcoex}), while $\theta_0$ and the $n$ coefficients $c_{i}$ are
determined by matching the small-$z$ expansion of $F(z)$. This means
that, for both schemes, in order to fix the $n$ coefficients $c_i$ we
need to know $n+1$ values of $r_{2j}$, i.e., $r_6,...r_{6+2n}$.  As
input parameters for our analysis we consider the estimates obtained
in this paper, i.e., $\alpha = -0.0146(8)$, $\eta=0.0380(4)$,
$r_6=1.950(15)$, $r_8=1.44(10)$, $r_{10}=-13(7)$.

Before presenting our results, we mention that the equation of state
has been recently studied by MC simulations of the standard XY model,
obtaining a fairly accurate determination of the scaling function
$f(x)$ \cite{EHMS-00}. In particular we mention the precise result
obtained for the universal amplitude ratio $R_\chi$ (see App.\ 
\ref{univra} for its definition), i.e., $R_\chi=1.356(4)$, and for the
constant $c_f$, i.e., $c_f=2.85(7)$, where $c_f$ is defined in Eq.\ 
(\ref{fxcc}).  In the following we will take into account these
results to find the best parametrization within our schemes (A) and
(B).

\begin{figure}[tp]
\vspace{0cm}
\centerline{\psfig{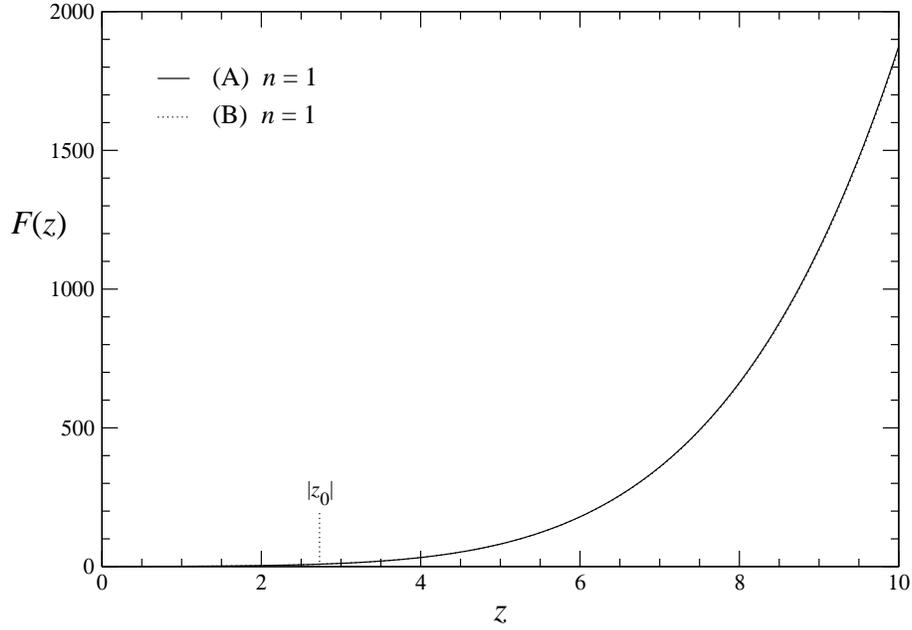}}
\vspace{0cm}
\caption{\label{figFzXY}
The scaling function $F(z)$.
}
\end{figure}

By using the few known coefficients $r_{2j}$---essentially $r_6$ and
$r_8$ because the estimate of $r_{10}$ is not very precise---one
obtains reasonably precise approximations of the scaling function
$F(z)$ for all positive values of $z$, i.e., for the whole HT phase up
to $t=0$.  In Fig.\ \ref{figFzXY} we show the curves obtained in
schemes (A) and (B) with $n=1$ that use the coefficients $r_6$ and
$r_8$.  The two approximations of $F(z)$ are practically
indistinguishable.  This fact is not trivial since the small-$z$
expansion has a finite convergence radius which is expected to be
$|z_0|=(R_4^+)^{1/2}\approx 2.7$. Therefore, the determination of
$F(z)$ on the whole positive real axis from its small-$z$ expansion
requires an analytic continuation, which turns out to be effectively
performed by the approximate parametric representations we have
considered.

\begin{figure}[tp]
\vspace{0cm}
\centerline{\psfig{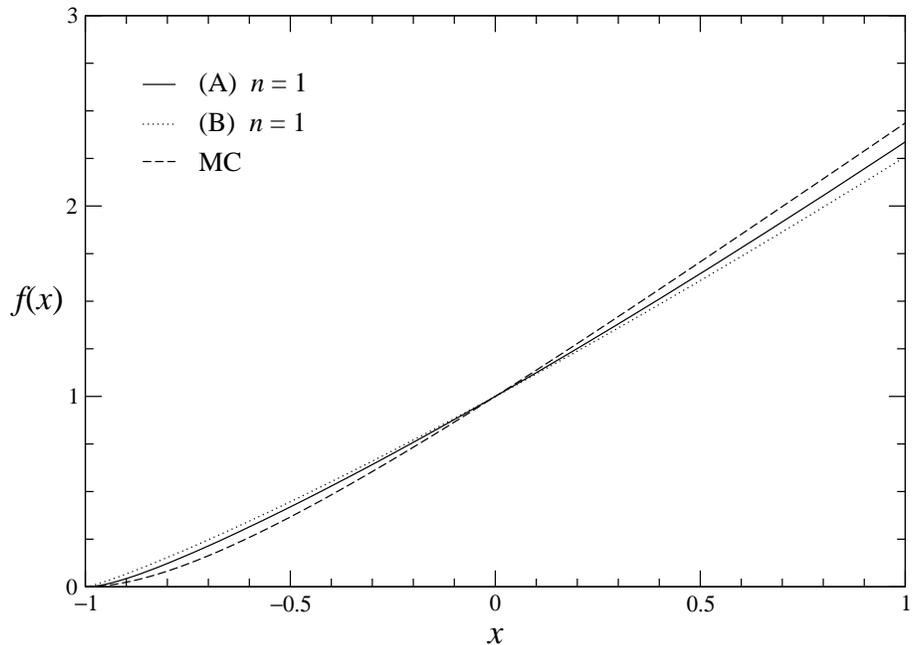}}
\vspace{0cm}
\caption{\label{figfxXY}
The scaling function $f(x)$. 
The MC curve is taken from Ref.\ \protect\cite{EHMS-00}.
}
\end{figure}

Larger differences between the approximations given by schemes (A)
and (B) for $n=1$ appear in the scaling function $f(x)$, which is
shown in Fig.\ \ref{figfxXY}, especially in the region $x<0$, which
corresponds to $t<0$ and $z$ imaginary.  Note that the sizeable
differences for $x>0$ are essentially caused by the normalization of
$f(x)$, which is performed at the coexistence curve $x=-1$ and at the
critical point $x=0$, by requiring $f(-1)=0$ and $f(0)=1$.  Although
the large-$x$ region corresponds to small values of $z$, the
difference between the two approximate schemes does not decrease in
the large-$x$ limit due to their slightly different estimates of
$R_\chi$ (see Table \ref{eqstresAB}). Indeed, for large values of $x$
\begin{equation}
f(x) \sim R_\chi^{-1} x^\gamma.
\end{equation}
In Fig.\ \ref{figfxXY} we also plot the curve obtained in Ref.\ 
\cite{EHMS-00} by fitting the MC data.

\begin{table}[tp]
\caption{\label{eqstresAB}
Universal ratios of amplitudes computed using $\alpha=-0.0146(8)$,
$\eta=0.0380(4)$, $r_6=1.950(15)$ and $r_8=1.44(10)$,
$r_{10}=-13(7)$. The numbers of the first four lines correspond to
central values of the input parameters. The errors reported are only
related to the uncertainty on the input parameters. Numbers marked
with an asterisk are inputs, not predictions.
}
\begin{tabular}{lcccc}
\multicolumn{1}{c}{}&
\multicolumn{1}{c}{[(A) $n=1$; $r_6,r_8$]}&
\multicolumn{1}{c}{[(B) $n=1$; $r_6,r_8$]}&
\multicolumn{1}{c}{[(A) $n=2$; $r_6,r_8,R_\chi$]}&
\multicolumn{1}{c}{[(B) $n=2$; $r_6,r_8,r_{10}$]}\\
\tableline \hline
$\rho$       & 2.22974    & 2.06825  & 2.23092      & 2.04 \\   
$\theta_0^2$ & 3.88383     & 2.93941  & 3.88686      & 2.70 \\  
$c_1$        &$-$0.0260296  & 0.0758028 & $-$0.0265055 & 0.11 \\
$c_2$        &0            & 0        & 0.0002163    & 0.01 \\
\hline
$A^+/A^-$    & 1.062(4)    & 1.064(4) & 1.062(3) & 1.062(5) \\
$R_\xi^+$    & 0.355(3)    & 0.350(1) & 0.355(2) & 0.354(5) \\
$R_c$        & 0.127(6)    & 0.115(2) & 0.126(2) & 0.119(8) \\
$R_\chi$     & 1.35(7)     & 1.50(2)  & $^*$1.356 & 1.45(8) \\
$R_4$        & 7.5(2)      &  7.92(8)  & 7.49(6) & 7.8(3) \\
$F_0^{\infty}$& 0.0302(3)  & 0.0300(2) & 0.0302(2) &  0.0302(4)\\
$r_{10}$     & $-$10(1)   & $-$11.9(1.4) & $-$10(1)  & $^*-$13(7) \\
$c_f$        & 4(2)    & 52(20)  &  4(2) & \\
\end{tabular}
\end{table}

In Table \ref{eqstresAB} we report the results for some universal
ratios of amplitudes. The notations are explained in App.\ 
\ref{univra}. The reported errors refer only to the uncertainty of the
input parameters and do not include the systematic error of the
procedure, which may be determined by comparing the results of the
various approximations.  Comparing the results for $R_\chi$ and $c_f$
with the MC estimates of Ref.\ \cite{EHMS-00}, we observe that the
parametrization (A) is in better agreement with the numerical data.
This is also evident from Fig.\ \ref{figfxXY}.

We also consider both schemes with $n=2$.  If we use $r_{10}$ to
determine the next coefficient $c_2$, scheme (A) is not particularly
useful because it is very sensitive to $r_{10}$, whose estimate has a
relatively large error \cite{CPRV-00-2}.  This fact was already
observed in Ref.\ \cite{CPRV-00-2}, and explained by considerations on
the more complicated analytic structure.  One may instead determine
$c_2$ by using the MC result $R_\chi=1.356(4)$.  The estimates of the
universal amplitude ratios obtained in this way are presented in Table
\ref{eqstresAB}.  They are very close to the $n=1$ case, providing
additional support to our estimates and error bars.  Scheme (B) is
less sensitive to $r_{10}$ and provides reasonable results if we use
$r_{10}$ to fix the coefficient $c_2$ in $h(\theta)$ and impose the
consistency condition $\theta_0 < \theta_l$.  The results are shown in
Table \ref{eqstresAB}, where one observes that they get closer to the
estimates obtained by using scheme (A).

As already mentioned, the most interesting quantity is the
specific-heat amplitude ratio $A^+/A^-$, because its estimate can be
compared with experimental result.  Our results for $A^+/A^-$ are
quite stable and insensitive to the different approximations of the
equation of state we have considered, essentially because they are
obtained from the function $g(\theta)$, which is not very sensitive to
the local behavior of the equation of state, see Eq.\ (\ref{pp1}).
From Table \ref{eqstresAB} we obtain the estimate
\begin{equation}
A^+/A^-=1.062(4).
\end{equation}

In Table \ref{summaryaa} we compare our result (denoted by IHT-PR)
with other available estimates.  Note that there is a marginal
disagreement with the result of Ref.\ \cite{CPRV-00-2}, i.e.,
$A^+/A^-=1.055(3)$, which was obtained using the same method but with
different input parameters: $\alpha = -0.01285(38)$ (the experimental
estimate of Ref.\ \cite{LSNCI-96}), $\eta=0.0381(3)$, $r_6=1.96(2)$,
$r_8=1.40(15)$ and $r_{10}=-13(7)$.  This discrepancy is mainly due to
the different value of $\alpha$, since the ratio $A^+/A^-$ is
particularly sensitive to it.  This fact is also suggested by the
phenomenological relation \cite{HAHS-76} $A^+/A^-\approx 1 - 4\alpha$.

We observe a discrepancy with the experimental result reported in
Refs.\ \cite{LSNCI-96,Lipa-etal-00}, $A^+/A^-=1.0442$.  However, we
note that in the analysis of the experimental data of
Ref.~\cite{LSNCI-96} the estimate of $A^+/A^-$ was strongly correlated
to that of $\alpha$; indeed $A^+/A^-$ was obtained by analyzing the
high- and low-temperature data with $\alpha$ fixed to the value
obtained from the low-temperature data alone. Therefore the slight
discrepancy for $A^+/A^-$ that we observe is again a direct
consequence of the differences in the estimates of $\alpha$.

For the other universal amplitude ratios we quote as our final
estimates the results obtained by using scheme (A) with $n=1$:
\begin{eqnarray}
&&R_\xi^+\equiv (A^+)^{1/3} f^+ = 0.355(3) ,\\
&& R_c\equiv {\alpha A^+ C^+\over B^2} = 0.127(6) ,\\
&&R_\chi \equiv {C^+ B^{\delta-1}\over (\delta C^c)^\delta} = 1.35(7) ,\\
&&R_4\equiv  - {C_4^+ B^2\over (C^+)^3} = |z_0|^2 = 7.5(2) ,\\
&&F_0^\infty\equiv\lim_{z\rightarrow\infty} z^{-\delta} F(z) = 0.0302(3),\\
&&c_f = 4(2).
\end{eqnarray}
These results are substantially equivalent to those reported in Ref.\ 
\cite{CPRV-00-2}.

\begin{table}[tp]
\caption{\label{summaryaa}
Estimates of $A^+/A^-$ obtained in different approaches.
}
\begin{tabular}{llll}
\multicolumn{1}{c}{IHT--PR}&
\multicolumn{1}{c}{$d$=3 exp.}&
\multicolumn{1}{c}{$\epsilon$-exp.}&
\multicolumn{1}{c}{experiments}\\
\tableline \hline
1.062(4) & 1.056(4)\cite{LMSD-98} & 1.029(13)\cite{Bervillier-86} &
           1.0442 \cite{LSNCI-96,Lipa-etal-00}$^{\ref{foot:exp}}$ \\
1.055(3)\cite{CPRV-00-2} & & &
           1.067(3) \cite{SA-84} \\
 & & &
           1.058(4) \cite{LC-83} \\
 & & &
           1.088(7) \cite{TW-82}
\end{tabular}
\end{table}

\section{The two-point function of the order parameter
         in the high-temperature phase}
\label{Gx}

The critical behavior of the two-point correlation function $G(x)$ of
the order parameter is relevant to the description of scattering
phenomena with light and neutron sources.

In the HT critical region, the two-point function $G(x)$ shows a
universal scaling behavior. For $k,m\to0$ ($m\equiv 1/\xi$ and $\xi$
is the second-moment correlation length) with $y\equiv k^2/m^2$ fixed,
we can write \cite{two-point}
\begin{equation}
g(y) = \chi/\widetilde{G}(k).
\end{equation}
The function $g(y)$ has a regular expansion in powers of $y$:
\begin{equation}
g(y)=1 + y + \sum_{i=2}^\infty c_i y^i.
\label{lexp}
\end{equation}
Two other quantities characterize the low-momentum behavior of $g(y)$:
they are given by the critical limit of the ratios
\begin{eqnarray}
S_M&\equiv&m_{\rm gap}^2/m^2,\label{SMdef}\\
S_Z&\equiv& \chi m^2/Z_{\rm gap},\label{SZdef}
\end{eqnarray}
where $m_{\rm gap}$ (the mass gap of the theory) and $Z_{\rm gap}$
determine the long-distance behavior of the two-point function:
\begin{equation}
G(x)\approx  {Z_{\rm gap}\over 4\pi |x|} e^{-m_{\rm gap}|x|}.
\label{largexbehavior}
\end{equation}
If $y_0$ is the negative zero of $g(y)$ that is closest to the origin,
then, in the critical limit, $S_M=-y_0$ and 
$S_Z= \left.\partial g(y) / \partial y \right|_{y=y_0}$.

The coefficients $c_i$ can be related to the critical limit of
appropriate dimensionless ratios of spherical moments of $G(x)$
\cite{CPRV-98,CPRV-99} and can be computed by analyzing the
corresponding HT series in the $\phi^4$ and in the dd-XY models, which
we have calculated to 20th order.  We report only our final estimates
of $c_2$ and $c_3$,
\begin{eqnarray}
&&c_2 = -3.99(4) \times 10^{-4},\label{resc2}\\ 
&&c_3 = 0.09(1) \times 10^{-4},\label{resc3}
\end{eqnarray}
and the bound
\begin{equation}
|c_4| < 10^{-6}.
\label{resc4}
\end{equation}
As already observed in Ref.\ \cite{CPRV-98}, the coefficients show the
pattern
\begin{equation}
|c_i|\ll |c_{i-1}|\ll...\ll |c_2| \ll 1
        \qquad\qquad {\rm for}\qquad i\geq 3.
\label{patternci}
\end{equation}
Therefore, a few terms of the expansion of $g(y)$ in powers of $y$
provide a good approximation in a relatively large region around
$y=0$, larger than $|y|\lesssim 1$. This is in agreement with the
theoretical expectation that the singularity of $g(y)$ closest to the
origin is the three-particle cut (see, e.g., Refs.\ 
\cite{FS-75,Bray-76,CPRV-98}).  If this is the case, the convergence
radius $r_g$ of the Taylor expansion of $g(y)$ is $r_g=9S_M$.  Since,
as we shall see, $S_M\approx 1$, at least asymptotically we should have
\begin{equation}
c_{i+1}\approx -{1\over 9}c_i.
\label{pattern-cip1-ci}
\end{equation}
This behavior can be checked explicitly in the large-$N$ limit of the
$N$-vector model \cite{CPRV-98}.

Assuming the pattern (\ref{patternci}), we may estimate $S_M$ and
$S_Z$ from $c_2$, $c_3$, and $c_4$.  We obtain
\begin{eqnarray}
S_M &=& 1 + c_2 - c_3  + c_4 + 2 c_2^2 + ... \label{SMest}\\
S_Z &=& 1 - 2 c_2 + 3 c_3 - 4 c_4 - 2 c_2^2 + ... \label{SZest}
\end{eqnarray}
where the ellipses indicate contributions that are negligible with
respect to $c_4$.  In Ref.\ \cite{CPRV-98} the relation (\ref{SMest})
has been confirmed by a direct analysis of the HT series of $S_M$.
From Eqs.\ (\ref{SMest}) and (\ref{SZest}) we obtain
\begin{equation}
S_M=0.999592(6),\qquad\qquad S_Z=1.000825(15).
\end{equation}
These results improve those obtained in Ref.\ \cite{CPRV-98} by using
HT methods in the standard XY model and field-theoretic methods, such
as the $\epsilon$ expansion and the fixed-dimension $g$ expansion.

For large values of $y$, the function $g(y)$ follows the Fisher-Langer
law \cite{FL-68}
\begin{equation}
g(y)^{-1} = {A_1\over y^{1 - \eta/2}} 
  \left(1 + {A_2\over y^{(1-\alpha)/(2 \nu)}} +
            {A_3\over y^{1/(2\nu)}}\right).
\label{FL-law}
\end{equation}
The coefficients have been computed in the $\epsilon$ expansion to
three loops \cite{Bray-76}, obtaining
\begin{equation}
A_1 \approx 0.92, \qquad A_2 \approx 1.8, \qquad A_3 \approx - 2.7.
\label{A-eps-exp}
\end{equation}
In order to obtain an interpolation that is valid for all values of
$y$, we will use a phenomenological function proposed by Bray
\cite{Bray-76}.  This approximation has the exact large-$y$ behavior
and its expansion for $y\to 0$ satisfies Eq.\ (\ref{pattern-cip1-ci}).
It requires the values of the exponents $\nu$, $\alpha$, and $\eta$,
and the sum of the coefficients $A_2 + A_3$. For the exponents we use
of course the estimates obtained in this paper, while the coefficient
$A_2 + A_3$ is fixed using the $\epsilon$-expansion prediction 
$A_2 + A_3 = -0.9$. Bray's phenomenological function predicts then the
constants $A_i$ and $c_i$. We obtain:
\begin{eqnarray}
A_1 \approx 0.915, \qquad\qquad A_2 &\approx& - 24.7, 
\qquad\qquad A_3 \approx 23.8, 
\label{A-Bray-approx} \\
c_2 \approx -4.4 \times 10^{-4}, \qquad\qquad
c_3 &\approx& 1.1 \times 10^{-5},  \qquad\qquad
c_4 \approx - 5 \times 10^{-7}. 
\end{eqnarray}
The results for $A_1$, $c_2$, $c_3$, and $c_4$ are in good agreement
with the above-reported estimates, while $A_2$ and $A_3$ differ
significantly from the $\epsilon$-expansion results (\ref{A-eps-exp}).
Notice, however, that, since $|\alpha|$ is very small, the relevant
quantity in Eq.  (\ref{FL-law}) is the sum $A_2 + A_3$ which is, by
construction, equal in Bray's approximation and in the $\epsilon$
expansion: in other words, the function (\ref{FL-law}) does not change
significantly if we use Eq.\ (\ref{A-eps-exp}) or Eq.\ 
(\ref{A-Bray-approx}) for $A_2$ and $A_3$. Thus, Bray's approximation
provides a good interpolation both in the large-$y$ and small-$y$
regions.

\appendix
\section{The Monte Carlo simulation}
\label{montecarloupdate}

\subsection{The Monte Carlo algorithm}
\label{update-cycle}

At present the best algorithm to simulate $N$-vector systems is the
cluster algorithm proposed by Wolff \cite{Wolff} (see Ref.\ 
\cite{CEPS-93} for a general discussion).  However, the cluster update
changes only the angle of the field.  Therefore, following Brower and
Tamayo \cite{BrTa}, we add a local update that changes also the
modulus of the field.

\subsubsection{The wall-cluster update}

We use the embedding algorithm proposed by Wolff \cite{Wolff} with two
major differences. First, we do not choose an arbitrary direction, but
we consider changes of the signs of the first and of the second
component of the fields separately.  Second, we do not use the
single-cluster algorithm to update the embedded model, but the
wall-cluster variant proposed in Ref.\ \cite{HPV-99}. In the
wall-cluster update, one flips at the same time all clusters that
intersect a plane of the lattice. In Ref.\ \cite{HPV-99} we found for
the 3D Ising model a small gain in performance compared with the
single-cluster algorithm.

Note that, since the cluster update does not change the modulus of the
field, identical routines can be used for the $\phi^4$ model and for
the dd-XY model.

\subsubsection{The local update of the $\phi^4$ model}

We sweep through the lattice with a local updating scheme. At each
site we perform a Metropolis step, followed by an overrelaxation step
and by a second Metropolis step.

In the Metropolis update, a proposal for a new field at site $x$ is
generated by
\begin{eqnarray}
\phi_x'^{(1)}  &=& \phi_x^{(1)} + c \, (r^{(1)}-0.5), \nonumber \\
\phi_x'^{(2)}  &=& \phi_x^{(2)} + c \, (r^{(2)}-0.5),
\end{eqnarray}
where $r^{(1)}$ and $r^{(2)}$ are random numbers that are uniformly
distributed in $[0,1)$.  The proposal is accepted with probability
\begin{equation}
A = \mbox{min}[1, \exp(-H'+H)] .
\end{equation}
The step size $c$ is adjusted so that the acceptance rate is
approximately 1/2.

The overrelaxation step is given by
\begin{equation}
\label{overrelax}
\vec{\phi}_x'  =  \vec{\phi}_x - 2 \, 
    \frac{(\vec{\phi}_x \cdot \vec{\phi}_n) \, 
        \vec{\phi}_n}{\vec{\phi}_n^2}\, ,
\end{equation}
where $\vec{\phi}_n = \sum_{y\in{\rm nn}(x)} \vec{\phi}_y$ and
${\rm nn}(x)$ is the set of the nearest neighbors of $x$.  Note that
this step takes very little CPU time. Therefore, it is likely that its
benefit out-balances the CPU cost.

\subsubsection{The local update of the dd-XY model}

We sweep through the lattice with a local updating scheme, performing
at each site one Metropolis update followed by the overrelaxation
update (\ref{overrelax}).

In the Metropolis update, the proposal for the field $\vec{\phi}_x$ at
the site $x$ is given by
\begin{eqnarray}
\label{propose}
\vec{\phi}_x' = (0,0)   \phantom{xxxxxxxx}  \phantom{xx} &
        \mbox{for} & \phantom{xx}
|\vec{\phi}_x|=1 \nonumber \\
\vec{\phi}_x' = (\cos(\alpha), \sin(\alpha))  \phantom{xx} &
        \mbox{for} & \phantom{xx} |\vec{\phi}_x|=0 ,
\end{eqnarray}
where $\alpha$ is a random number with a uniform distribution in
$[0,2\pi)$.  This proposal is accepted with probability
\begin{equation}
\label{accept}
A(\vec{\phi}_x',\vec{\phi}_x)  = \min[1,\exp(-H'+H) ] =
\min[1,\exp(\beta \, \vec{\phi}_n \cdot  (\vec{\phi}_x'-\vec{\phi}_x) 
+ D (\vec{\phi}_x'^2 - \vec{\phi}_x^2) )],
\end{equation}
where $\vec{\phi}_n = \sum_{y\in{\rm nn}(x)} \vec{\phi}_y$ is the sum
of the nearest-neighbor spins. We will prove in Sec.
\ref{proof-stat-ddXY} that this update leaves the Boltzmann
distribution invariant.

\subsubsection{The update cycle}

Finally, we summarize the complete update cycle:

local update sweep;

global field rotation, in which the angle is taken from a
uniform distribution in $[0,2\pi)$;

6 wall-cluster updates.

The sequence of the 6 wall-cluster updates is given by the wall in
1-2, 1-3 and 2-3 plane. In each of the three cases, we update
separately each component of the field.

\subsubsection{Proof of the stationarity
               of the local update of the dd-XY model} 
\label{proof-stat-ddXY}

Stationarity means that the update leaves the Boltzmann distribution
invariant. Since our update is local, it is sufficient to consider the
conditional distribution of a single spin for given neighbors
\begin{equation}
P_b(\vec{\phi}_x) = 
\frac{1}{z} \exp(\beta\,\vec{\phi}_x\cdot\vec{\phi}_n + D \vec{\phi}_x^2) ,
\end{equation}
where 
\begin{equation}
z=\int d \mu(\vec{\phi}_x)
\exp(\beta \, \vec{\phi}_x\cdot  \vec{\phi}_n + D \vec{\phi}_x^2)
\end{equation}
and $d \mu(\vec{\phi}_x)$ is defined in Eq.\ (\ref{lmeasure}).  We
have to prove that
\begin{equation}
\label{stab}
P_b(\vec{\phi}_x')  =  
\int  d  \mu(\vec{\phi}_x) \,
W(\vec{\phi}_x',\vec{\phi}_x) \, P_b(\vec{\phi}_x) 
\end{equation}
is satisfied, where $W$ is the update probability defined by Eqs.\ 
(\ref{propose}) and (\ref{accept}).

Using Eq.\ (\ref{lmeasure}), the right-hand side of Eq.\ (\ref{stab})
can be rewritten as
\begin{equation}
\label{stab2}
W(\vec{\phi}_x',0) \, P_b(0) + 
\int d \Omega({\phi}_x) \,
W(\vec{\phi}_x',\vec{\phi}_x) \, P_b(\vec{\phi}_x) ,
\end{equation}
where $d \Omega({\phi}_x)$ is the normalized measure on the unit
circle.

Let us first consider the case $|\vec{\phi}_x'|=1$. Then, we have
\begin{equation}
W(\vec{\phi}_x',0) \, P_b(0) =\frac{1}{z} \,  A(\vec{\phi}_x',0),
\end{equation}
and 
\begin{eqnarray}
\int d \Omega({\phi}_x) \,
W(\vec{\phi}_x',\vec{\phi}_x) \, P_b(\vec{\phi}_x) &=& 
    W(\vec{\phi}_x',\vec{\phi}_x') \, P_b(\vec{\phi}_x') = 
   (1 - A(0,\vec{\phi}_x')) \, P_b(\vec{\phi}_x')  
\nonumber \\
&=& P_b(\vec{\phi}_x') - \frac{1}{z} \,  A(\vec{\phi}_x',0),
\end{eqnarray}
where we have used the property
\begin{equation}
A(0,\vec\psi ) P_b(\vec\psi) = A(\vec\psi, 0) \, P_b(0),
\label{det-balance}
\end{equation}
valid for $|\vec{\psi}| = 1$. Summing the two terms we obtain Eq.\ 
(\ref{stab}) as required.

For $|\vec{\phi}_x'|=0$, we have 
\begin{equation} 
W(0,0) P_b(0) = {1\over z} 
\left(1 - \int d \Omega({\psi}) \, A(\vec{\psi},0) \right),
\end{equation}
and 
\begin{equation}
\int d \Omega({\phi}_x) \,
W(0,\vec{\phi}_x) \, P_b(\vec{\phi}_x) = 
{1\over z} \int d \Omega({\psi}) \, A(\vec{\psi},0)
\end{equation}
where we have used again Eq.\ (\ref{det-balance}).  Summing the two
terms we obtain Eq.\ (\ref{stab}) as required.

\subsection{Measuring $Z_a/Z_p$}
\label{zazpapp}

One of the phenomenological couplings that we have studied is the
ratio $Z_a/Z_p$ of the partition function $Z_a$ of a system with
anti-periodic boundary conditions (a.b.c.)  in one direction and the
partition $Z_p$ with periodic boundary conditions (p.b.c.) in all
three directions. A.b.c.\ are obtained by multiplying the term
$\vec{\phi}_x \vec{\phi}_y$ in the Hamiltonian by $-1$ for all
$x=(L_1,x_2,x_3)$ and $y=(1,x_2,x_3)$.  This ratio can be obtained
using the so-called boundary-flip algorithm, applied in Ref.\ 
\cite{Ha-93} to the Ising model and generalized in Ref.\ \cite{GH-94}
to general $O(N)$-invariant non-linear $\sigma$ models.

In the boundary-flip algorithm, one considers fluctuating boundary
conditions, i.e., a model with partition function
\begin{equation}
Z_{\rm fluct}  =  Z_p + Z_a  = 
\sum_{J_b=\pm 1} \int \mbox{D}[\phi] \, \exp\left[
        \beta \sum_{\left<xy\right>} J_{\left<xy\right>} 
        \vec{\phi}_x \vec{\phi}_y + ... \right],
\end{equation}
where $J_{\left<xy\right>}=J_b $  for 
$x=(L_1,x_2,x_3)$ and $y=(1,x_2,x_3)$, and $J_{\left<xy\right>}=1 $
otherwise.  $J_b=1$ and $J_b=-1$ correspond to p.b.c.\ and a.b.c.\ 
respectively.

In this notation, the ratio of partition functions is given by
\begin{equation}
\frac{Z_a}{Z_p} = \frac{\langle\delta_{J_b,-1}\rangle}{
\langle\delta_{J_b,1}\rangle} ,
\end{equation}
where the expectation value is taken with fluctuating boundary
conditions.

In order to simulate these boundary conditions, we need an algorithm
that easily allows flips of $J_b$.  This can be done with a special
version of the cluster algorithm.  For both components of the field we
perform the freeze (delete) operation for the links with probability
\begin{equation}
p_d = \mbox{min}[1,
    \exp(-2 \beta J_{\left<xy\right>} \phi_x^{(p)} \phi_y^{(p)})] ,
\end{equation}
where $\phi_x^{(p)}$ is the chosen component of $\phi_x$.  The sign of
$J_b$ can be flipped if there exists, for the first as well as the
second component of the field, no loop of frozen links with odd
winding number in the first direction.  In Ref.\ \cite{Habil} it is
discussed how this can be implemented.  For a more formal and general
discussion, see Ref.\ \cite{CGN-00}.  Note that for $J_b=-1$ the flip
can always be performed. Hence, as F.~Gliozzi and A.~Sokal have
remarked \cite{G-S-pc}, the boundary flip needs not to be performed in
order to determine $Z_a/Z_p$. It is sufficient to use p.b.c.\ and
check if the flip to a.b.c.\ is possible.  Setting $b=1$ if the
boundary can be flipped and $b=0$ otherwise, we have
\begin{equation}
\frac{Z_a}{Z_p} = \langle b\rangle ,
\end{equation}
where the expectation value is taken with p.b.c.

\subsection{Checks of the program}

\subsubsection{Schwinger-Dyson equations}

The properties of the integration measure allow to derive an infinite
set of nontrivial equations among observables of the model. Here we
have used two such equations to test the correctness of the programs
and the reliability of the random-number generator.  For a more
general discussion of such tests, see Ref.\ \cite{BMM-98}.

For the $\phi^4$ model, the partition function remains unchanged when,
at site $x$, the first component of the field is shifted by $\psi$:
$\phi_x^{(1)} \rightarrow \phi_x^{(1)} + \psi$. We obtain, using also
the ${\rm O}(2)$ invariance
\begin{eqnarray}
\label{phicheck}
0 &=& \left. \frac{1}{Z} 
  \frac{\partial \langle\phi_x^{(1)} + \psi\rangle_{\psi}}{\partial \psi}
  \right |_{\psi=0} \nonumber \\
  &=& \beta \sum_{\mu} \langle\vec{\phi}_x \vec{\phi}_{x+\hat\mu}\rangle
   - \left\langle \vec{\phi}_x^2
   + 2 \lambda (\vec{\phi_x}^2 -1) \vec{\phi}_x^2 \right\rangle
   + 1 ,
\end{eqnarray}
where $\langle...\rangle_{\psi}$ indicates that the Boltzmann factor
is taken with the shifted field at the site $x$.  A second equation,
which is valid for the $\phi^4$ model as well as for the dd-XY model,
can be derived from the invariance of the measure under rotations:
\begin{eqnarray}
\phi^{(1)}_x &\rightarrow& 
\cos(\alpha) \, \phi^{(1)}_x - \sin(\alpha) \, \phi^{(2)}_x \, ,
\nonumber \\
\phi^{(2)}_x &\rightarrow& 
\sin(\alpha) \, \phi^{(1)}_x + \cos(\alpha) \, \phi^{(2)}_x \, .
\end{eqnarray}
Taking the second derivative of the partition function with respect to
$\alpha$ yields
\begin{equation} 
\label{rotate}
0 = \left.\frac{1}{Z} \frac{\partial^2 Z(\alpha)}{\partial \alpha^2} 
    \right |_{\alpha=0} = 
   \beta^2 \left\langle \left( \sum_{y\in{\rm nn}(x)}
   [\phi_x^{1} \phi_y^{2}  - \phi_x^{2} \phi_y^{1}] 
   \right)^2 \right \rangle
   - \beta \sum_{y\in{\rm nn}(x)} 
   \langle \vec{\phi}_x \vec{\phi}_y \rangle ,
\end{equation}
where $y\in{\rm nn}(x)$ indicates that the sum runs over the six
nearest neighbors of $x$.

We checked Eq.\ (\ref{phicheck}) for all our simulations of the
$\phi^4$ model. In the simulation we averaged Eq.\ (\ref{phicheck})
over all sites of the lattice to reduce the error bars.  For most of
the simulations the right-hand side of Eq.\ (\ref{phicheck}) differed
by zero by less than 2 standard deviations. Only in one case ($L=8$,
$\lambda=2.2$) the difference was about 3 standard deviations.  The
weighted average of the right-hand side of Eq.\ (\ref{phicheck}) over
all our simulations is $-8(9) \times 10^7$.  Hence, there is no
indication for a program error or a problem with the random-number
generator.

We implemented Eq.\ (\ref{rotate}) in all simulations of the standard
XY model and, unfortunately, only in the most recent simulations of
the dd-XY model.  For $L=96$, $D=1.02$, and $\beta=0.56379$ we found
$\beta_m=0.5637896(24)$ from 575,000 measurements, where
\begin{equation}
\beta_m =
\frac{2 \sum_{\left<xy\right>} \langle\vec{\phi}_x\vec{\phi}_y\rangle}
   {\sum_x \left\langle \left( \sum_{y\in{\rm nn}(x)}
  [\phi_x^{1} \phi_y^{2}  - \phi_x^{2} \phi_y^{1}]
 \right)^2 \right \rangle} .
\end{equation}

We have measured $\beta_m$ in all our simulations of the XY model. In
most cases the deviation of $\beta_m$ from $\beta=0.454165$ was less
than one standard deviation. The largest deviation was
$\beta_m-\beta=-0.0000049(22)$ for $L=28$.  The weighted average over
all simulations is $\beta_m-\beta=-0.00000024(52)$.

Also this check does not indicate a program error or a problem with
the random-number generator.

\subsubsection{Checks of the Taylor expansion}

As a test of the MC program and of the analysis software, we simulated
the $\phi^4$ model for $L=4$ and $\lambda=2.1$ at the following values
of $\beta$: $\beta=0.485$, $0.490$, $0.495$, $0.500$, $0.505$,
$0.510$, $0.515$, and $0.520$.

We computed $\bar{R}$ with $R_{1,f}=(Z_a/Z_{p})_f=0.3202$ and
$R_{1,f}=(\xi_{\rm 2nd}/L)_f=0.5925$ and $R_2=U_4$ and $R_2=U_6$ for
all these simulations, using a third-order Taylor expansion.  The
results for $\bar{R}$ are summarized in Table \ref{test}.  They show
that there is a large interval in which the method works: indeed, the
results for $\bar{R}$ for $\beta_s=0.505$, $0.510$ and $0.515$ agree
within two standard deviations, although the variation of $U_4$ and
$U_6$ at $\beta_s$ is several hundred standard deviations.  In
addition we have gained information about the range of $\beta$ where
the extrapolation works with the desired accuracy:
\begin{equation}
|\beta_s-\beta_f| < 0.005 \times (L/4)^{1/\nu} .
\end{equation}
The factor $(L/4)^{1/\nu}$ takes care of the fact that the slope of
the couplings $R$ scales like $L^{1/\nu}$.  We have carefully checked
that this requirement is always fulfilled in our simulations.
Therefore, we are confident that the extrapolation in $\beta$, using
the Taylor expansion, is implemented correctly.

\begin{table}[tp]
\caption{\label{test}
Test of the Taylor expansion. Simulations with $L=4$,
$\beta_f\approx0.50773$ for $(Z_a/Z_{p})_f$, and
$\beta_f\approx0.50994$ for $(\xi_{\rm 2nd}/L)_f$. In the first column
we give the value $\beta_s$ where the simulations have been performed
(i.e., the configurations have been generated with a weight
proportional to the Boltzmann factor that corresponds to
$\beta_s$). In the columns 2 up to 5 we give the results for four
choices of $\bar{R}$. These choices are labelled by $R_2$ at
$R_{1,f}$. Finally in columns 6 and 7 we give $U_4$ and $U_6$ at
$\beta_s$ for comparison.
}
\begin{tabular}{crrrrrr}
\multicolumn{1}{c}{$\beta_s$} &
\multicolumn{1}{c}{$U_4$ at $(Z_a/Z_{p})_f$} &
\multicolumn{1}{c}{$U_6$ at $(Z_a/Z_{p})_f$} &
\multicolumn{1}{c}{$U_4$ at $(\xi_{\rm 2nd}/L)_f$}&
\multicolumn{1}{c}{$U_6$ at $(\xi_{\rm 2nd}/L)_f$}&
\multicolumn{1}{c}{$U_4$ at $\beta_s$}      &
\multicolumn{1}{c}{$U_6$ at $\beta_s$} \\
\hline
0.485&1.249912(46)&1.76951(17)&1.239726(58)&1.73513(21)&1.360445(71)&2.17525(28)\\
0.490&1.249589(42)&1.76749(15)&1.239863(51)&1.73434(18)&1.334375(67)&2.07444(25)\\
0.495&1.249431(40)&1.76671(14)&1.239866(46)&1.73396(16)&1.309184(63)&1.98000(23)\\
0.500&1.249343(39)&1.76634(14)&1.239805(43)&1.73364(15)&1.284661(60)&1.89063(22)\\
0.505&1.249365(39)&1.76642(13)&1.239824(41)&1.73370(14)&1.261574(57)&1.80882(20)\\
0.510&1.249458(40)&1.76678(14)&1.239885(40)&1.73394(13)&1.239660(54)&1.73318(18)\\
0.515&1.249373(42)&1.76641(14)&1.239802(39)&1.73359(13)&1.218859(50)&1.66302(17)\\
0.520&1.249313(45)&1.76616(15)&1.239804(39)&1.73358(13)&1.199535(47)&1.59940(15)\\
\end{tabular}
\end{table}

\subsection{CPU-time, acceptance rate, cluster size,
            and autocorrelation times} 

We have used ANSI C to implement our simulation programs. We have used
our own implementation of the G05CAF random-number generator from the
NAG-library. The G05CAF is a linear congruential random-number
generator with modulus $m=2^{59}$, multiplier $a=13^{13}$ and
increment $c=0$. Most of our simulations have been performed on 450
MHz Pentium III PCs running the Linux operating system.

For $D=1.03$, the average size of the clusters per volume decreases
from about $0.52$ for $L=5$ to about $0.14$ for $L=80$.  For large
lattices the behavior of the size of the wall clusters per volume is
roughly given by $1.198 \times L^{-0.488}$.  For $\lambda=2.1$ we get
$1.276 \times L^{-0.488}$.

The acceptance rate of the local update of the dd-XY model at $D=1.03$
is about $0.273$ for $L=5$.  It increases slowly and reaches about
$0.309$ for $L=80$.  At the same time the density of
$\vec{\phi}=(0,0)$ spins increases from about $0.150$ to $0.16628$.

The acceptance rate of the local update of the $\phi^4$ model at
$\lambda=2.1$ is about $0.395$ for $L=5$ and increases to $0.4098$ for
$L=80$. We have chosen a step size $c=2.0$ throughout.

For the dd-XY model for $ 16 \le L \le 80$ one cycle of the update
plus the measurement takes about $L^3 \times 3.3 \times 10^{-6}$
seconds on a 450 MHz Pentium III CPU.  In this range of lattice sizes
the decrease of the wall-cluster size per volume and the increase of
memory access times seems to cancel almost exactly (by chance).

Roughly $2/3$ of the CPU-time is spent with the cluster update (which
incorporates the measurements needed for $Z_a/Z_p$). Again about $2/3$
of the remaining time is spent in the local update.

The CPU time required for the update of the $\phi^4$ model is slightly
larger due to a different local update and due to slightly larger wall
clusters than in the dd-XY model.

In total about 5 years on a single 450 MHz Pentium III CPU were used
for our study. (We had 8 CPUs available for the study.)

We performed additional runs of the $\phi^4$ model at $\lambda=2.1$
and $\beta=0.50915$ and the dd-XY model at $D=1.03$ and
$\beta=0.5628$, where we had stored all measurements of the square of
the local field, the energy, the magnetization, $F$, and the boundary
variable. Hence, we could compute autocorrelation functions and times
from these data.  We simulated lattices of size $L=8$ up to $L=64$ for
this purpose.  In all cases we performed 200,000 measurements.  A
measurement was performed after each update cycle.

Our results for the integrated autocorrelation times are summarized in
Table \ref{autoco}. All autocorrelation times are given in units of
measurements.  We see that the autocorrelation times slightly increase
with increasing $L$. At a given lattice size, the autocorrelation
times for the $\phi^4$ model are a little smaller than those for the
dd-XY model.  Likely, this is caused by the particular implementation
of the local updates.  Among the observables that we have studied, the
energy has the largest autocorrelation time.  It is interesting to
note that the autocorrelation time of $\sum_x\vec{\phi_x}^2$ shows no
sign of severe critical slowing down, despite the fact that
$\sum_x\vec{\phi_x}^2$ is only changed by the local update.

\begin{table}[tp]
\caption{\label{autoco}
Integrated autocorrelation times for the $\phi^4$ model at
$\lambda=2.1$ and $\beta=0.50915$, and the dd-XY model at $D=1.03$ and
$\beta=0.5628$. We have always performed 200,000 cycles. The
autocorrelation times are given in units of update cycles. $\tau_E$ is
the integrated autocorrelation time of $\sum_{\left<xy\right>}
\vec{\phi}_x \vec{\phi}_y$, $\tau_{\phi^2}$ of 
$\sum_x \vec{\phi}^2_x$, $\tau_{\chi}$ of the magnetic susceptibility,
$\tau_F$ of the Fourier transform of the field at minimal nonvanishing
momentum, and $\tau_b$ of the boundary variable. ``trunc'' gives the
time at which the summation of the autocorrelation function was
truncated. Note that trunc was chosen to be the same  for all
observables.
}
\begin{tabular}{rlllllc}
\multicolumn{1}{c}{$L$} &
\multicolumn{1}{c}{$\tau_E$} &
\multicolumn{1}{c}{$\tau_{\phi^2}$} &
\multicolumn{1}{c}{$\tau_{\chi}$} &
\multicolumn{1}{c}{$\tau_F$} &
\multicolumn{1}{c}{$\tau_b$} &
\multicolumn{1}{c}{trunc}  \\
\hline
\multicolumn{7}{c}{$\phi^4$ model} \\
\hline
 8 &  2.61(5)  & 2.36(4)  & 2.30(4)  & 0.79(1) & 1.67(3) &   15  \\ 
12 &  3.12(6)  & 2.72(5)  & 2.68(5)  & 0.86(2) & 1.90(4) &   18  \\
16 &  3.39(7)  & 2.85(6)  & 2.81(6)  & 0.89(2) & 1.99(4) &   20  \\
24 &  4.06(9)  & 3.25(7)  & 3.31(7)  & 0.94(2) & 2.29(5) &   25  \\  
32 &  4.54(11) & 3.51(9)  & 3.57(9)  & 0.94(2) & 2.53(6) &   30  \\
48 &  5.65(15) & 4.22(11) & 4.23(11) & 1.01(2) & 2.93(8) &   35  \\
64 &  5.89(17) & 4.36(12) & 4.26(12) & 1.01(3) & 3.08(9) &   40  \\
\hline
\multicolumn{7}{c}{dd-XY model} \\
\hline
 8 &  2.80(4)  & 1.78(3)  & 2.47(4)  & 0.78(1) & 1.79(3) &   15  \\
12 &  3.21(6)  & 2.01(4)  & 2.74(5)  & 0.81(2) & 1.91(3) &   18  \\  
16 &  3.57(7)  & 2.25(5)  & 2.97(5)  & 0.88(2) & 2.13(4) &   20  \\
24 &  4.23(10) & 2.66(6)  & 3.44(8)  & 0.93(2) & 2.42(5) &   25  \\ 
32 &  4.92(12) & 3.14(7)  & 3.83(10) & 0.99(2) & 2.75(6) &   30  \\
48 &  5.81(15) & 3.65(10) & 4.17(11) & 0.97(3) & 2.95(8) &   35  \\
64 &  6.58(19) & 4.24(12) & 4.55(13) & 1.01(3) & 3.22(9) &   40  \\ 
\end{tabular}
\end{table}

\section{Analysis of the high-temperature expansions}
\label{seriesanalysis}

In this appendix we report a detailed discussion of our HT analyses.
This detailed description should allow the reader to understand how we
determined our estimates and the reliability of the errors we report,
which are to some extent subjective.

\subsection{Definitions and HT series}
\label{HTexp}

Using the linked-cluster expansion technique, we computed the
20th-order HT expansion of the magnetic susceptibility
\begin{equation}
\chi = \sum_x \langle \phi_{\alpha}(0) \phi_{\alpha}(x) \rangle,
\label{chi}
\end{equation}
of the second moment of the two-point function
\begin{equation}
m_2 = \sum_x x^2 \langle \phi_{\alpha}(0) \phi_{\alpha}(x) \rangle,
\end{equation}
and therefore, of the second-moment correlation length
\begin{equation}
\xi^2 = {m_2\over 6\chi}.
\end{equation}
Moreover, we calculated the HT expansion of the zero-momentum
connected $2j$-point Green's functions $\chi_{2j}$
\begin{equation}
\chi_{2j} = \sum_{x_2,...,x_{2j}}
    \langle \phi_{\alpha_1}(0) \phi_{\alpha_1}(x_2) ...
        \phi_{\alpha_j}(x_{2j-1}) \phi_{\alpha_j}(x_{2j})\rangle_c
\end{equation}
($\chi = \chi_2$). More precisely, we computed $\chi_4$ to 18th order,
$\chi_6$ to 17th order, $\chi_8$ to 16th order, and $\chi_{10}$ to
15th order.  The series for the $\phi^4$ Hamiltonian with
$\lambda=2.07$ and the dd-XY model with $D=1.02$ are reported in
Tables \ref{HTexpansionsphi4} and \ref{HTexpansionsddXY}.  The HT
series of the zero-momentum four-point coupling $g_4$ and of the
coefficients $r_{2j}$ that parametrize the equation of state can be
computed using their definitions in terms of $\chi_{2j}$ and $\xi^2$,
i.e.,
\begin{table}[tp]
\caption{\label{HTexpansionsphi4}
Coefficients of the HT expansion of $m_2$, $\chi$, $\chi_4$,
$\chi_6$,, $\chi_8$, and $\chi_{10}$. They have been computed using
the $\phi^4$ Hamiltonian with $\lambda = 2.07$.
}
\begin{tabular}{cccc}
$i$& \multicolumn{1}{c}{$\mu_2$} &
     \multicolumn{1}{c}{$\chi_2$} &
     \multicolumn{1}{c}{$\chi_4$} \\
\hline
 0 &                          0 &  0.82195468340525626553069  &  
     $-$0.18234682673209145113698 \\ 
 1 & 1.01341425235775258955047  &  2.02682850471550517910095  &  
     $-$1.79856993883835218414950 \\ 
 2 & 4.99788354573054604609481  &  4.39836023278442865137831  &  
     $-$10.1455394390643744635266 \\ 
 3 & 17.0521428795050699809477  &  9.45608303384639636168661  &  
     $-$44.4379274874486579169902 \\ 
 4 & 50.2802232377961742954821  &  19.7509483421835061143003  &  
     $-$166.911247402827967481247 \\ 
 5 & 136.081955345771888598704  &  41.0953224517675210993058  &  
     $-$566.231817578861031037728 \\ 
 6 & 349.014986520228913452360  &  84.3784681474012731739965  &  
     $-$1783.81074566131415934408 \\ 
 7 & 861.072204516234675501406  &  172.831419974338990927678  &  
     $-$5314.73425458572561425498 \\ 
 8 & 2065.11115559310738635122  &  351.412467687273478895617  &  
     $-$15153.8054671243945401658 \\ 
 9 & 4843.65801296852863958594  &  713.327141375620774069366  &  
     $-$41706.6393990107185189143 \\ 
10 & 11163.1410843891753269547  &  1441.29537543678062498269  &  
     $-$111480.350225534613432415 \\ 
11 & 25357.2828059634753398090  &  2908.62198285250042113961  &  
     $-$290779.553330813045449802 \\ 
12 & 56914.8160479537305800002  &  5851.22700855642135604880  &  
     $-$742792.316757751368697227 \\ 
13 & 126448.957588634753901037  &  11759.8313973078166535503  &  
     $-$1863674.70918289087066236 \\ 
14 & 278504.794338270260511244  &  23580.6969253332410696568  &  
     $-$4603313.91229513488724197 \\ 
15 & 608775.044038619834901124  &  47248.9211902315365277418  &  
     $-$11214943.2818348822933596 \\ 
16 & 1321948.27688188944339553  &  94508.3475046413360663308  &  
     $-$26991439.1725436301683188 \\ 
17 & 2853823.94933643220583008  &  188924.811397083165660961  &  
     $-$64258568.1525362515312215 \\ 
18 & 6128960.82003386821732524  &  377150.816225492759953104  &  
     $-$151492730.104215500664052 \\ 
19 & 13101467.5362982920867821  &  752534.725866450821199491  &  
      \\ 
20 & 27889129.6761627014637264  &  1499898.13514730628043402  &  
      \\ 
\hline\hline
$i$& \multicolumn{1}{c}{$\chi_6$} &
     \multicolumn{1}{c}{$\chi_8$} &
     \multicolumn{1}{c}{$\chi_{10}$} \\
\hline
 0 & 0.41197816889670188853628  &     
     $-$2.04244608438484921752518  &  17.4966181373390399118217 \\ 
 1 & 8.09031485009238719234751  &  
     $-$65.5324533503954629382215  &  827.907846284122870554505 \\ 
 2 & 80.5579033480687931078788  &  
     $-$993.099544110386544944672  &  17583.1558423294442666385 \\ 
 3 & 569.040955700888943670354  &  
     $-$10125.9430216092445934264  &  242162.737636025061958065 \\ 
 4 & 3235.75380473965447264355  &  
     $-$79870.4410385967305759821  &  2508010.66864972545742050 \\ 
 5 & 15823.8724636340283846359  &  
     $-$524859.679867495864300189  &  21148750.0733186453907577 \\ 
 6 & 69189.8243270623365490494  &  
     $-$3005237.58408599956662949  &  152451568.855786252573448 \\ 
 7 & 277430.904691458150896928  &  
     $-$15443601.8321215251366505  &  970471959.824139222337835 \\ 
 8 & 1038008.04844006612521747  &  
     $-$72717890.9115560213984880  &  5582186363.05878798571648 \\ 
 9 & 3669720.02487068188513696  &  
     $-$318510657.462768013289640  &  29507970388.4631661994252 \\ 
10 & 12374182.7602063550019508  &  
     $-$1312668767.17043587616989  &  145203489229.461390006077 \\ 
11 & 40084338.4384102309965122  &  
     $-$5135470803.97668705323439  &  671866031736.491009927361 \\ 
12 & 125446404.887628912440242  &  
     $-$19206423870.3058981804587  &  2946715041148.83369711661 \\ 
13 & 381003987.146313729964559  &  
     $-$69057827061.9724837730205  &  12330196249913.8306690405 \\ 
14 & 1127156412.88952101681812  &  
     $-$239824993302.080594778804  &  49488831955055.0830738902 \\ 
15 & 3257906807.50792443987963  &  
     $-$807540440278.527392223891  &  191378645936343.575972510 \\ 
16 & 9223412391.97191941389912  &  
     $-$2645013087720.90303565763  &   \\ 
17 & 25631282620.7774958190658  &  
       &   \\ 
\end{tabular}
\end{table}
\begin{table}[tp]
\caption{\label{HTexpansionsddXY}
Coefficients of the HT expansion of $m_2$, $\chi$,
$\chi_4$, $\chi_6$,, $\chi_8$, and $\chi_{10}$. They have been
computed using the dd-XY Hamiltonian with $D = 1.02$.
}
\begin{tabular}{cccc}
$i$& \multicolumn{1}{c}{$\mu_2$} &
     \multicolumn{1}{c}{$\chi_2$} &
     \multicolumn{1}{c}{$\chi_4$} \\
\hline
 0 &                          0 &  0.73497259946651881066155  &  
  $-$0.12952381667498436104661 \\ 
 1 & 0.81027708294985783008938  &  1.62055416589971566017876  &  
  $-$1.14235747481325709236407 \\ 
 2 & 3.57318872366283058263222  &  3.19240289872507821851086  &  
  $-$5.86566452750871250322137 \\ 
 3 & 11.0006482367650529136739  &  6.24412164584056694746990  &  
  $-$23.4964691362678107734860 \\ 
 4 & 29.3868020693268986850123  &  11.8078991080110710061038  &  
  $-$80.6562687454557047131467 \\ 
 5 & 72.0537292098720051000559  &  22.2452167397051997713027  &  
  $-$249.761461812242731292855 \\ 
 6 & 167.391323293772704647731  &  41.3083108337880277068933  &  
  $-$717.220908333053642342640 \\ 
 7 & 373.956365385479749391430  &  76.5360051001560916440577  &  
  $-$1945.68221342894368675090 \\ 
 8 & 811.915341705911027278243  &  140.679660885119763690958  &  
  $-$5046.15791369304901147071 \\ 
 9 & 1723.52034635335232533650  &  258.178251744079668880881  &  
  $-$12622.3575801936426660898 \\ 
10 & 3594.34333565606104441219  &  471.478755943757026593209  &  
  $-$30642.5601039304090869577 \\ 
11 & 7386.64345245316356464020  &  860.010504587591344158158  &  
  $-$72548.9796473949259402584 \\ 
12 & 14997.4679478666032080397  &  1563.48753262138558216805  &  
  $-$168135.618026803729643010 \\ 
13 & 30136.9726456923668836816  &  2839.86987807993137143111  &  
  $-$382565.790192788611922604 \\ 
14 & 60029.1887398932204011553  &  5145.84503638874703629194  &  
  $-$856629.943580595646246982 \\ 
15 & 118656.316956262327168223  &  9317.67206591240832596126  &  
  $-$1891351.24076972180490224 \\ 
16 & 232979.699867454031093529  &  16841.0660076130187494565  &  
  $-$4124166.11180668899412336 \\ 
17 & 454746.664171304150538747  &  30421.5573167338465805825  &  
  $-$8893532.37374656560556900 \\ 
18 & 882960.924794534410812953  &  54875.4729390613106530869  &  
  $-$18987953.9690154439591383 \\ 
19 & 1706330.67007276458833100  &  98938.9870168970865371838  &  
      \\ 
20 & 3283569.77023650242548276  &  178182.095750601905570976  &  
      \\ 
\hline\hline
$i$& \multicolumn{1}{c}{$\chi_6$} &
     \multicolumn{1}{c}{$\chi_8$} &
     \multicolumn{1}{c}{$\chi_{10}$} \\
\hline
 0 & 0.19923804166181643967757  &  
  $-$0.67796674603175173967960  &  4.06523419917732615691360 \\ 
 1 & 3.64240617023376584468207  &  
  $-$20.6297276122379010898516  &  182.870340299506381670926 \\ 
 2 & 33.7843552454418076210054  &  
  $-$295.585300853935935335485  &  3697.42581896510915833415 \\ 
 3 & 221.913477156277495814452  &  
  $-$2838.14010017098792994188  &  48330.0439419608161526720 \\ 
 4 & 1169.53606787476515583468  &  
  $-$20984.8625853355842977237  &  472993.887593911146392516 \\ 
 5 & 5284.11768819259530573126  &  
  $-$128723.340916169008103149  &  3752687.19462636307477999 \\ 
 6 & 21286.1074543663282451107  &  
  $-$685456.078166050751079896  &  25350793.0882363704732840 \\ 
 7 & 78446.3667728434480782029  &  
  $-$3265581.01409539304748618  &  150699914.609020169056435 \\ 
 8 & 269224.069430361947223693  &  
  $-$14216412.9210915918573320  &  806981988.823436545704567 \\ 
 9 & 871590.289814654747926984  &  
  $-$57438991.7197952962871671  &  3960575199.72383885451240 \\ 
10 & 2687478.50894078411160055  &  
  $-$217927113.188388756606030  &  18052318528.0576344450433 \\ 
11 & 7951104.44962467048194209  &  
  $-$783550202.710253766125616  &  77211058033.5079429953492 \\ 
12 & 22703108.4838172682261707  &  
  $-$2689172405.44915075129067  &  312454827813.125173762158 \\ 
13 & 62855482.2396162632890035  &  
  $-$8861507312.08886862026490  &  1204410146707.71423350686 \\ 
14 & 169374342.873715022154662  &  
  $-$28171979758.2629139055303  &  4446806950469.81858484485 \\ 
15 & 445613055.420361814905254  &  
  $-$86751935057.8910676591399  &  15798640107921.6847945606 \\ 
16 & 1147647339.61141521962212  &  
  $-$259625721060.742527022681  &   \\ 
17 & 2899724764.50550187290555  &  
       &   \\ 
\end{tabular}
\end{table}
\begin{equation}
g_4 = - {3N\over N+2} {\chi_4\over \chi^2 \xi^3},
\label{grdef}
\end{equation}
and
\begin{eqnarray}
r_6 =&& 10 - {5(N+2)\over 3(N+4)}{\chi_6\chi_2\over \chi_4^2}, 
    \label{r2jgreen}\\
r_8 =&& 280 - {280 (N+2)\over 3(N+4)}{\chi_6\chi_2\over \chi_4^2} 
+{35(N+2)^2\over 9(N+4)(N+6)}{\chi_8\chi_2^2\over \chi_4^3}, \nonumber\\
r_{10} =&& 
15400  
-{7700  (N + 2)\over (N + 4)} {\chi_6 \chi_2\over \chi_4^2}        
+{ 350  (N + 2)^2\over(N + 4)^2} {\chi_6^2 \chi_2^2\over \chi_4^4} 
\nonumber \\ 
 &&+\;{1400 (N + 2)^2\over 3(N + 4)(N + 6)} {\chi_8\chi_2^2\over\chi_4^3}
    - {35 (N + 2)^3\over 3(N + 4) (N + 6) (N + 8)} 
                   {\chi_{10} \chi_2^3\over \chi_4^4}.
\nonumber
\end{eqnarray}
The formulae relevant for the XY universality class are obtained
setting $N=2$.

\subsection{Critical exponents}
\label{exponents}

In order to determine the critical exponents $\gamma$ and $\nu$ from
the HT series of $\chi$ and $\xi^2/\beta$ respectively, we used
quasi-diagonal first-, second-, and third-order integral approximants
(IA1's, IA2's and IA3's respectively).  Since the most precise results
are obtained by using the MC estimates of $\beta_c$ to bias the
approximants, we shall report only the results of the biased analyses.
The values of $\beta_c$ used in the analyses are reported in Table
\ref{betacMC}.

\begin{table}[tp]
\caption{\label{betacMC}
MC estimates of $\beta_c$.
}
\begin{tabular}{cr@{}l}
\multicolumn{1}{c}{$$}&
\multicolumn{2}{c}{$\beta_c$}\\
\tableline \hline
$\lambda=2.00$ & 0&.5099049(15) \\
$\lambda=2.07$ & 0&.5093853(24) \\
$\lambda=2.10$ & 0&.5091507(13)  \\
$\lambda=2.20$ & 0&.5083366(24)  \\
\hline
$D=0.90$   & 0&.5764582(24) \\
$D=1.02$   & 0&.5637972(21) \\
$D=1.03$   & 0&.5627975(14) \\
$D=1.20$   & 0&.5470377(26) \\
\end{tabular}
\end{table}

Given an $n$th-order series $f(\beta)= \sum_{i=0}^n c_i \beta^i$, its
$k$th-order integral approximant $[m_k/m_{k-1}/.../m_0/l]$ IA$k$ is a
solution of the inhomogeneous $k$th-order linear differential equation
\begin{equation}
P_k(\beta) f^{(k)}(\beta) + P_{k-1}(\beta) f^{(k-1)}(\beta) + ... 
+ P_1(\beta)f^{(1)}(\beta)+ P_0(\beta)f(\beta)+R(\beta)= 0,
\label{IAkdef}
\end{equation}
where the functions $P_i(\beta)$ and $R(\beta)$ are polynomials of
order $m_i$ and $l$ respectively, which are determined by the known
$n$th-order small-$\beta$ expansion of $f(\beta)$ (see, e.g., Ref.\
\cite{Guttrev}).

We consider three types of biased IA$k$'s:

(i) The first kind of biased IA$k$'s, which will be denoted by
bIA$k$'s, is obtained by setting
\begin{equation}
P_k(\beta) = \left( 1 - \beta/\beta_c \right) p_k(\beta),
\end{equation}
where $p_k(\beta)$ is a polynomial of order $m_k-1$. 

(ii) Since on bipartite lattices $\beta=-\beta_c$ is also a singular
point associated to the antiferromagnetic critical behavior
\cite{Fisher-62}, we consider IA$k$'s with
\begin{equation}
P_k(\beta) = \left( 1 - \beta^2/\beta_c^2 \right) p_k(\beta),
\end{equation}
where $p_k(\beta)$ is a polynomial of order $m_k-2$.  We will denote
them by b$_{\pm}$IA$k$'s.

(iii) Following Fisher and Chen \cite{FC-85}, we also consider IA$k$'s
where the polynomial associated with the highest derivative of
$f(\beta)$ is even, i.e., it is a polynomial in $\beta^2$.  In this
case $m_k$ is the order of the polynomial $P_k$ as a function of
$\beta^2$, i.e., $P_k\equiv \sum_{j=0}^{m_k} c_j \beta^{2j}$.  Thus,
in order to bias the singularity at $\beta_c$, we write
\begin{equation}
P_k(\beta) = \left( 1 - \beta^2/\beta_c^2 \right) p_k(\beta^2),
\end{equation}
where $p_k(\beta)$ is a polynomial in $\beta^2$ of order $m_k-1$.  We
will denote them by bFCIA$k$'s.

In our analyses we consider diagonal or quasi-diagonal approximants,
since they are expected to give the most accurate results.  Below, we
give the rules we used to select the quasi-diagonal approximants.  We
introduce a parameter $q$ that determines the degree of
off-diagonality allowed (see below).  In order to check the stability
of the results with respect to the order of the series, we also
perform analyses in which we average over the results obtained with
series of different length. For this purpose, we introduce a parameter
$p$ and perform analyses in which we use all approximants obtained
from series of $\bar{n}$ terms with $n\geq \bar{n}\geq n-p$.

We consider the following sets of IA$k$'s:

(a) $[m_1/m_0/k]$ bIA1's with
\begin{eqnarray}
&&n\geq m_1+m_0+k+1\geq n-p,\nonumber \\
&&{\rm Max}\left[\lfloor (n-1)/3 \rfloor -q, 3\right]
  \leq m_1,m_0,k \leq \lceil (n-1)/3\rceil +q.
\label{bIA1}
\end{eqnarray}

(b) $[m_1/m_0/k]$ b$_{\pm}$IA1's with
\begin{eqnarray}
&&n\geq m_1+m_0+k\geq n-p,\nonumber \\
&&{\rm Max}\left[\lfloor (n-1)/3 \rfloor -q, 3\right]
  \leq m_1,m_0,k \leq \lceil (n-1)/3\rceil +q.
\label{bbIA1}
\end{eqnarray}

(c) $[m_1/m_0/k]$ bFCIA1's with
\begin{eqnarray}
&&n\geq m_1+m_0+k+1\geq n-p,\nonumber \\
&&{\rm Max}\left[\lfloor (n-1)/3 \rfloor -q, 3\right]
  \leq m_1,m_0,k \leq \lceil (n-1)/3\rceil +q.
\label{bFCIA1}
\end{eqnarray}

(d) $[m_2/m_1/m_0/k]$ bIA2's  with 
\begin{eqnarray}
&&n\geq m_2+m_1+m_0+k+3\geq n-p,\nonumber \\
&&{\rm Max}\left[\lfloor (n-3)/4 \rfloor -q, 2\right]\leq m_2-1,m_1,m_0,k 
    \leq \lceil (n-3)/4\rceil +q.
\label{bIA2}
\end{eqnarray}

(e) $[m_2/m_1/m_0/k]$ b$_{\pm}$IA2's  with 
\begin{eqnarray}
&&n\geq m_2+m_1+m_0+k+2\geq n-p,\nonumber \\
&&{\rm Max}\left[\lfloor (n-3)/4 \rfloor -q, 2\right]\leq m_2-2,m_1,m_0,k 
    \leq \lceil (n-3)/4\rceil +q.
\label{bbIA2}
\end{eqnarray}

(f) $[m_2/m_1/m_0/k]$ bFCIA2's  with 
\begin{eqnarray}
&&n\geq m_2+m_1+m_0+k+3\geq n-p,\nonumber \\
&&{\rm Max}\left[\lfloor (n-3)/4 \rfloor -q, 2\right]\leq m_2-1,m_1,m_0,k 
    \leq \lceil (n-3)/4\rceil +q.
\label{bFCIA2}
\end{eqnarray}

(g) $[m_3/m_2/m_1/m_0/k]$ bIA3's  with 
\begin{eqnarray}
&&n\geq m_3+m_2+m_1+m_0+k+5\geq n-p,\nonumber \\
&&{\rm Max}\left[\lfloor (n-5)/5 \rfloor -q, 2\right]
    \leq m_3-1,m_2,m_1,m_0,k 
    \leq \lceil (n-5)/5\rceil +q.
\label{gaapprox}
\end{eqnarray}

In the following we fix $q=3$ for the IA1's and $q=2$ for the IA2's
and IA3's.

For each set of IA$k$'s we calculate the average of the values
corresponding to all nondefective IA$k$'s listed above. Approximants
are considered defective when they have singularities close to the
real $\beta$ axis near the critical point.  More precisely, we
consider defective those approximants that have singularities in the
rectangle
\begin{equation}
x_{\rm min} \leq {\rm Re} \, \beta/\beta_c \leq x_{\rm max},\qquad 
| {\rm Im} \, \beta/\beta_c | \leq y_{\rm max}.
\label{filter}
\end{equation}
The values of $x_{\rm min}$, $x_{\rm max}$ and $y_{\rm max}$ are fixed
essentially by stability criteria, and may differ in the various
analyses.  One should always check that the results depend very little
on the chosen values of $x_{\rm min}$, $x_{\rm max}$, and 
$y_{\rm max}$, by varying them within a reasonable and rather wide
range of values.  The domain (\ref{filter}) cannot be too large,
otherwise only few approximants are left. In this case the analysis
would be less robust and therefore less reliable.  We introduce a
parameter $s$ such that
\begin{equation}
x_{\rm min} = 1-s,\qquad
x_{\rm max} = 1+s,\qquad
y_{\rm max} = s,
\end{equation}
and we present results for various values of $s$.  We also discard
some nondefective IA's---we call them outliers---whose results are far
from the average of the other approximants. Such approximants are
eliminated algorithmically: first, we compute the average $A$ and the
standard deviation $\sigma$ of the results using all nondefective
IA's. Then, we discard those IA's whose results differ by more than
$n_\sigma\sigma$ from $A$ with $n_\sigma=2$.  We repeat the procedure
on the remaining IA's, by calculating the new $A$ and $\sigma$, but
now eliminating the IA's whose results differ by more than
$n_\sigma\sigma$ with $n_\sigma=3$. The procedure is again repeated,
increasing $n_\sigma$ by one at each step. This procedure converges
rapidly and, as we shall see, the outliers so determined are always a
very small part of the selected nondefective IA's.

In the Tables \ref{gammaphi4} and \ref{nuphi4}, we present the results
for the critical exponents $\gamma$ and $\nu$ respectively, obtained
from the HT analysis of the $\phi^4$ Hamiltonian.  In the Tables
\ref{gammaspin1} and \ref{nuspin1} we report the results for the HT
analysis of the dd-XY model.  There we also quote the ``approximant
ratio'' $r_a\equiv(g-f)/t$, where $t$ is the total number of
approximants in the given set, $g$ is the number of nondefective
approximants, and $f$ is the number of outliers which are discarded
using the above-presented algorithm; $g-f$ is the number of ``good''
approximants used in the analysis; notice that $g \gg f$, and $g-f$ is
never too small.  For each analysis, beside the corresponding
estimate, we report two numbers.  The number in parentheses, $e_1$, is
basically the spread of the approximants for $\beta_c$ fixed at the MC
estimate.  It is the standard deviation of the results obtained from
all ``good'' IA's divided by the square root of $r_a$, i.e.,
$e_1=\sigma/\sqrt{r_a}$.  Such a definition of $e_1$ is useful to
compare results obtained from different subsets of approximants of the
same type, obtained imposing different constraints.  The number in
brackets, $e_2$, is related to the uncertainty on the value of
$\beta_c$ and it is estimated by varying $\beta_c$ in the range
$[\beta_c-\Delta\beta_c,\beta_c+\Delta\beta_c]$.

\begin{table}[tp]
\caption{\label{gammaphi4}
Results for $\gamma$ obtained from the analysis of the 20th-order HT
series of $\chi$ for the $\phi^4$ Hamiltonian. The number $n$ of terms
used in the analysis is indicated explicitly when it is smaller than
the number of available terms ($n=20$). $p=0$ when its value is not
explicitly given.
}
\begin{tabular}{clcr@{}l}
\multicolumn{1}{c}{$\lambda$}&
\multicolumn{1}{c}{approximants}&
\multicolumn{1}{c}{$r_a$}&
\multicolumn{2}{c}{$\gamma$}\\
\tableline \hline
2.00
& bIA1$_{s=1/2}$ & $ (28-2)/48$      & 1&.31755(11)[19]  \\
& bIA2$_{s=1/2}$ & $ (82-5)/115$     & 1&.31749(9)[17] \\
\hline

2.07
& bIA1$_{s=1/4}$ & $ (35-3)/48$      & 1&.31786(14)[30]  \\
& bIA1$_{s=1/2}$ & $ (28-2)/48$      & 1&.31785(10)[29]  \\
& bIA1$_{s=1}$   & $ (19-1)/48$      & 1&.31784(10)[28]  \\
& bIA1$_{p=3,s=1/2}$ & $ (103-1)/172$ & 1&.31766(21)[24]  \\

& b$_{\pm}$IA1$_{s=1/4}$   & $ (36-2)/48$ & 1&.31789(20)[28]  \\
& b$_{\pm}$IA1$_{s=1/2}$   & $ (21-1)/48$ & 1&.31789(22)[28]  \\

& bFCIA1$_{s=1/4}$   & $ (36-4)/48$ & 1&.31775(10)[28]  \\
& bFCIA1$_{s=1/2}$   & $ (35-4)/48$ & 1&.31775(9)[28]  \\

& bIA2$_{s=1/8}$ & $ (99-7)/115$      & 1&.31780(9)[27] \\
& bIA2$_{s=1/4}$ & $ (93-4)/115$      & 1&.31780(9)[27] \\
& bIA2$_{s=1/2}$ & $ (87-4)/115$      & 1&.31780(8)[28] \\
& bIA2$_{s=1}$   & $ (60-2)/115$      & 1&.31781(7)[27]  \\
& bIA2$_{n=19,s=1/2}$ & $ (48-6)/70$      & 1&.31777(10)[28] \\
& bIA2$_{n=18,s=1/2}$ & $ (53-4)/62$      & 1&.31768(9)[28] \\
& bIA2$_{p=3,s=1/2}$ & $ (277-18)/345$ & 1&.31772(14)[25] \\
& bIA2$_{p=3,s=1}$ & $ (192-5)/345$    & 1&.31773(14)[25] \\
& b$_{\pm}$IA2$_{s=1/2}$ & $ (46-3)/100$ & 1&.31781(29)[23] \\
& bFCIA2$_{s=1/2}$ & $ (91-2)/140$ & 1&.31780(11)[29]   \\

& bIA3$_{s=1/2}$ & $ (56-4)/61$ & 1&.31787(8)[31]   \\
\hline

2.10
& bIA1$_{s=1/2}$ & $(29-2)/48$  & 1&.31777(9)[16]   \\
& bIA2$_{s=1/2}$ & $(92-2)/115$ & 1&.31773(6)[15]   \\
& bIA2$_{p=3,s=1/2}$ & $(295-17)/345$ & 1&.31769(10)[14]   \\
& b$_\pm$IA2$_{s=1/2}$ & $(49-5)/100$ & 1&.31774(20)[15]   \\
& bFCIA2$_{s=1/2}$ & $(92-5)/140$ & 1&.31772(15)[17]   \\
\hline

2.20
& bIA1$_{s=1/2}$ & $(31-3)/48$  & 1&.31809(7)[30]   \\
& bIA2$_{s=1/2}$ & $(94-6)/115$ & 1&.31807(3)[27]   \\
\end{tabular}
\end{table}

\begin{table}[tp]
\caption{\label{nuphi4}
Results for $\nu$ obtained from the analysis of the 19th-order HT
series of $\xi^2/\beta$ for the $\phi^4$ Hamiltonian. The number $n$
of terms used in the analysis is indicated explicitly when it is
smaller than the number of available terms ($n=19$). $p=0$ when its
value is not explicitly given.
}
\begin{tabular}{clcr@{}l}
\multicolumn{1}{c}{$\lambda$}&
\multicolumn{1}{c}{approximants}&
\multicolumn{1}{c}{$r_a$}&
\multicolumn{2}{c}{$\nu$}\\
\tableline \hline
2.00
  & bIA1$_{s=1/2}$ & $36/37$ & 0&.67140(2)[9] \\
  & bIA2$_{s=1/2}$ & $(63-6)/70$ & 0&.67141(4)[8]   \\
\hline

2.07
  & bIA1$_{s=1/2,1}$ & $36/37$ & 0&.67161(2)[13] \\
  & bIA1$_{n=18,s=1/2}$ & $30/36$ & 0&.67160(4)[13] \\
  & bIA1$_{n=17,s=1/2}$ & $(30-1)/33$ & 0&.67163(11)[12] \\
  & bIA1$_{p=3,s=1/2}$ & $(124-5)/134$ & 0&.67162(5)[12]   \\
  & bIA1$_{p=3,s=1}$   & $(120-6)/134$ & 0&.67162(5)[12]   \\
  & b$_{\pm}$IA1$_{s=1/2}$ & $(33-1)/36$ & 0&.67161(2)[13] \\
  & bFCIA1$_{s=1/2}$ & $(29-3)/37$ & 0&.67158(10)[12] \\

  & bIA2$_{s=1/2}$ & $(66-5)/70$ & 0&.67161(4)[12]   \\
  & bIA2$_{s=1}$   & $(50-3)/70$ & 0&.67162(4)[12]   \\
  & bIA2$_{n=18,s=1/2}$ & $(44-3)/62$ & 0&.67162(5)[12]   \\
  & bIA2$_{n=17,s=1/2}$ & $(38-2)/49$ & 0&.67166(4)[11]   \\
  & bIA2$_{p=3,s=1/2}$ & $(180-6)/215$ & 0&.67164(6)[12]   \\
  & bIA2$_{p=3,s=1}$ & $(145-7)/215$ & 0&.67164(5)[12]   \\
  & b$_{\pm}$IA2$_{s=1/2}$ & $(55-3)/55$ & 0&.67161(3)[13]  \\
  & bFCIA2$_{s=1/2}$ & $(60-4)/85$ & 0&.67161(11)[14]  \\

  & bIA3$_{s=1/2}$ & $(17-1)/34$ & 0&.67159(6)[14]  \\
\hline

2.10
  & bIA1$_{s=1/2}$ & $36/37$ & 0&.67160(2)[8]   \\
  & bIA2$_{s=1/2}$ & $(63-5)/70$ & 0&.67161(4)[8]   \\
\hline

2.20
  & bIA1$_{s=1/2}$ & $36/37$ & 0&.67182(3)[14]   \\
  & bIA2$_{s=1/2}$ & $(60-4)/70$ & 0&.67183(7)[14]   \\
\end{tabular}
\end{table}

\begin{table}[tp]
\caption{\label{gammaspin1}
Results for $\gamma$ obtained from the analysis of the 20th-order HT
series of $\chi$ for the dd-XY  model. $p=0$ when its value is not
explicitly given.
}
\begin{tabular}{clcr@{}l}
\multicolumn{1}{c}{$D$}&
\multicolumn{1}{c}{approximants}&
\multicolumn{1}{c}{$r_a$}&
\multicolumn{2}{c}{$\gamma$}\\
\tableline \hline
0.90
& bIA1$_{s=1/2}$ & $(35-1)/48$  & 1&.31685(29)[24] \\
& bIA2$_{s=1/2}$ & $(66-3)/115$ & 1&.31693(28)[27] \\
\hline

1.02 
& bIA1$_{s=1/4}$ & $(46-2)/48$  & 1&.31745(20)[21] \\
& bIA1$_{s=1/2}$ & $(41-3)/48$  & 1&.31746(17)[22] \\
& bIA1$_{s=1}$   & $(24-1)/48$  & 1&.31748(15)[23]  \\
& bIA1$_{p=3,s=1/2}$ & $(162-9)/172$ & 1&.31733(35)[20] \\
& b$_{\pm}$IA1$_{s=1/2}$ & $(35-2)/48$  & 1&.31735(13)[22]  \\
& bFCIA1$_{s=1/2}$ & $(31-4)/48$  & 1&.31745(21)[22] \\

& bIA2$_{s=1/4}$ & $(103-3)/115$ & 1&.31748(25)[23]  \\
& bIA2$_{s=1/2}$ & $(68-1)/115$  & 1&.31748(16)[22] \\
& bIA2$_{s=1}$   & $(22-1)/115$     & 1&.31754(20)[20]  \\
& bIA2$_{p=3,s=1/2}$ & $(259-7)/345$ & 1&.31730(26)[19]  \\
& b$_{\pm}$IA2$_{s=1/2}$    & $(74-3)/100$ & 1&.31754(26)[22]  \\
& bFCIA2$_{s=1/2}$ & $(69-4)/140$  & 1&.31738(38)[18]  \\

& bIA3$_{s=1/2}$   & $(41-1)/61$     & 1&.31776(19)[24]  \\
\hline

1.03
& bIA1$_{s=1/2}$ & $(40-3)/48$  & 1&.31747(15)[14] \\
& bIA2$_{s=1/2}$  & $(71-1)/115$ & 1&.31751(13)[15]  \\
& bIA2$_{p=3,s=1/2}$ & $(263-11)/345$ & 1&.31733(22)[13]  \\
& b$_{\pm}$IA2$_{s=1/2}$    & $(73-2)/100$ & 1&.31756(24)[13]  \\
& bFCIA2$_{s=1/2}$ & $(72-4)/140$ & 1&.31744(27)[11] \\
& bIA3$_{s=1/2}$ & $(41-2)/61$ & 1&.31776(16)[16]  \\
\hline

1.20
& bIA1$_{s=1/2}$ & $(43-1)/48$   & 1&.31867(20)[28]   \\
& bIA2$_{s=1/2}$ & $(99-4)/115$  & 1&.31868(10)[25] \\
\end{tabular}
\end{table}

\begin{table}[tp]
\caption{\label{nuspin1}
Results for $\nu$ obtained from the analysis of the 19th-order HT
series of $\xi^2/\beta$ for the dd-XY model.  $n=19$ and $p=0$ when
not explicitly given.
}
\begin{tabular}{clcr@{}l}
\multicolumn{1}{c}{$D$}&
\multicolumn{1}{c}{approximants}&
\multicolumn{1}{c}{$r_a$}&
\multicolumn{2}{c}{$\nu$}\\
\tableline \hline
0.90
& bIA1$_{s=1/2}$ & $(33-2)/37$  & 0&.67091(6)[12]  \\
& bIA2$_{s=1/2}$ & $(62-3)/70$  & 0&.67092(10)[12]  \\
\hline

1.02
& bIA1$_{s=1/2}$ & $(35-3)/37$  & 0&.67146(7)[10]  \\
& bIA1$_{s=1}$   & $(30-1)/37$  & 0&.67147(5)[10]  \\
& bIA1$_{n=18,s=1/2}$ & $(32-1)/36$  & 0&.67148(15)[10]  \\
& bIA1$_{n=17,s=1/2}$ & $(31-2)/33$  & 0&.67132(28)[9]  \\
& bIA1$_{p=3,s=1/2}$ & $(124-11)/134$  & 0&.67143(12)[10]   \\
& bIA1$_{p=3,s=1}$ & $(103-9)/134$  & 0&.67145(10)[10]   \\
& b$_{\pm}$IA1$_{s=1/2}$ & $(34-1)/36$  & 0&.67143(5)[11]   \\
& bFCIA1$_{s=1/2}$ & $(33-3)/37$  & 0&.67136(12)[10]   \\

& bIA2$_{s=1/2}$ & $(64-1)/70$  & 0&.67144(5)[10]  \\
& bIA2$_{s=1}$ & $(55-3)/70$  & 0&.67145(4)[10]  \\
& bIA2$_{n=18,s=1/2}$ & $(54-2)/62$  & 0&.67145(11)[10]  \\
& bIA2$_{n=17,s=1/2}$ & $(48-5)/49$  & 0&.67137(3)[9]  \\
& bIA2$_{p=3,s=1/2}$ & $(198-9)/215$  & 0&.67141(8)[9]  \\
& b$_{\pm}$IA2$_{s=1/2}$ & $(53-9)/55$  & 0&.67144(2)[10]   \\
& bFCIA2$_{s=1/2}$ & $(78-8)/85$  & 0&.67140(6)[11]   \\

& bIA3$_{s=1/2}$ & $(34-4)/34$  & 0&.67149(5)[10]   \\
\hline

1.03 
& bIA1$_{s=1/2}$ & $(34-2)/37$  & 0&.67149(7)[8]   \\
& bIA2$_{s=1/2}$ & $(67-4)/70$  & 0&.67147(5)[7]   \\
\hline

1.20
& bIA1$_{s=1/2}$ & $(30-1)/37$  & 0&.67231(12)[13]   \\
& bIA1$_{s=1/2}$ & $(64-4)/70$  & 0&.67236(7)[12]   \\
\end{tabular}
\end{table}

The results of the analyses are quite stable: all sets of IA's give
substantially consistent results.  The comparison of the results
obtained using all available terms of the series with those using less
terms (in the Tables the number of terms is indicated explicitly when
it is smaller than the number of available terms) and those obtained
for $p=3$ (i.e., using $n,n-1,n-2$, and $n-3$ terms in the series)
shows that the results are also stable with respect to the order of
the HT series.  Therefore, we do not need to perform problematic
extrapolations in the number of terms, or rely on phenomenological
arguments, typically based on other models, suggesting when the number
of terms is sufficient to provide a reliable estimate.

From the intermediate results reported in Tables \ref{gammaphi4},
\ref{nuphi4}, \ref{gammaspin1} and \ref{nuspin1} (which, we stress,
are determined algorithmically once chosen the set of IA$k$'s),
we obtain the estimates of $\gamma$ and $\nu$.

From the analyses for the $\phi^4$ Hamiltonian at $\lambda= 2.07$, we
obtain
\begin{eqnarray}
&&\gamma = 1.31780(10)[28] + 0.003 (\lambda - 2.07),\label{gaphi41}\\
&&\nu    = 0.67161(5)[12] + 0.002 (\lambda - 2.07).\label{nuphi41}
\end{eqnarray}
As before, the number between parentheses is basically the spread of
the approximants at $\lambda=2.07$ using the central value of
$\beta_c$, while the number between brackets gives the systematic
error due to the uncertainty on $\beta_c$. Eqs.\ (\ref{gaphi41}) and
(\ref{nuphi41}) show also the dependence of the results on the chosen
value of $\lambda$.  The coefficient is estimated from the
results for $\lambda=2.2$ and $\lambda=2.0$, i.e., from the ratio
\begin{equation}
{Q(\lambda=2.2) - Q(\lambda=2.0)\over 0.2},
\end{equation}
where $Q$ represents the quantity at hand.  Using $\lambda^*=2.07(5)$,
we obtain finally
\begin{equation}
\gamma = 1.31780(10)[28]\{15\},\qquad\qquad \nu    = 0.67161(5)[12]\{10\},
\label{phi4exp}
\end{equation}
where the error due to the uncertainty on $\lambda^*$ is reported
between braces.

Since for $\lambda = 2.10$ a more precise estimate of $\beta_c$ is
available, it is interesting to perform the same analysis, using the
HT series of the $\phi^4$ model at $\lambda = 2.10$.  We obtain
\begin{eqnarray}
&&\gamma = 1.31773(10)[15] + 0.003 (\lambda - 2.10),\label{gaphi41b}\\
&&\nu    = 0.67160(5)[8] + 0.002 (\lambda - 2.10),\label{nuphi41b}
\end{eqnarray}
which, using $\lambda^*=2.07(5)$, give
\begin{equation}
\gamma = 1.31764(10)[15]\{15\},\qquad\qquad \nu    = 0.67154(5)[8]\{10\},
\label{phi4expb}
\end{equation}
in perfect agreement with the results obtained at $\lambda=2.07$.  The
slight difference of the central values is essentially due to the
independent estimates of $\beta_c$.

From the analyses for the dd-XY model at $D= 1.02$, we have
\begin{eqnarray}
&&\gamma = 1.31748(20)[22] + 0.006  (D-1.02),\label{gas1}\\
&&\nu    = 0.67145(10)[10] + 0.005  (D-1.02),\label{nus1}
\end{eqnarray}
where the coefficient determining the dependence of the results on $D$
is estimated by computing
\begin{equation}
{Q(D=1.2) - Q(D=0.9)\over 0.3}.
\end{equation}
Since $D^*=1.02(3)$, we obtain the final estimates
\begin{equation}
\gamma = 1.31748(20)[22]\{18\},\qquad\qquad \nu    = 0.67145(10)[10]\{15\}.
\label{spin1exp}
\end{equation}
Since for $D=1.03$ a more precise estimate of $\beta_c$ is available,
it is worthwhile to repeat the analysis using the series at this value
of $D$. We have
\begin{eqnarray}
&&\gamma = 1.31751(20)[15] + 0.006 (D - 1.03),\label{gas41b}\\
&&\nu    = 0.67148(10)[8] + 0.005 (D- 1.03),\label{nus41b}
\end{eqnarray}
which, for $D^*=1.02(2)$, give
\begin{equation}
\gamma = 1.31745(20)[15]\{18\},\qquad\qquad \nu    = 0.67143(10)[8]\{15\},
\label{ddexpb}
\end{equation}
in good agreement with the results (\ref{spin1exp}).

Consistent, although significantly less precise, results are obtained
from IHT analyses that do not make use of the MC estimate of
$\beta_c$.  For example, by analyzing the HT series for the $\phi^4$
Hamiltonian at $\lambda=2.07$, we find $\beta_c=0.509385(8)$,
$\gamma=1.3178(8)\{3\}$, $\nu=0.6716(4)\{1\}$, where the error in
parentheses is the spread of the approximants and the error between
braces corresponds to the uncertainty on $\lambda^*$.  Here, we
determine $\beta_c$ and $\gamma$ from the analysis of $\chi$, using
IA2's, FCIA2's, and IA3's, and $\nu$ from the analysis of $\xi^2$
using bIA2's biased with the estimate of $\beta_c$ obtained in the HT
analysis of $\chi$.

From the results for $\gamma$ and $\nu$, one can obtain $\eta$ by the
scaling relation $\gamma=(2-\eta)\nu$. This gives $\eta=0.0379(10)$,
where the error is estimated by considering the errors on $\gamma$ and
$\nu$ as independent, which is of course not true.  We can obtain an
estimate of $\eta$ with a smaller, yet reliable, error using the
so-called critical-point renormalization method (CPRM) (see Ref.\ 
\cite{int-appr-ref} and references therein).  In the CPRM, given two
series $D(x)$ and $E(x)$ that are singular at the same point $x_0$,
$D(x)=\sum_i d_i x^i\sim (x_0-x)^{-\delta}$ and 
$E(x)=\sum_i e_i x^i\sim (x_0-x)^{-\epsilon}$, one constructs a new
series $F(x)=\sum_i (d_i/e_i)x^i$.  The function $F(x)$ is singular at
$x=1$ and for $x\to 1$ behaves as $F(x)\sim (1-x)^{-\phi}$, where
$\phi = 1+ \delta - \epsilon$.  Therefore, the difference
$\delta-\epsilon$ can be obtained by analyzing the expansion of $F(x)$
by means of biased approximants with a singularity at $x_c=1$.  In
order to check for possible systematic errors, we applied the CPRM to
the series of $\xi^2/\beta$ and $\chi$ (analyzing the corresponding
19th-order series) and to the series of $\xi^2$ and $\chi$ (analyzing
the corresponding 20th-order series).  We used IA's biased at $x_c=1$.
In Table \ref{eta} we present the results of several sets of IA's.
For the $\phi^4$ model at $\lambda=2.07$ we obtain
\begin{equation}
\eta\nu = 0.02550(20) + 0.0013 (\lambda-2.07). 
\end{equation}
Thus, taking into account that $\lambda^*=2.07(5)$, we find
\begin{equation}
\eta\nu = 0.02550(20)\{7\},
\end{equation}
where the first error is related to the spread of the IA's and the
second one to the uncertainty on $\lambda^*$, evaluated as before.
Analogously, for the dd-XY model we find
\begin{equation}
\eta\nu = 0.02550(40) + 0.004 (D-1.02), 
\end{equation}
and therefore, using $D^*=1.02(3)$,
\begin{equation}
\eta\nu = 0.02550(40)\{12\},
\end{equation}
where again the first error is related to the spread of the IA's,
while the second one is related to the uncertainty on $D^*$.

\begin{table}[tp]
\caption{\label{eta}
Results for $\eta$ obtained using the CPRM:
$(a)$ applied to $\xi^2/\beta$ and $\chi$ (19 orders);
$(b)$ applied to $\xi^2$ and $\chi$ (20 orders).
}
\begin{tabular}{clcr@{}l}
\multicolumn{1}{c}{$$}&
\multicolumn{1}{c}{approximants}&
\multicolumn{1}{c}{$r_a$}&
\multicolumn{2}{c}{$\eta\nu$}\\
\tableline \hline
$\lambda=2.00$
& (a) bIA1$_{s=1/2}$ & $33/37$  & 0&.02547(7)  \\
& (a) bIA2$_{s=1/2}$ & $(47-1)/70$  & 0&.0256(2)  \\
& (b) bIA2$_{s=1/2}$ & $(99-9)/115$  & 0&.0251(3)  \\
\hline

$\lambda=2.07$
& (a) bIA1$_{s=1/2}$ & $37/37$  & 0&.02555(7)  \\
& (a) bIA2$_{s=1/2}$ & $47/70$  & 0&.0257(2)  \\
& (a) bIA3$_{s=1/2}$ & $(20-1)/34$  & 0&.0255(2)  \\

& (b) bIA2$_{s=1/2}$ & $(96-8)/115$  & 0&.0252(3)  \\
& (b) bIA3$_{s=1/2}$ & $(51-2)/61$  & 0&.0253(5)  \\
\hline

$\lambda=2.20$
& (a) bIA1$_{s=1/2}$ & $33/37$  & 0&.02573(7)  \\
& (a) bIA2$_{s=1/2}$ & $(49-2)/70$  & 0&.0259(3)  \\
& (b) bIA2$_{s=1/2}$ & $(95-11)/115$  & 0&.0252(3)  \\
\hline\hline

$D=0.90$
& (a) bIA2$_{s=1/2}$ & $(45-1)/70$  & 0&.0252(3)  \\
& (b) bIA2$_{s=1/2}$ & $(84-3)/115$  & 0&.0248(9)  \\
\hline

$D=1.02$
& (a) bIA1$_{s=1/2}$ & $(22-1)/37$  & 0&.0256(9)  \\
& (a) bIA2$_{s=1/2}$ & $(37-1)/70$  & 0&.0257(3)  \\
& (a) bIA3$_{s=1/2}$ & $(23-2)/34$  & 0&.0252(5)  \\

& (b) bIA2$_{s=1/2}$ & $(93-3)/115$ & 0&.0252(8)  \\
& (b) bIA3$_{s=1/2}$ & $(59-3)/61$  & 0&.0253(4)  \\
\hline

$D=1.20$
& (a) bIA2$_{s=1/2}$ & $(33-2)/70$  & 0&.0263(3)  \\
& (b) bIA2$_{s=1/2}$ & $(96-4)/115$  & 0&.0259(8)  \\
\end{tabular}
\end{table}

\subsection{Amplitude ratios}
\label{ratioofamp}

In the following we describe the analysis method we employed to
evaluate zero-momentum renormalized couplings, such as $g_4$ and
$r_{2j}$.  In the case of $g_4$ we analyzed the series $\beta^{3/2}
g_4 =\sum_{i=0}^{17} a_i\beta^i$.

Consider an amplitude ratio $A$ which, for
$t\equiv\beta_c/\beta-1\to0$, behaves as
\begin{equation}
A(t) = A^* + c_1 t^\Delta + c_2 t^{\Delta_2} + ... \, .
\end{equation}
In order to determine $A^*$ from the HT series of $A(t)$, we consider
biased IA1's, whose behavior at $\beta_c$ is given by (see, e.g.,
Ref.\ \cite{int-appr-ref})
\begin{equation}
{\rm IA1} \approx 
f(\beta) \left(1 - \beta/\beta_c\right)^{\zeta} + g(\beta),
\label{IA1bh}
\end{equation}
where $f(\beta)$ and $g(\beta)$ are regular at $\beta_c$, except when
$\zeta$ is a non-negative integer.  In particular,
\begin{equation}
\zeta = {P_0(\beta_c)\over P_1'(\beta_c)},\qquad\qquad 
    g(\beta_c) = - {R(\beta_c)\over P_0(\beta_c)}.
\label{IA1bhf}
\end{equation}
In the case we are considering, $\zeta$ is positive and therefore,
$g(\beta_c)$ provides an estimate of $A^*$.  Moreover, for improved
Hamiltonians we expect $\zeta =\Delta_2 \approx 2 \Delta$ and 
$\Delta \approx 0.5$.  In our analyses we consider bIA1's and
b$_{\pm}$IA1's (see Eqs.\ (\ref{bIA1}) and (\ref{bbIA1})) and impose
various constraints on the value of $\zeta$ by selecting bIA1's with
$\zeta$ larger than a given nonnegative value.

In Table \ref{g4} we report the results obtained for $g_4$ using
different sets of approximants.  In this case the variation due to the
uncertainty of $\beta_c$ is negligible. Therefore, we report only the
average of the results of the ``good'' IA1's and their standard
deviation (divided by $\sqrt{r_a}$) calculated at $\beta_c$. In Table
\ref{g4} we also report the value of $\zeta$ obtained from the
selected IA1's.  The comparison of the results for different values of
$\lambda$ and $D$ shows that the errors due to uncertainty on
$\lambda^*$ and $D^*$ are small and negligible.

\begin{table}[tp]
\caption{\label{g4}
Results for $g_4$, obtained from the analysis of the 17th-order series
of $\beta^{-3/2}g_4(\beta)$.
}
\begin{tabular}{clcr@{}lr@{}l}
\multicolumn{1}{c}{}&
\multicolumn{1}{c}{approximants}&
\multicolumn{1}{c}{$r_a$}&
\multicolumn{2}{c}{$g_4$}&
\multicolumn{2}{c}{$\zeta$}\\
\tableline \hline
$\lambda=2.00$
& bIA1$_{s=1/4,\zeta > 0}$ & $(41-1)/43$  & 21&.16(7)  & 1.&1(5) \\
& b$_{\pm}$IA1$_{s=1/4,\zeta >0}$ & $(40-2)/44$  & 21&.14(6)  & 1.&3(8) \\
\hline

$\lambda=2.07$
& bIA1$_{s=1/10,1/4,\zeta > 0}$ & $(41-2)/43$  & 21&.17(6)  & 1.&2(5) \\
& bIA1$_{s=1/4,\zeta > 0.5}$ & $(38-2)/43$  & 21&.16(6)  & 1.&2(5) \\
& bIA1$_{s=1/4,1.3> \zeta > 0.7}$ & $23/43$  & 21&.19(5)  & 1.&0(2) \\
& bIA1$_{p=2,s=1/4,\zeta > 0}$ & $(105-4)/118$  & 21&.14(7)  & 1.&7(8) \\

& b$_{\pm}$IA1$_{s=1/10,1/4,\zeta > 0}$ & $(40-3)/44$ & 21&.14(5) & 1.&4(9) \\
& b$_{\pm}$IA1$_{s=1/4,\zeta > 0.5}$ & $(39-2)/44$  & 21&.14(5)  & 1.&4(9) \\
& b$_{\pm}$IA1$_{s=1/4,1.3>\zeta > 0.7}$ & $(20-1)/44$ & 21&.16(3) & 1.&1(1) \\
& b$_{\pm}$IA1$_{p=2,s=1/4,\zeta > 0}$ & $(80-3)/97$ & 21&.13(7) & 2&(2) \\
\hline

$\lambda=2.10$
& bIA1$_{s=1/4,\zeta > 0}$ & $(41-2)/43$  & 21&.17(6)  & 1.&2(5) \\
& bIA1$_{s=1/4,\zeta > 0.5}$ & $(38-1)/43$  & 21&.17(6)  & 1.&2(5) \\
& bIA1$_{s=1/4,1.3 > \zeta > 0.7}$ & $23/43$  & 21&.19(5)  & 1.&0(2) \\
& bIA1$_{p=2,s=1/4,\zeta > 0}$ & $(105-4)/118$ & 21&.14(7) & 1.&7(8) \\

& b$_{\pm}$IA1$_{s=1/4,\zeta > 0}$ & $(39-2)/44$ & 21&.16(6) & 1.&4(6) \\
& b$_{\pm}$IA1$_{p=2,s=1/4,\zeta > 0}$ & $(79-3)/97$ & 21&.14(7) & 2&(2) \\
\hline

$\lambda=2.20$ 
& bIA1$_{s=1/4,\zeta >0}$ & $(40-2)/43$  & 21&.19(5)  & 1.&3(6) \\
& b$_{\pm}$IA1$_{s=1/4,\zeta >0}$ & $(39-2)/44$ & 21&.17(5) & 1.&5(9) \\
\hline\hline

$D=0.90$
& bIA1$_{s=1/4,\zeta > 0}$ & $(32-1)/43$  & 21&.07(12)  & 2&(2) \\
& b$_{\pm}$IA1$_{s=1/4,\zeta > 0}$ & $(35-5)/44$ & 21&.07(9) & 1.&1(4) \\
\hline

$D=1.02$
& bIA1$_{s=1/10,\zeta > 0}$ & $(31-2)/43$  & 21&.16(10)  & 1.&5(1.4) \\
& bIA1$_{s=1/4,\zeta > 0}$ & $(30-2)/43$  & 21&.16(10)  & 1.&5(1.4) \\
& bIA1$_{s=1/4,\zeta > 0.5}$ & $(24-1)/43$  & 21&.13(6)  & 1.&7(1.4) \\
& bIA1$_{s=1/4,1.3>\zeta > 0.7}$ & $9/43$  & 21&.16(7)  & 0.&9(2) \\
& bIA1$_{p=2,s=1/4,\zeta > 0}$ & $(69-8)/118$  & 21&.2(3)  & 1.&6(1.3) \\

& b$_{\pm}$IA1$_{s=1/4,\zeta >0}$ & $(36-3)/44$ & 21&.13(7) & 1.&5(1.0) \\
& b$_{\pm}$IA1$_{s=1/4, \zeta > 0.5}$ & $(32-4)/44$ & 21&.11(5) & 1.&5(8) \\
& b$_{\pm}$IA1$_{s=1/4,1.3>\zeta>0.7}$ & $(16-1)/44$ & 21&.13(3) & 1.&1(1) \\
& b$_{\pm}$IA1$_{p=2,s=1/4,\zeta > 0}$ & $(66-7)/97$ & 21&.13(12) & 2&(2) \\
\hline

$D=1.03$
& bIA1$_{s=1/4,\zeta > 0}$ & $(32-3)/43$  & 21&.17(12)  & 1.&5(1.5) \\
& b$_{\pm}$IA1$_{s=1/4,\zeta>0}$ & $(36-3)/44$ & 21&.13(6) & 1.&6(1.1) \\
\hline

$D=1.20$
& bIA1$_{s=1/4,\zeta > 0}$ & $(34-4)/43$  & 21&.23(3)  & 3&(3) \\
& b$_{\pm}$IA1$_{s=1/4,\zeta > 0}$ & $(33-2)/44$  & 21&.23(3)  & 2&(2) \\
\end{tabular}
\end{table}

From the results of Table \ref{g4} we derive the estimates 
\begin{equation}
g_4 = 21.15(6),\qquad\qquad g_4 = 21.13(7),
\end{equation}
respectively for the $\phi^4$ Hamiltonian and the dd-XY model.  We
note that these results are slightly larger than the estimates
reported in Ref.\ \cite{CPRV-00-2}.  The difference is essentially due
to the different analysis employed.  There, the analysis was based on
Pad\'e (PA), Dlog-Pad\'e (DPA) and IA1's, selecting those without
singularities in a neighborhood of $\beta_c$ and evaluating them at
$\beta_c$.  However, by analyzing the longer series that are now
available for the Ising model \cite{Campostrini-00}, we have realized
that such procedure is not very accurate and that the analyses using
bIA1's are more reliable when a sufficiently large number of terms is
available.  Moreover, when the series is sufficiently long, most (and
eventually all) PA's, DPA's and IA1's become defective.  Indeed, the
functions we are considering do {have} singularities at $\beta_c$,
although with a positive exponent.

In the analysis of $r_{2j}$, we also consider PA's and DPA's.  We
indeed expect that, when the series is not sufficiently long to be
asymptotic, the approximants obtained by biasing the singularity at
$\beta_c$ may not provide a robust analysis. For comparison, we also
use quasi-diagonal Pad\'e approximants (PA's) and Dlog-Pad\'e
approximants (DPA's), evaluating them at $\beta_c$.

We consider:

$[l/m]$ PA's with 
\begin{eqnarray}
&&l+m \geq n-2, \nonumber \\
&&{\rm Max}\left[ n/2-q,4\right] \leq l,m \leq  n/2 +q,
\label{paapprox}
\end{eqnarray}
where $l,m$ are the orders of the polynomials respectively in the
numerator and denominator of the PA.  As estimate from the PA's we
take the average of the values at $\beta_c$ of the nondefective
approximants using all the available terms of the series and
satisfying the condition (\ref{paapprox}) with $q=3$.  The error we
quote is the standard deviation (divided by $\sqrt{r_a}$) of the
results from all the nondefective approximants listed above.  We
consider defective those PA's that have singularities in the rectangle
defined in Eq.\ (\ref{filter}) with: $x_{\rm min} = 0.9$, 
$x_{\rm max} = 1.01$, and $y_{\rm max} = 0.1$ for $r_6$ and $r_8$;
$x_{\rm min}=0$, $x_{\rm max} = 1.1$, and $y_{\rm max} = 0.1$ for
$r_{10}$.

$[l/m]$ DPA's with
\begin{eqnarray}
&&l+m \geq n-3,\nonumber\\
&&{\rm Max}\left[(n-1)/2-q,4\right] \leq l,m \leq  (n-1)/2+q,
\end{eqnarray}
where $l,m$ are the orders of the polynomials respectively in the
numerator and denominator of the PA of the series of its logarithmic
derivative.  We again fix $q=3$.  The estimate with the
corresponding error is obtained as in the case of PA's. We consider
defective those DPA's that have singularities in the rectangle
(\ref{filter}) with $x_{\rm min} = 0$, $x_{\rm max} = 1.01$, and
$y_{\rm max} = 0.1$.

For $r_6$ and $r_8$ the above PA's and DPA's give results
substantially consistent with those of bIA1's, see Tables \ref{r6} and
\ref{r8}.  Therefore the systematic error due to the use of PA's
and DPA's should be smaller than the final quoted errors.  From
Tables~\ref{r6} and \ref{r8} we derive the estimates reported in
Table~\ref{finalres2}.  For $r_{10}$ we obtain only very rough
estimates using essentially PA's: $r_{10}=-13(7)$ from the $\phi^4$
Hamiltonian and $r_{10}=-11(14)$ from the dd-$XY$ model.

\begin{table}[tp]
\caption{\label{r6}
Results for $r_6$ from the analysis of the corresponding 17th-order
HT series.
}
\begin{tabular}{clcr@{}l}
\multicolumn{1}{c}{}&
\multicolumn{1}{c}{approximants}&
\multicolumn{1}{c}{$r_a$}&
\multicolumn{2}{c}{$r_6$}\\
\tableline \hline
$\lambda=2.00$
& PA & $19/21$  & 1&.945(20)  \\
& DPA & $12/18$  & 1&.959(27)  \\
\hline

$\lambda=2.07$
& PA & $19/21$  & 1&.947(22)  \\
& DPA & $12/18$  & 1&.962(26)  \\

& bIA1$_{s=1/10,\zeta>0}$ & $(21-1)/43$  & 1&.98(10)  \\
& bIA1$_{s=1/10,\zeta>0.5}$ & $(17-1)/43$  & 1&.955(9)  \\
& bIA1$_{p=2,s=1/10,\zeta> 0}$ & $(58-3)/118$  & 1&.99(9) \\
& bIA1$_{p=2,s=1/10,\zeta> 0.5}$ & $(42-2)/118$  & 1&.956(10) \\
\hline  

$\lambda=2.10$
& PA & $19/21$  & 1&.948(23)  \\
& DPA & $12/18$  & 1&.963(26)  \\

& bIA1$_{s=1/10,\zeta> 0 }$ & $(21-1)/43$  & 1&.98(7) \\
& bIA1$_{s=1/10,\zeta>0.5}$ & $(18-1)/43$  & 1&.959(19)  \\
\hline

$\lambda=2.20$
& PA & $19/21$  & 1&.951(30)  \\
& DPA & $13/18$  & 1&.967(26)  \\
\hline\hline

$D=0.90$
& PA & $21/21$  & 1&.934(15)  \\
& DPA & $16/18$  & 1&.936(11)  \\
\hline

$D=1.02$
& PA & $20/21$  & 1&.945(13)  \\
& DPA & $12/18$  & 1&.952(13)  \\

& bIA1$_{s=1/10,\zeta>0}$    & $(28-1)/43$   & 1&.948(11)   \\ 
& bIA1$_{s=1/10,\zeta>0.5}$ & $(27-1)/43$   & 1&.947(8)   \\ 
& bIA1$_{p=2,s=1/10,\zeta > 0}$ & $(69-1)/118$   & 1&.947(17)  \\
& bIA1$_{p=2,s=1/10,\zeta >0.5}$ & $(65-1)/118$   & 1&.948(16)  \\
\hline  

$D=1.03$
& PA & $20/21$  & 1&.946(8)  \\
& DPA & $12/18$  & 1&.954(16)  \\
\hline

$D=1.20$
& PA & $21/21$  & 1&.963(6)  \\
& DPA & $16/18$  & 1&.963(12)  \\
\end{tabular}
\end{table}

\begin{table}[tp]
\caption{\label{r8}
Results for  $r_8$ from the analysis of the corresponding 16-th order
HT series.
}
\begin{tabular}{clcr@{}l}
\multicolumn{1}{c}{}&
\multicolumn{1}{c}{approximants}&
\multicolumn{1}{c}{$r_a$}&
\multicolumn{2}{c}{$r_8$}\\
\tableline \hline
$\lambda=2.00$
& PA & $17/18$  & 1&.32(14)  \\
& DPA & $14/21$  & 1&.34(22)  \\
\hline

$\lambda=2.07$
& PA & $17/18$  & 1&.36(14)  \\
& DPA & $13/21$  & 1&.37(15)  \\

& bIA1$_{s=1/10,\zeta>0}$ & $(10-1)/33$  & 1&.39(13)  \\
& bIA1$_{p=2,s=1/10,\zeta>0}$ & $(33-2)/91$  & 1&.30(15)  \\
\hline

$\lambda=2.10$
& PA & $18/18$  & 1&.35(12)  \\
& DPA & $13/21$  & 1&.38(12)  \\
\hline

$\lambda=2.20$
& PA & $18/18$  & 1&.41(12)  \\
& DPA & $13/21$  & 1&.42(17)  \\
\hline\hline

$D=0.90$
& PA & $16/18$  & 1&.34(18)  \\
& DPA & $18/21$  & 1&.28(10)  \\
\hline

$D=1.02$
& PA & $17/18$  & 1&.50(14)  \\
& DPA & $15/21$  & 1&.47(6)  \\

& bIA1$_{s=1/10,\zeta>0}$ & $(15-1)/33$  & 1&.45(12) \\
& bIA1$_{p=2,s=1/10,\zeta>0}$ & $(45-5)/91$  & 1&.50(33) \\
\hline

$D=1.03$
& PA & $17/18$  & 1&.52(13)  \\
& DPA & $15/21$  & 1&.54(8)  \\
\hline

$D=1.20$
& PA & $16/18$  & 1&.64(9)  \\
& DPA & $15/21$  & 1&.65(13)  \\
\end{tabular}
\end{table}

\section{Universal amplitude ratios}
\label{univra}

We give here the definitions of the amplitude ratios that are used in
the text.  They are expressed in terms of the amplitudes derived from
the singular behavior of the specific heat
\begin{equation}
C_H = A^{\pm} |t|^{-\alpha},
\label{sphamp}
\end{equation}
the magnetic susceptibility in the high-temperature phase
\begin{equation}
\chi = {1\over 2} C^{+} t^{-\gamma},
\label{chiamp}
\end{equation}
the zero-momentum four-point connected correlation function in the
high temperature phase
\begin{equation}
\chi_4 = {8\over 3} C_4^+ t^{-\gamma-2\beta\delta},
\label{chinamp}
\end{equation}
the second-moment correlation length in the high-temperature phase
\begin{equation}
\xi = f^{+} t^{-\nu},
\label{xiamp}
\end{equation}
the spontaneous magnetization on the coexistence curve
\begin{equation}
M = B |t|^{\beta},
\label{magamp}
\end{equation}
and of the susceptibility along the critical isotherm
\begin{equation}
\chi_L = C^c |H|^{-{\gamma/\beta\delta}}. \label{chicris}
\end{equation}
We consider the following universal amplitude ratios:
\begin{eqnarray}
&& R_c \equiv {\alpha A^+ C^+\over B^2} ,\\
&& R_4 \equiv  - {C_4^+ B^2\over (C^+)^3} = |z_0|^2, \\
&& R_\chi \equiv {C^+ B^{\delta-1}\over (\delta C^c)^\delta} ,\\
&& R_\xi^+ \equiv (A^+)^{1/3} f^+= \left( {R_4 R_c\over g_4}\right)^{1/3}.
\end{eqnarray}

\end{document}